\documentclass[fleqn,usenatbib]{mnras}

\usepackage{natbib}
\usepackage{newtxtext,newtxmath}

\usepackage[T1]{fontenc}
\usepackage{ae,aecompl}

\usepackage[switch, modulo]{lineno}

\usepackage{graphicx}	
\usepackage{hyperref}
\usepackage{footmisc}

\graphicspath{ {images/} }

\def\blue{\color{black}}

\title[PAUS image photometry]{The PAU Survey: Narrow-band image photometry}

\author[Serrano et al.]{
S.~Serrano$^{1,2,3}$\thanks{E-mail: serrano@satlantis.com},
E.~Gazta\~naga$^{1,2,4}$,
F.~J.~Castander$^{1,2}$,
M.~Eriksen$^{5}$\thanks{Also at Port d'Informaci\'{o} Cient\'{i}fica (PIC), Campus UAB, C. Albareda s/n, 08193 Bellaterra (Cerdanyola del Vall\`{e}s), Spain},
R.~Casas$^{1,2}$,
\newauthor
D.~Navarro-Giron\'es$^{1,2}$,
A.~Alarcon$^{6}$,
A.~Bauer$^{1,2}$,
L.~Cabayol$^{5}$,
J.~Carretero$^{5}$\footnotemark[2],
\newauthor
E.~Fernandez$^{5}$,
C.~Neissner$^{5}$\footnotemark[2],
P.~Renard$^{1,2,7}$,
P.~Tallada-Cresp\'i$^{5}$\footnotemark[2],
N.~Tonello$^{8}$\footnotemark[2],
\newauthor
I.~Sevilla-Noarbe$^{9}$,
M.~Crocce$^{1,2}$,
J.~Garc\'ia-Bellido$^{10}$,
H.~Hildebrandt$^{11}$,
H.~Hoekstra$^{12}$,
\newauthor
B.~Joachimi$^{13}$,
R.~Miquel$^{5}$,
C.~Padilla$^{5}$,
E.~S\'anchez$^{9}$
and J.~de~Vicente$^{9}$
\\ 
$^{1}$Institute of Space Sciences (ICE, CSIC), Campus UAB, Carrer de Can Magrans, s/n, 08193 Barcelona, Spain\\
$^{2}$Institut d'Estudis Espacials de Catalunya (IEEC), 08193 Barcelona, Spain\\
$^{3}$Satlantis, University Science Park, Sede Bld 48940, Leioa-Bilbao, Spain \\
$^{4}$Institute of Cosmology \& Gravitation, University of Portsmouth, Dennis Sciama Building, Burnaby Road, Portsmouth PO1 3FX, UK \\
$^{5}$Institut de F\'{\i}sica d'Altes Energies (IFAE), The Barcelona Institute of Science and Technology, 08193 Bellaterra (Barcelona), Spain \\
$^{6}$HEP Division, Argonne National Laboratory, Lemont, IL 60439\\
$^{7}$Department of Astronomy, Tsinghua University, Beijing 100084, China\\
$^{8}$Barcelona Supercomputing Center (BSC), Pla\c ca Eusebi G\"uell 1-3, 08034-Barcelona, Spain\\
$^{9}$Centro de Investigaciones Energ\'eticas, Medioambientales y Tecnol\'ogicas (CIEMAT), Madrid, Spain\\
$^{10}$Instituto de Fisica Teorica UAM/CSIC, Universidad Autonoma de Madrid, 28049 Madrid, Spain\\
$^{11}$Ruhr University Bochum, Faculty of Physics and Astronomy, Astronomical Institute (AIRUB)\\ 
German Centre for Cosmological Lensing, 44780 Bochum, Germany\\
$^{12}$Leiden Observatory, Leiden University, Niels Bohrweg 2, 2333 CA, Leiden, the Netherlands\\
$^{13}$Department of Physics and Astronomy, University College London, Gower Street, London WC1E 6BT, UK\\
}

\date{Accepted XXX. Received YYY; in original form ZZZ}

\pubyear{2022}

\begin{document}
\label{firstpage}
\pagerange{\pageref{firstpage}--\pageref{lastpage}}
\maketitle

\begin{abstract}
PAUCam is an innovative optical narrow-band imager mounted at the William Herschel Telescope built for the Physics of the Accelerating Universe Survey (PAUS). Its set of 40 filters results in images that are complex to calibrate, with specific instrumental signatures that cannot be processed with traditional data reduction techniques. In this paper we present two pipelines developed by the PAUS data management team with the objective of producing science-ready catalogues from the uncalibrated raw images. The \textsc{Nightly} pipeline takes care of all image processing, with bespoke algorithms for photometric calibration and scatter-light correction. The Multi-Epoch and Multi-Band Analysis (\textsc{MEMBA}) pipeline performs forced photometry over a reference catalogue to optimize the photometric redshift performance. We verify against spectroscopic observations that the current approach delivers an inter-band photometric calibration of 0.8\% across the 40 narrow-band set. The large volume of data produced every night and the rapid survey strategy feedback constraints require operating both pipelines in the Port d'Informaci\'o Cientifica data centre with intense parallelization. While alternative algorithms for further improvements in photo-z performance are under investigation, the image calibration and photometry presented in this work already enable state-of-the-art photometric redshifts down to 
{\blue $i_{\rm AB}=23.0$.}

\end{abstract}




\section{Introduction}
\label{s:intro}

Current cosmological studies have been increasing their volume and complexity of data up to a point that traditional analysis methods are not practical anymore. In the past, the data obtained by an astronomer could be processed at the telescope itself or by a personal computer at the research institute. Today, the massive volume and complex analysis require processing in a data center, or even in a grid of computing centers. In 2000 when the Sloan Digital Sky Survey (SDSS) \citep{sdss-York2000} started their observations, the volume of data produced in a single week was larger than all previous data collected in the history of astronomy. The whole SDSS data will be negligible compared to the 60 PB of raw observations that will be produced by the Vera Rubin Observatory \citep{lsst-Ivezi2019}. 

Fortunately, the increased volume of observations came with new and more powerful computing technologies that enabled its necessary analysis. The era of big data arrived in time providing the data management tools we need. The data reduction techniques used in previous surveys cannot be simply applied, and new scalable algorithms had to be designed that meet the more strict needs of today's studies. Higher level languages such as Python enable fast and versatile program development not possible with older programming languages like \textsc{Fortran} or \textsc{C}. Even standard astronomy libraries such as IRAF \citep{IRAF-tody1986} that have been used for decades are becoming obsolete with more flexible astronomical Python libraries like Astropy \citep{astropy:2018} or Photutils \citep{photutils-bradley2020}. The combination of parallel processing in High Throughput Computing data centers with these new advanced libraries changed the paradigm in which data reduction pipeline are being built. 

In this paper we describe the particular image processing and analysis required for a large-scale narrow-band cosmology survey, the Physics of the Accelerating Universe Survey (PAUS). To achieve the scientific goals of the survey, we built PAUCam \citep{paucam-padilla2019}, a large field of view camera equipped with a large set of narrow-band filters that enables low resolution spectra for all the sources in the sky. This massive camera was mounted in the prime focus of the William Herschel Telescope (WHT). Its 4.2m diameter mirror allows to observe fainter objects through the reduced transmission of the particular narrow-band filters of PAUS. Both camera and data reduction system are designed to optimize the photometric redshift precision, delivering complete and homogeneous galaxy catalogues down to a total magnitude of $i_{\rm AB}$ or $I_{auto}(AB)=23.0$.
\footnote{{\blue Note that we have extend  our science analysis to $i_{\rm AB}=23.0$, or even to $i_{\rm AB}=24.0$ in some deep fields, as we have verified that our pipeline produces interesting science results for those magnitudes. But note that for the standard exposure times and number of exposures used in the wide fields, the $S/N$ is usually below 3 for narrow bands with $i_{\rm AB}<22.5$. Because we use forced photometry over galaxies selected with deeper exposures in broad band filters,  even  $S/N<3$  fluxes are useful to measure photometric redshifts (or constrain SEDs) with the 40 narrow bands,  e.g. see Fig.\ref{fig:ValidateExp}.}}

This paper describes the imaging data set we are dealing with (\S \ref{s:data}), the raw instrumental detrending (\S \ref{s:detrending}), the photometric (\S \ref{s:photocal}) and astrometric calibration (\S \ref{s:astrocal}), the particular forced photometry process (\S \ref{s:photometry}) that enables the science ready catalogues and the validation process (\S \ref{s:validation}). {\blue In \S\ref{s:photo-z} we present a validation of our pipeline using photometric redshift estimation, which has a direct impact on the science applications already published or plan in separate papers. In Appendix \ref{s:app_operation} we present the Operation and technical performance of the pipeline, including the data flow and orchestration. Appendix \ref{s:dbschema} gives details of the database schema.
Appendix \ref{s:app_synth} compares some example reference spectra to PAUS data, while Appendix \ref{s:flags} provides the list of flags used. }
We emphasise in the specific challenges of processing narrow-band images that prevent us from using generic software. We also describe the validation procedures and intensive operations at the computing center.
{\blue This paper is part of a series of scientific analysis which have been already published (\citealt{galaxyfeatures-stothert2018,bcnz-eriksen2019,paudm-ops-tonello2019,bkgnet-cabayol2020,Eriksen:2020,lumos-cabayol2021,photoz-alarcon2021,photoz-soo2021,galaxyprops-tortorelli2021,lymanalpha-renard2021,pau-ia-johnston2021,kids-photoz,Renard4000,PAUS-Euclid2022}) or are in preparation, which all used the data reductions presented here.}

\section{Imaging Data}
\label{s:data}

Here we describe the image data used in this work. First in \S \ref{s:data_raw} we describe the uncalibrated raw data that comes directly from the camera and secondly, in \S \ref{s:data_red} we define the reduced data produced by the PAUS data management system (PAUdm) after the instrumental detrending process that will be detailed in \S \ref{s:detrending}.

\subsection{PAUCam Raw exposures}
\label{s:data_raw}

PAUCam is an 18-detector imager camera \citep{paucam-padilla2019} with 40 narrow-band filters, covering the range from 450nm to 850nm in steps of 10nm. After passing the optical system, the incident photon fluxes in the Charge-Coupled Device (CCD) detectors are converted into photo-electrons which are then stored in the individual CCD pixel potential wells. During the read-out process of the detectors, the charge in each pixel is passed sequentially (clocked) to an on-chip amplifier that converts the charge into a voltage and amplifies this voltage before sending it to an Analog-to-Digital Converter (ADC). To reduce read-out time, PAUCam CCD's have four output amplifiers, one for each read-out region consisting of 4096 rows and 512 columns.

The data produced by the mosaic array is packed into multi-extension FITS (MEF) files \citep{fits-wells1981} containing the pixel imaging data for the various types of frames, with associated metadata information in the header of each extension. Separate header-data units (HDU) are created for each amplifier, resulting into a 72-extension MEF file of about 670 MiB. 

During the afternoon, bias and dome flats are observed for calibration purposes. In the twilight, when the sky is too bright for science observations, a high-altitude standard star is observed with each detector-filter set for photometric calibration and total system throughput calibrations. Once the sky is dark enough, the main scientific field observations are taken.

\begin{figure}
  \includegraphics[width=\linewidth]{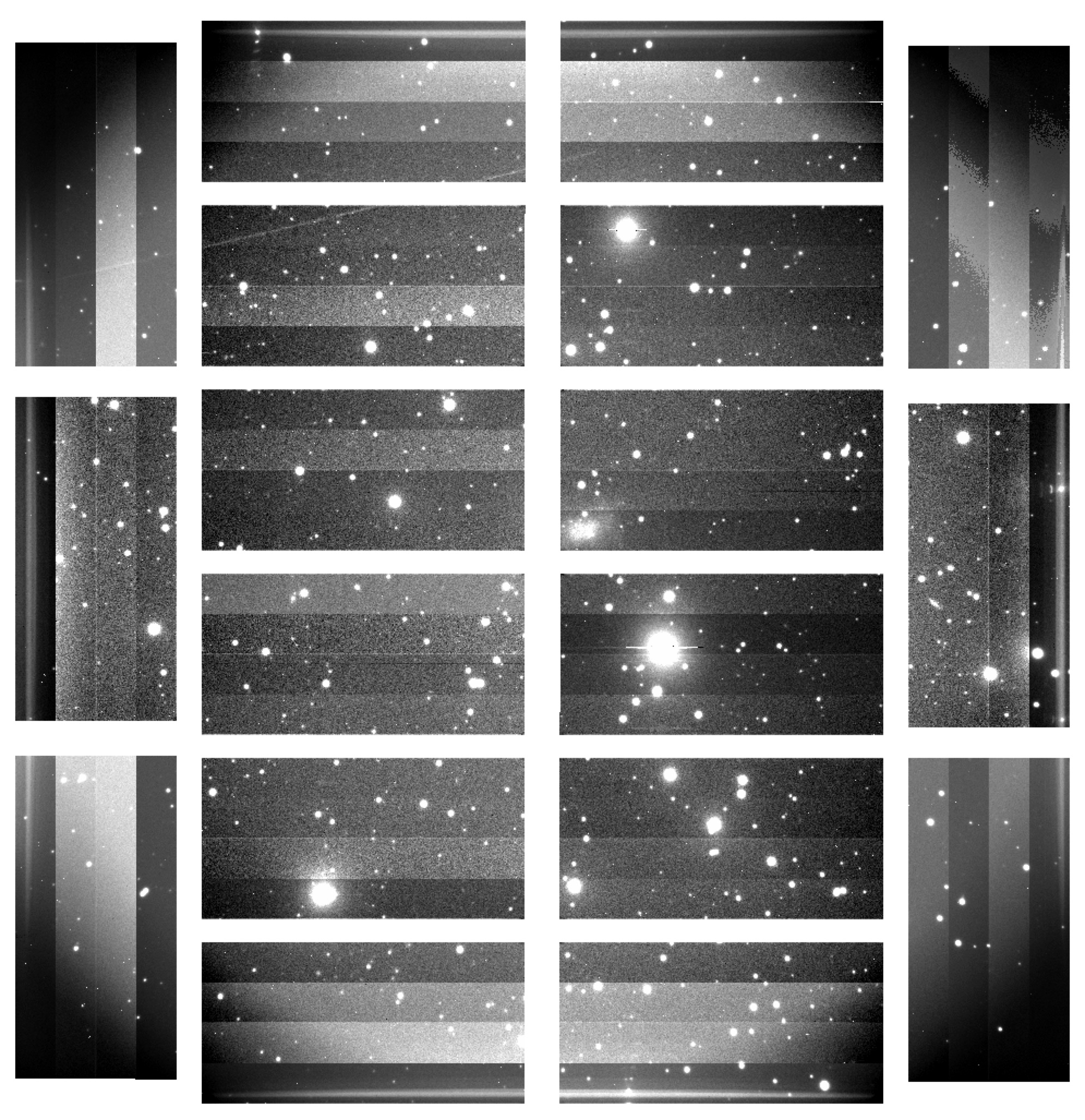}
  \caption{A sky exposure of the full 18-detector PAUCam mosaic. As a raw image, all instrumental signatures are present, and the 72 amplifiers can be identified as well as the vignetting from the WHT prime focus corrector.}
  \label{fig:focal_plane_raw}
\end{figure}

The PAUCam raw data are organized in observation sets, that group exposures from a single night. Typically for an observing night, the calibration frames (bias and dome flat-fields) and the science exposures are stored in separate observation sets. Observation sets allow also to separate observations between PAUS and community observations in shared nights. 

The three main types of raw exposures are

\begin{itemize}
  \item \textbf{Bias frames}: zero exposure time images for electronic pattern calibration.
  \item \textbf{Flat-fields frames}: images of a screen with a flat illumination at the dome for total throughput calibration.
  \item \textbf{Science frames}: the scientific exposures with the target sources on sky. An example of a raw science exposure is shown in Figure \ref{fig:focal_plane_raw}.
\end{itemize}

There are also other types of exposures (e.g. stacked focus or Photon Transfer Curve (PTC) sequences) but these are not part of the regular data reduction process. So far we have observed 240 nights at the WHT, producing a total of 68000 raw exposures.

\subsection{PAUdm reduced images}
\label{s:data_red}

After the image processing that will be described in the following sections, five more types of images are created; 

\begin{itemize}
  \item \textbf{Master Bias}: Stack of corrected bias images electronic bias calibration (described in \S \ref{s:detrending_overscan}).
  \item \textbf{Master Flat}: Stack of processed flat-field frames for throughput variations compensation (described in \S \ref{s:detrending_sl_flats}).
  \item \textbf{Reduced Science}: the instrumentally calibrated science images (described in \S \ref{s:detrending})
  \item \textbf{Reduced Mask}: the associated mask image with flag values per pixel (described in \S \ref{s:detrending_mask})
  \item \textbf{Reduced Weight}: the associated weight image (described in \S \ref{s:detrending_mask})
\end{itemize}

All reduced images contain only 8 extensions, one per detector on the 8 most central and illuminated CCDs. Additionally, the Point Spread Function (PSF) models and astrometry catalogues are stored in the archive, together with other associated data that are recorded in the PAUS database such as image zero-points and single-epoch detections.

\section{Instrumental Detrending}
\label{s:detrending}

Raw images contain significant instrumental signatures that need to be neutralized or masked before carrying out any photometric measurement. This section describes the correction flow for all raw science exposures from PAUCam.  

\subsection{Gain calibration and PTC analysis}
\label{s:detrending_gain}

During the assembly of the PAUCam instrument \citep{detector-charact-casas2012}, we performed the first Photon Transfer Curve (PTC) analysis with the method described in \cite{ptc-janesick2001} to estimate the gain for each of the four amplifiers in all detectors. Once on the mountain, with PAUCam mounted in the Prime Focus of the William Herschel Telescope under a stable installation, we repeated the test as the environmental conditions were different (low temperature, pressure, and humidity) and with the interaction with other devices and a different grounding. We decided to repeat this test on each observing run since 2015.

In order to process the PTC we need to obtain dome flats with a broad band filter, usually $g$ or $r$, and obtain pairs of images with scaled exposure times, from 1 to 30 seconds, covering the range between very low counts until saturation. Due to the strong vignetting caused by the WHT prime focus optical system, PAUCam allocated narrow-bands only on the eight central detectors. The external detectors use broad-band filters for calibration and guiding purposes. For this work we will focus on the central detectors with narrow-band filters. An analysis of the gain for the full focal plane (the 18 detector set) is described in \cite{paucam-padilla2019}. Each $2K \times 4K$ Hamamatsu CCD of PAUCam has four outputs, and each one must be analysed independently for the PTC analysis, so we analyse 32 regions from the eight central CCDs.

\begin{figure}
  \includegraphics[width=\linewidth]{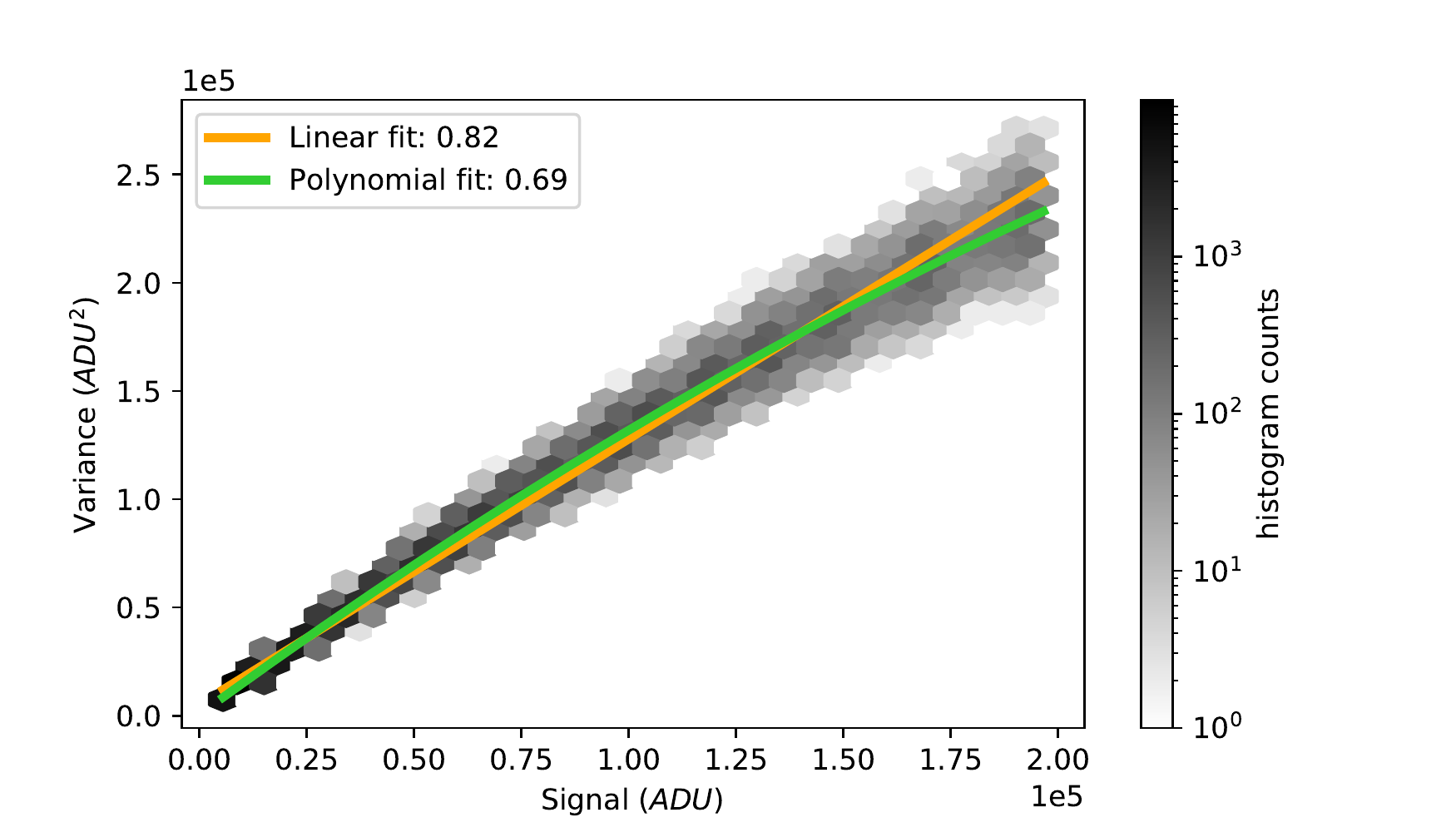}
  \caption{The relation between the mean flux value and its variance used to infer the gain as described in the PTC process for the first amplifier of CCD 1 in a test during the 19A period. The gain is measured as the inverse of the first order component of the polynomial fit (green line). A linear fit over the whole flux range (yellow line) cannot be used to estimate the gain as it delivers a biased value due to the non linear response of the detector.}
  \label{fig:ptc}
\end{figure}

To compute the PTC we subtract the median overscan value for each image and average the pairs of images with the same exposure time to remove possible patterns and reduce the noise. A random choice of small squares of $100 \times 100$ pixels in the subtracted and averaged images are used to determine the mean signal (in Analog-to-Digital Units or ADU) and the variance (in ${\rm ADU}^2$). The PTC represents the variance as a function of the signal as shown in Figure \ref{fig:ptc}. {\blue The gain can be evaluated by the fit between the variance and the signal. The inverse gain can be estimated as the slope between the variance and its signal, in units of $e^{-}$/ADU. As the detectors do not respond linearly in the high end of the flux range, as shown in Figure \ref{fig:ptc}, we have measured the slope in the fully linear regime of the detector range (<20.000 ADU) where the polynomial and linear fit estimate the same value.}

Since the first installation of PAUCam in the Prime Focus of the WHT in 2015 we ran 16 PTC tests. The value of the gain varies from amplifier to amplifier but has remained stable in time across the various observing runs.

\subsection{Overscan, gain, bias and flat-field correction}
\label{s:detrending_overscan}

The electronic readout process adds an artificial pedestal signal across the whole image, biasing the values of each pixel by a constant number. In order to estimate this bias value that slightly varies from one amplifier to each other, we use the overscan section where only the bias signal is present. Then we estimate the value of the bias in each amplifier, computing the median row by row in the overscan region. To correct for low frequency oscillations in the vertical readout, a 8-row Gaussian filtering is applied, allowing a more accurate subtraction of the electronic bias as the single row median did not deliver accurate statistics of the varying bias value. The gain estimated from the PTC analysis is applied to each amplifier array, and ADUs are transformed to electrons in this early stage of the process. To compensate for readout patterns that are present in all exposures, a master bias frame is produced, combining around 10 individual zero-exposure bias frames. As we have identified residual patterns in the first 2 or 3 exposures after a full readout system restart {\blue(typically once per day)}, the individual bias frames are analyzed and those with abnormal levels of noise are removed from the median average. Finally, a master flat is produced from the individual dome-flat exposures that are taken every afternoon, prior to the night-time observations. We perform a median average of at least 5 individual flat-field exposures, to reduce noise fluctuations and cosmic ray hits. Scatter-light residuals found in the raw flat images needed to be removed, as this signature in the flat images is an additive component of the light, while the flat-field needs to contain only multiplicative factors of the main optical path. The process of removing the scatter-light will be described in \S \ref{s:detrending_sl_flats}, together with the correction of the scientific sky images. Once a clean master flat has been produced, we use it to divide the science exposures from the same night, flattening the response across the field of view.

\subsection{Cross-talk calibration and correction}
\label{s:detrending_ct}

Since the four amplifiers are read in parallel it is possible that current is induced through magnetic fields between the different channels read at the same time, an effect that is commonly known as crosstalk \citep{crosstalk-Freyhammer2001}. The charges from the same row are read simultaneously in all four amplifiers of each detector and in all 18 detectors. Amplifiers 1 and 3 read the pixels from right to left while amplifiers 2 and 4 read in the opposite direction, due to its disposition in the detector. This allows us to identify unambiguously the pixels that are read simultaneously. 

\begin{figure}
  \includegraphics[width=\linewidth]{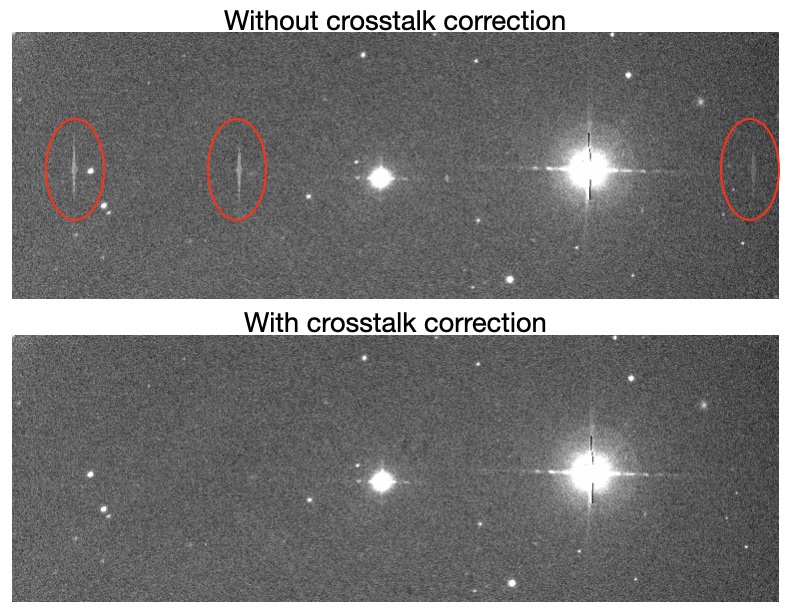}
  \caption{\emph{Top:} An image crop showing an 8.5 magnitude star that saturates the detector producing bleeding and a visible crosstalk signal on the mirrored positions of the remaining three amplifiers. \emph{Bottom:} The same crop with crosstalk correction enabled. The 3 ghost signals from the saturated star mirrored on the other amplifiers are completely removed after applying the corresponding ratios and subtraction.}
  \label{fig:crosstalk_correction}
\end{figure}

Another relevant instrumental effect occurs when too many electrons in a given pixel cause electrons to overflow to the nearby pixel wells in the same column. The average capacity of these detectors is $\sim$ 210.000 e$^-$. This effect is known as saturation bleeding, producing an elongated shape around very bright stars. If the pixel signal is above 18-bit (current depth of PAUCam ADCs), it will saturate at $N^{\rm sat} = 2^{18}-1$ (or 262,143) ADUs. With an average gain of 0.7, the final saturation value is limited by the ADC conversion limit and not by the full well. {\blue Saturation starts approximately on a $G_{\rm AB}\sim10$ stars, affecting approximately 200 pixels for a $G_{\rm AB}\sim9$ star due to bleeding.} In the top panel of Figure \ref{fig:crosstalk_correction} one can see ghost images parallel to the elongated star image due to the crosstalk effect. Even though we can only see the ghost image from the saturated pixels, the crosstalk effect happens at all levels but is only visible when the signal is strong enough to stand out from the noise in the background.

To remove the crosstalk signal we need to estimate the induction ratios $r_{\rm xy\_ij}$ between each pair of amplifiers from detector $x$ amplifier $y$ to detector $i$ amplifier $j$. Therefore, the signal of each amplifier would be the sum of its direct integration signal $I^{\rm int}$ plus the signal of all the remaining amplifiers in the mosaic being read out at the same time scaled by the induction ratio 
\begin{equation} 
I^{\rm tot}_{\rm xy} = I^{\rm int}_{\rm xy} + \sum^{18}_{\rm i} \sum^{4}_{\rm j} I^{\rm int}_{\rm ij} r_{\rm xy\_ij}  
\end{equation}			
for a given set of detector $x$ and amplifier $y$.

Although crosstalk analyses were made at the facility lab in Barcelona, the crosstalk ratios may vary once the camera is mounted in the telescope and thus we had to estimate them with sky data obtained when the instrument was mounted in the prime focus. In order to estimate the induction ratios between amplifier $i$ and $j$ of exposure $z$, we measured the average image background level in the target amplifier $bg_{\rm i}$ and subtracted it from the average level in the mirrored positions of the saturated pixels $f_{\rm j}$, looking for any change in flux with respect to the rest of the image background due to crosstalk such as:
\begin{equation} 
r_{\rm ij}^z = \text{median}\{f_{\rm j}(x,y) - bg_{\rm i}\}\,.
\end{equation}
where ${x,y}$ are saturated pixels in amplifier image $j$ of exposure $z$. If the mirrored position contains sources or is not flat enough, we discard the measurement. We combine the $i,j$ ratio measured in all available images, weighting by the number of saturated pixels available in the each ratio as:
\begin{equation} 
r_{\rm ij} = \frac{\sum_{\rm z} r_{\rm ij}^{\rm z} N_{\rm sat}^{\rm z}}{\sum_{\rm z} N_{\rm sat}^{\rm z}}\,.
\end{equation}

We compute the ratios for the 8 central detectors (where narrow-band filters are located) against the remaining 18 detectors and 4 amplifiers, resulting into a total of 2272 pairs. To remove the crosstalk signal, each amplifier needs to subtract the flux of the other 71 amplifiers by the corresponding ratio. Figure \ref{fig:crosstalk_correction} illustrates how the ghost produced by the bright star on the right is now removed after applying the calibrated crosstalk ratios.

\begin{figure}
  \includegraphics[width=\linewidth]{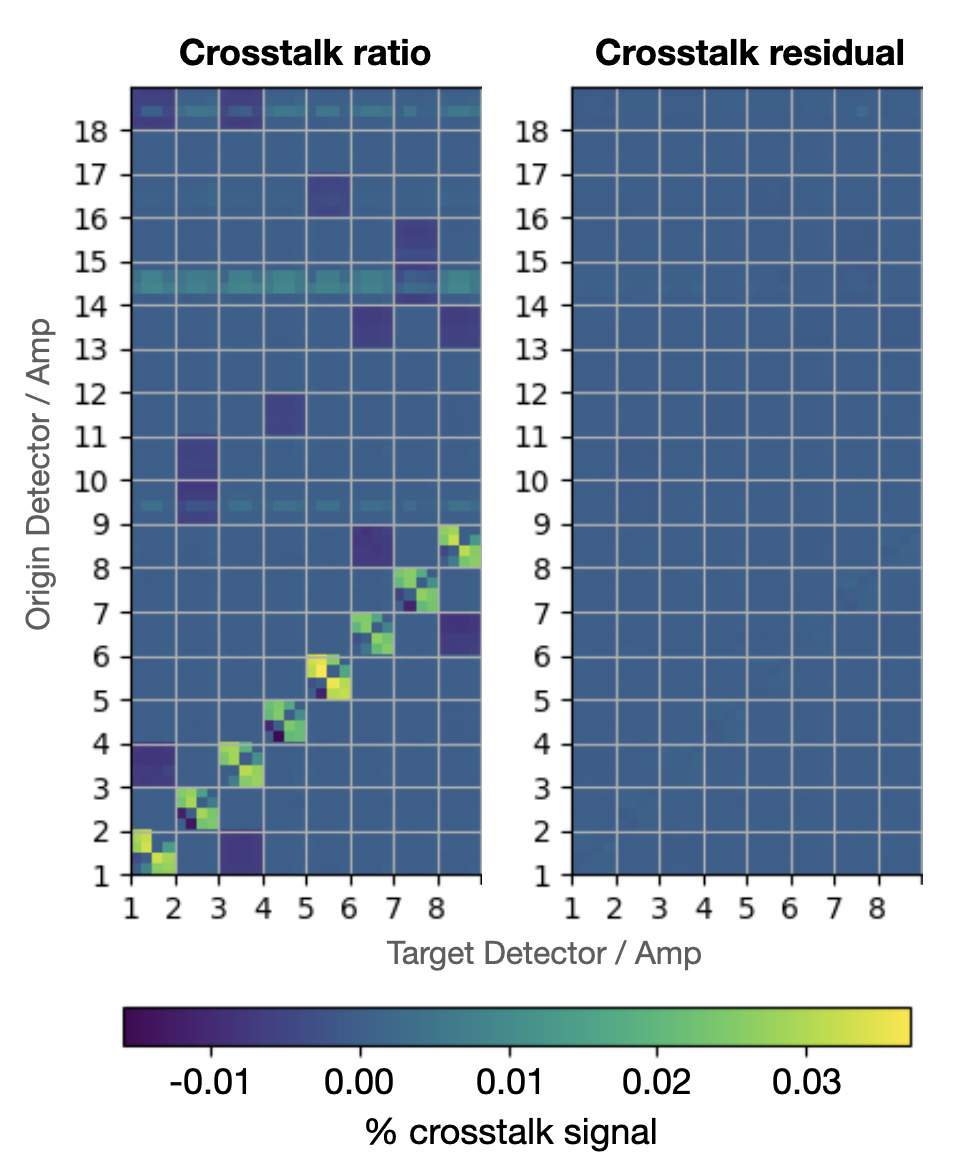}
  \caption{\emph{Left:} The crosstalk induction ratios used to apply the correction. Each box contains the 4x4 amplifiers of each detector against another detector. \emph{Right:} The residual crosstalk ratios measured on a different dataset after crosstalk correction.}
  \label{fig:crosstalk_ratios}
\end{figure}

To verify the correct implementation of the crosstalk calibration, we have measured the same crosstalk ratios over a dataset that has been already crosstalk corrected. Figure \ref{fig:crosstalk_ratios} illustrates the matrix of crosstalk ratios. It can be seen how the intra-detector crosstalk ratios are significantly higher than the inter-detector ones, as expected due to further distance of wiring in the electronics. We measure the crosstalk ratios in a large dataset with more than 430,000 images, and apply those ratios to a different dataset (as applying it over the same dataset would give a zero residual by construction). We can see how the crosstalk ratios reduce from $\sim$ 0.04\% to less than 0.002\% once the crosstalk calibration has been applied. 

\subsection{Scatter-light correction}
\label{s:detrending_sl}

The large narrow-band set of PAUS required engineering an efficient design to include all 46 filters, as described in \cite{paucam-padilla2019}.  
{\blue The filter exchange system has a total of 14 filter tray slots. Only 5 trays are use for narrow-bands. There are also a standard ugrizY broad band filters. These are larger and they cover the full field of view each. So one tray per broad band filter.  Note nevertheless that for the science fields in the PAU Survey we chose regions of the sky which already have deep broad band ugrizY observations (from CFHTLS, COSMOS or KIDS) so we do not need to repeat those broad band observations with PAUcam.}

\begin{figure}
  \centering
  \includegraphics[width=\linewidth]{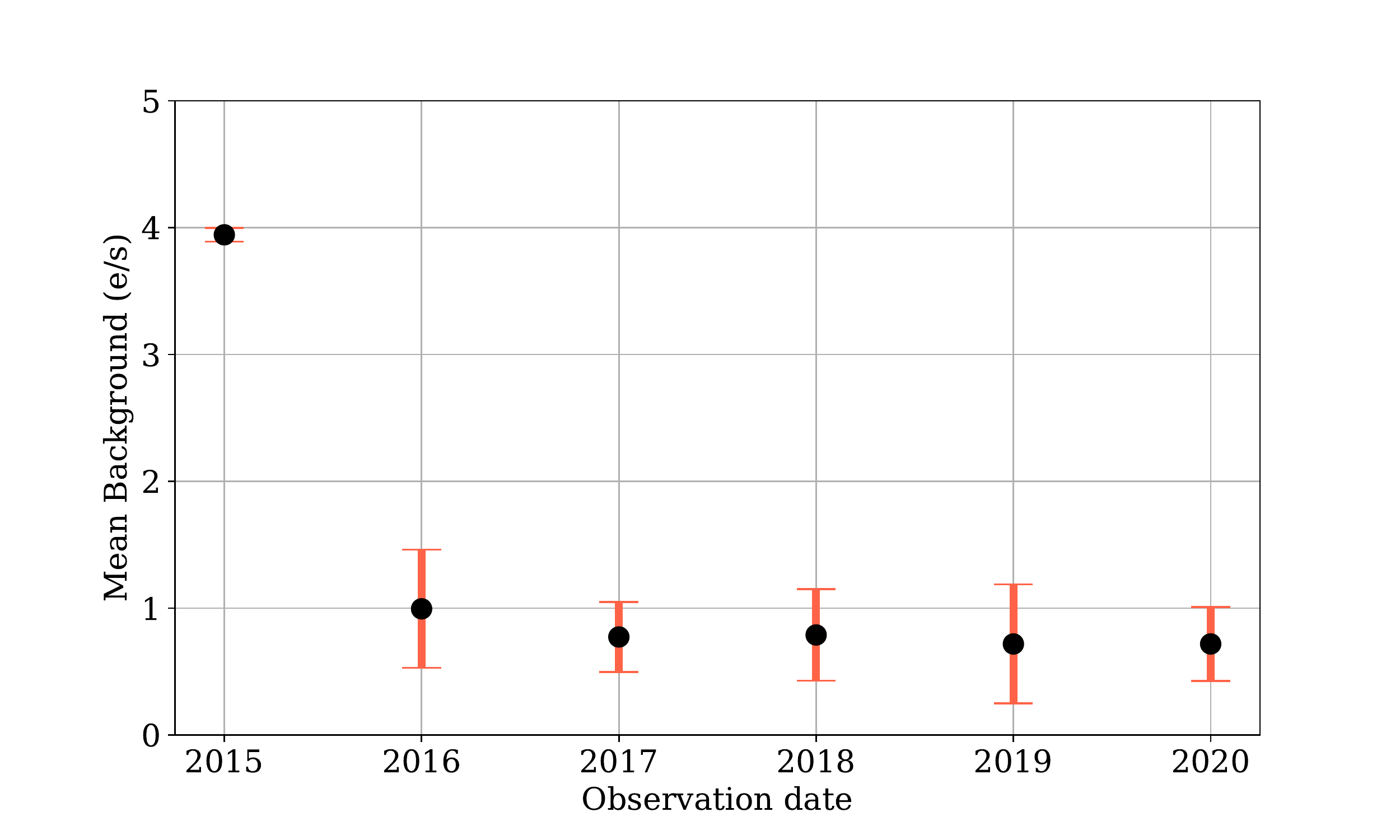}
  \caption{The mean background in the reduced PAUS images. With the original filter tray design, a substantial fraction of the light was reflected at the edges of the filter pieces and filter tray, causing severe levels of scatter-light in the science images. An intervention was carried out in mid 2016 and the mean background was reduced from $\sim 4~e^-/s$ to less than $1~e^-/s$, as can be seen after 16B observations and beyond.}
  \label{fig:background_evolution}
\end{figure}

The final solution for narrow bands was to distribute multiple small filters, each covering a single detector in filter trays, such that with 5 trays and 8 narrow-bands in each covering the central most illuminated detectors, we could cover the entire wavelength range. This design caused a side effect in the quality of the images, as significant reflections from the lateral edge of the filter and the filter trays themselves created localized scatter-light at the edge of each detector image (see the top panels of Fig. \ref{fig:SL_science} and \ref{fig:SL_science_extended}). This effect was identified quickly and the camera was reopened in mid 2016 to redesign the filter trays and minimize the problem. After the camera intervention, the background caused by scatter-light was reduced by a factor of 4 as it can be seen in Figure \ref{fig:background_evolution}. The scatter-light was not fully eliminated and specific processing (described in the following sections) had to be developed to correct the areas affected by this issue, without compromising the light from the sources we need to measure.

\subsubsection{Implementation in flat-fields}
\label{s:detrending_sl_flats}

\begin{figure}
  \centering
  \includegraphics[width=0.45\linewidth]{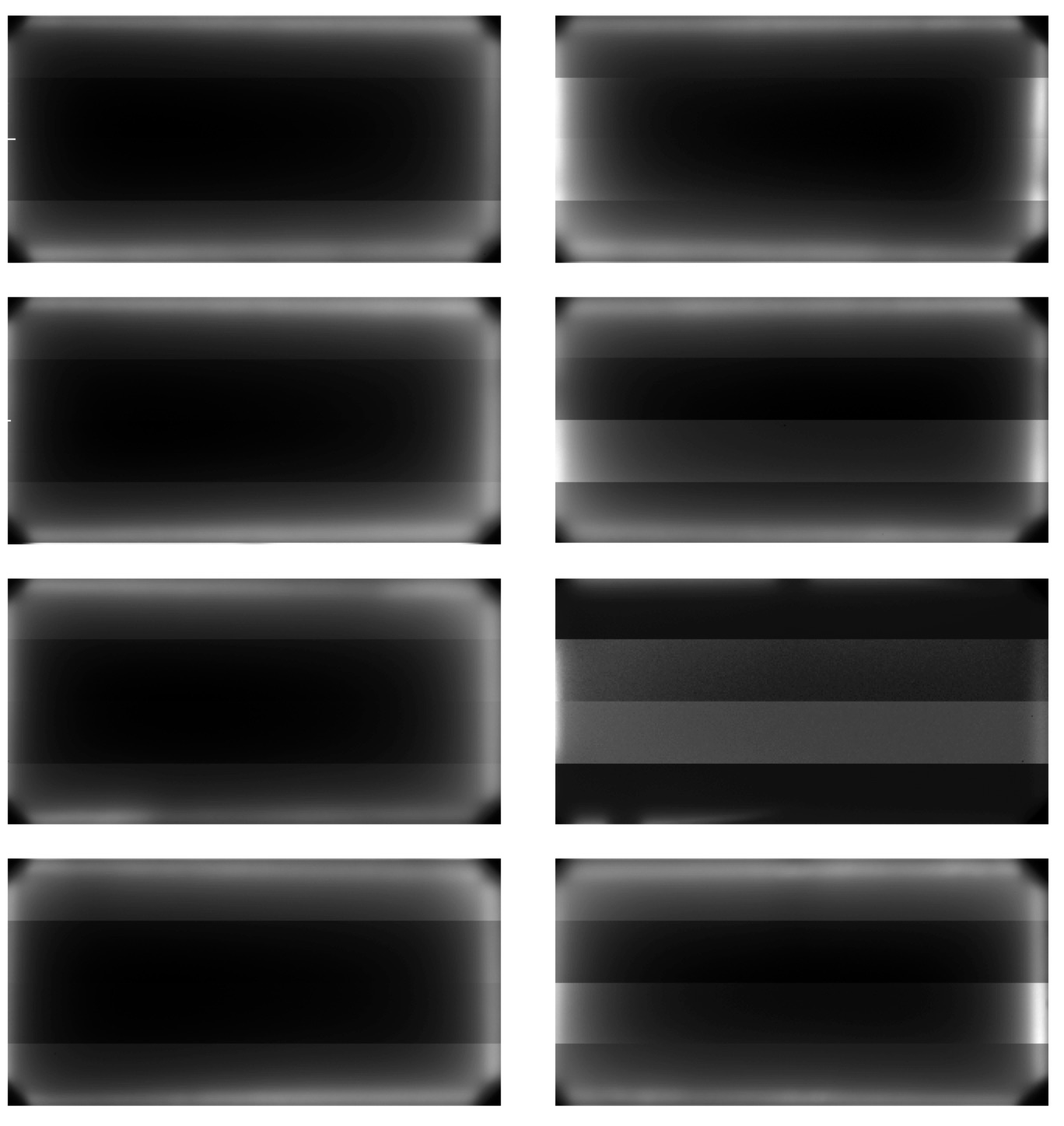}
  \hspace{0.2cm}
  \includegraphics[width=0.45\linewidth]{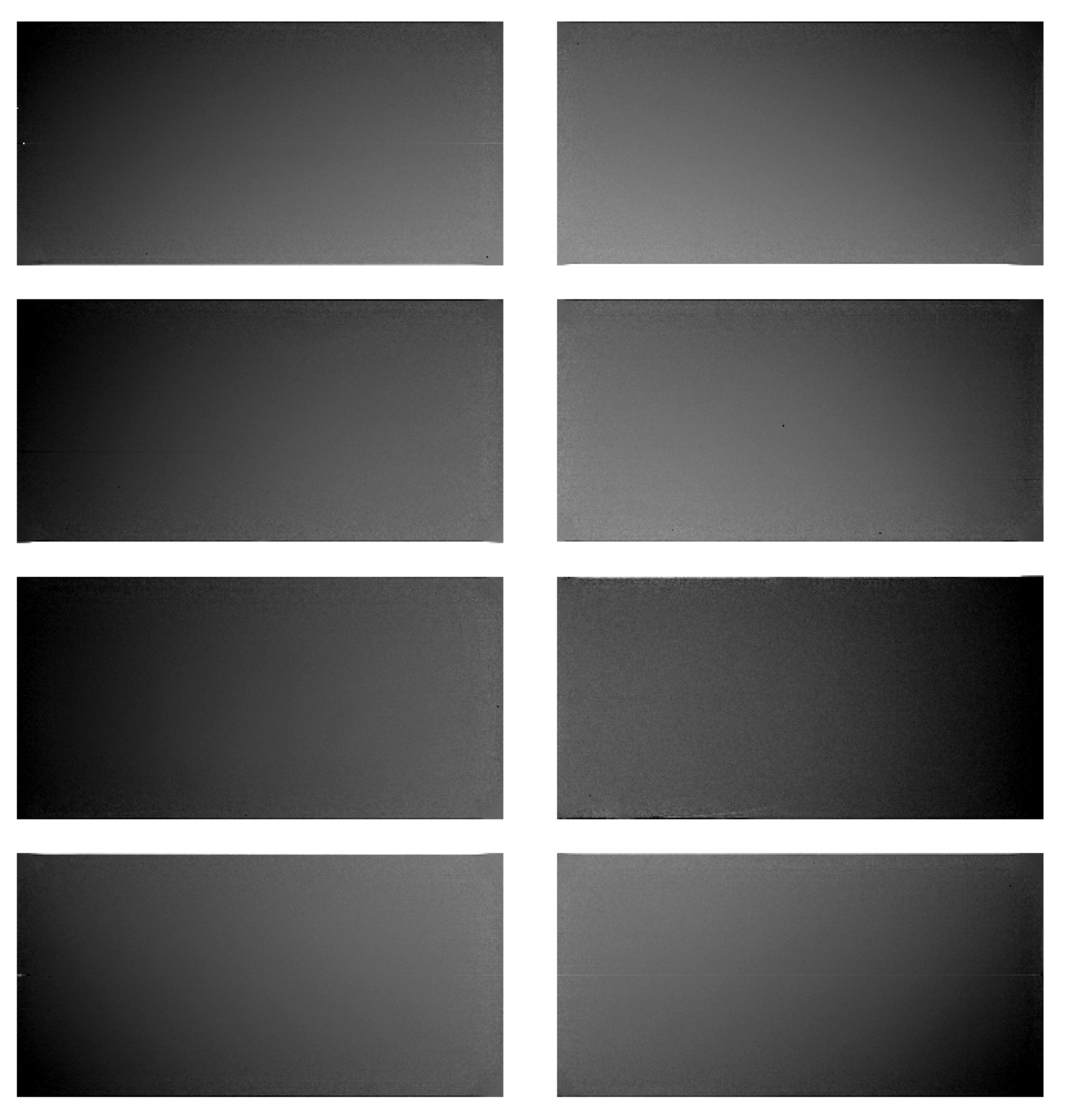}
  \caption{\emph{Left:} The raw flat-field exposure with severe scatter-light signal. \emph{Right:} The scatter-light corrected flat-fields preserving both the low frequency component of the vignetting and the high frequency of the pixel-to-pixel variations.}
  \label{fig:SL_flat}
\end{figure}

\begin{figure}
  \centering
  \includegraphics[width=\linewidth]{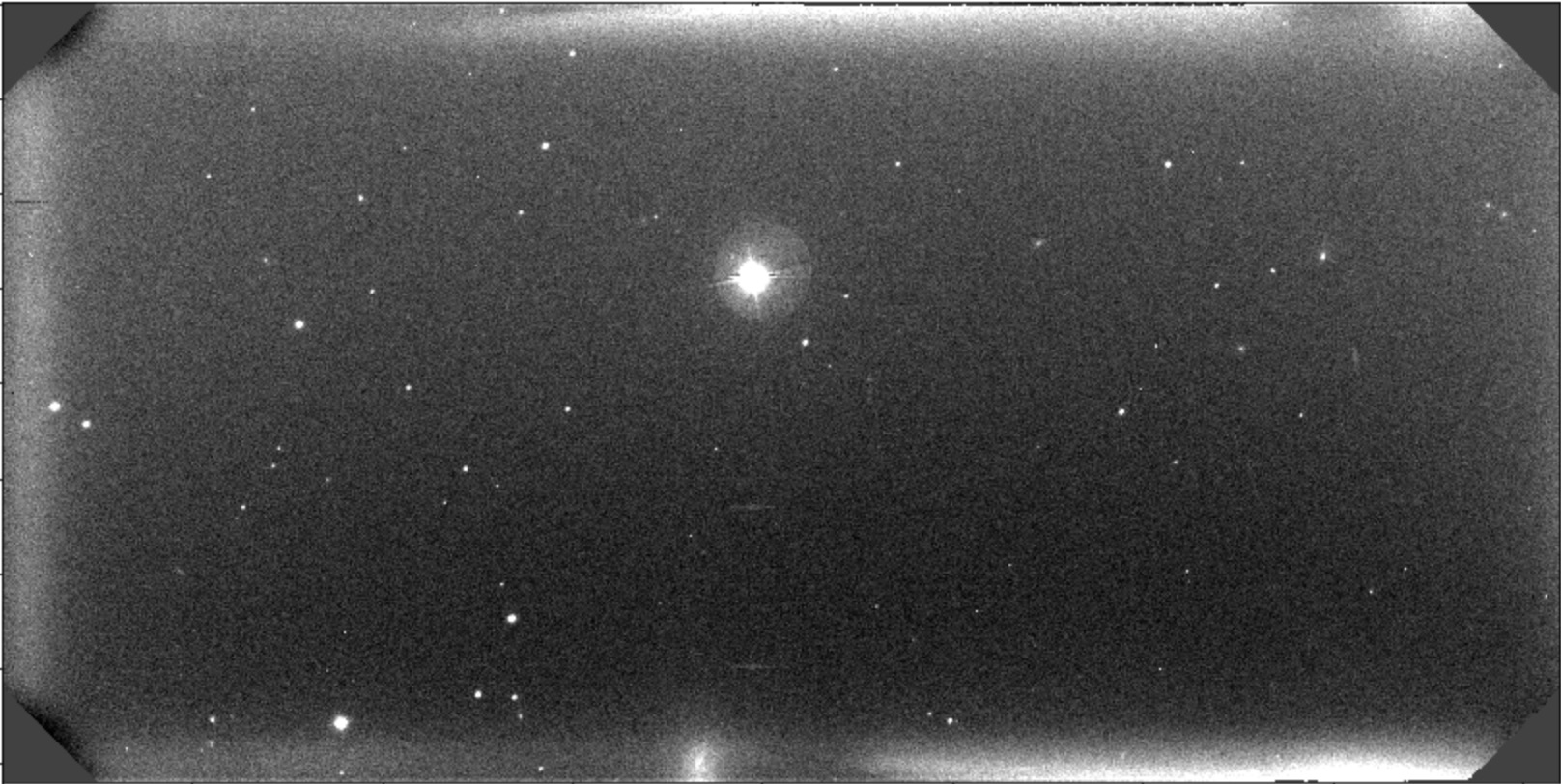}
  \includegraphics[width=\linewidth]{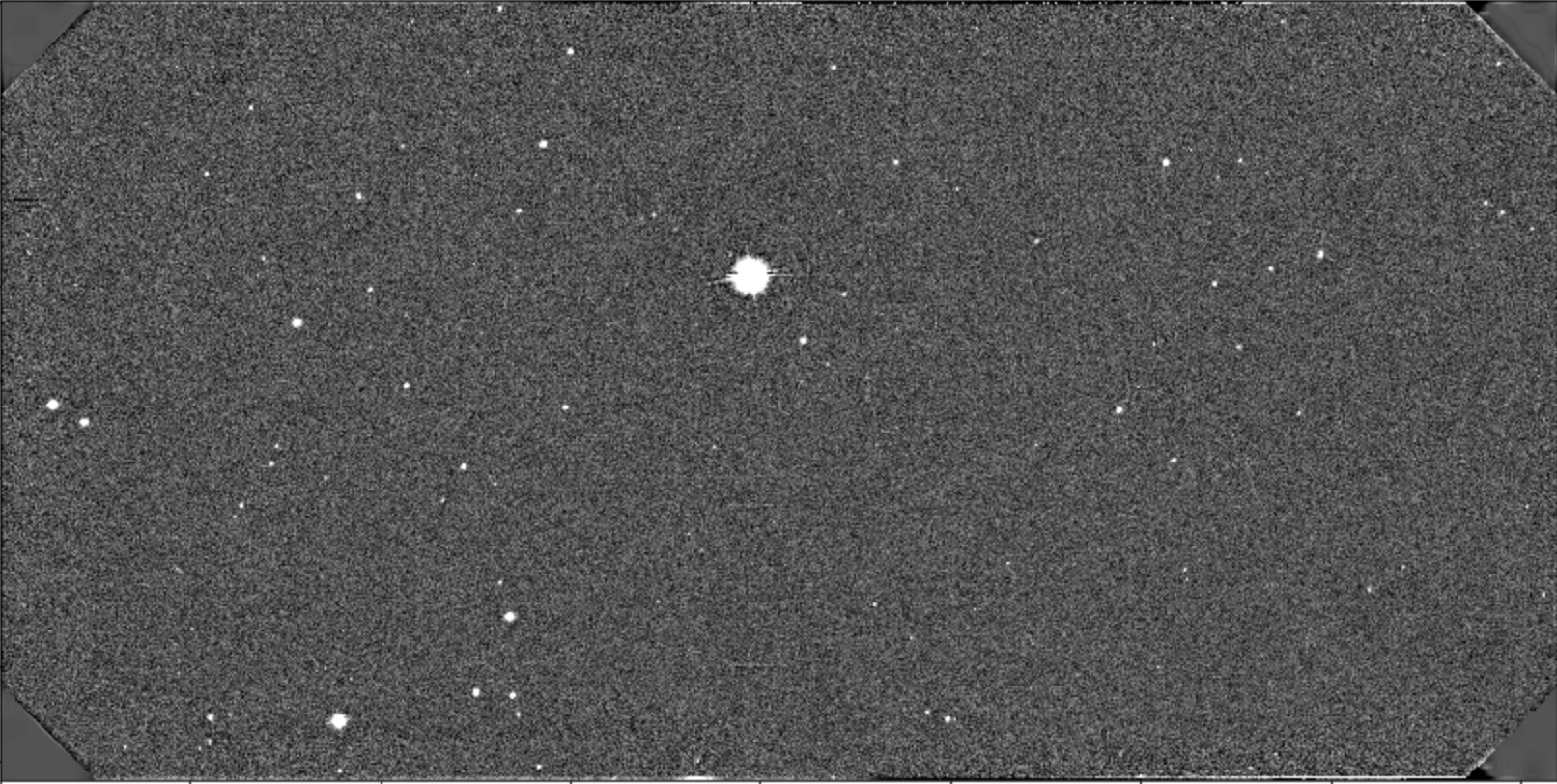}
  \caption{\emph{Top:} A typical reduced science frame without scatter-light correction. \emph{Bottom:}  The flat-field science image preserving the flux from the stars and small galaxies, but affecting the larger extended sources and halos.}
  \label{fig:SL_science}
\end{figure}

Flat-fields can be divided into two main frequency components across the focal plane: a low-pass band due to vignetting and a high pass pixel due to pixel variations (i.e. dust, dead and hot pixels). Fortunately the scatter-light is in between with a mid-size frequency variation, so we are able to isolate and correct for it. First we use the broad-band flat-fields made of a single large filter, that does not contain scatter-light, to construct a vignette image, mostly caused by the prime focus corrector optics. With a low-pass filtering we could isolate the vignette component and dismiss the high frequency variations. Dividing the narrow-band flats with the vignette profile flattens the image and leaves the scatter-light as the lowest component in the image, ensuring that high-pass filtering would leave the scatter-light component only, which can then be subtracted from the flat. The resulting correction can be seen in Figure \ref{fig:SL_flat}.

\subsubsection{Implementation in science images}
\label{s:detrending_sl_science}

In the case of science images, the process is more complicated as large extended sources (i.e. large galaxies or nearby objects) may have a similar spatial frequency as the scatter-light. In the cases where the target sources are distant and small galaxies, a low-pass filtering, sigma-clipping the sources, isolates the scatter-light, without affecting the photometry of the small sources. This is similar to background subtraction techniques, with the difference that the codes used take into account the preferable direction of the scatter-light across the edges of the detector-filter system. 

To preserve the flux from extended sources, a sky-flat correction was the most effective solution. To produce the sky flats, a large set of images were combined following a median average stack. We identified that around 50 images were enough to provide an accurate model of the background, including scatter-light. A combination of multiplicative factors (residuals from the dome flat) and additive components (mostly scatter-light) are present in the sky flats and therefore need to be separated to subtract and divide the images to detrend the effects properly. The complication of this method is that it requires multiple epochs to detrend an image, and the images stacked need to have similar background levels, which might not always be possible with shorter observations. However, when there are enough exposures from the same filter tray under the similar sky conditions, this method provides a more accurate illumination correction than the traditional dome flats.

\begin{figure}
  \centering
  \includegraphics[width=\linewidth]{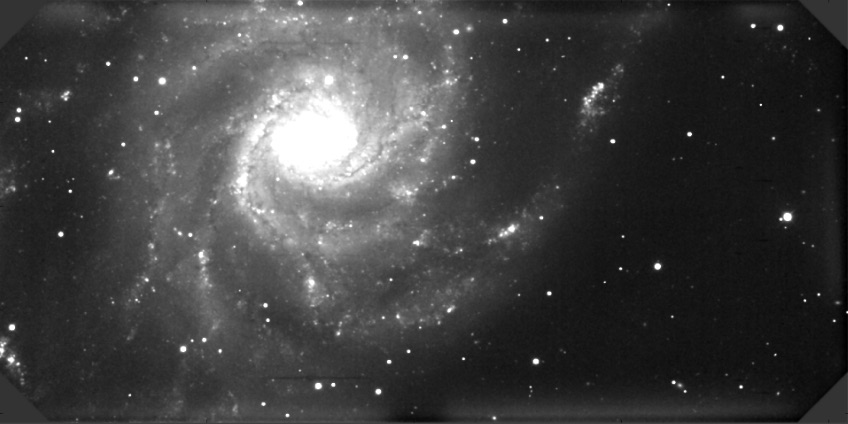}
  \includegraphics[width=\linewidth]{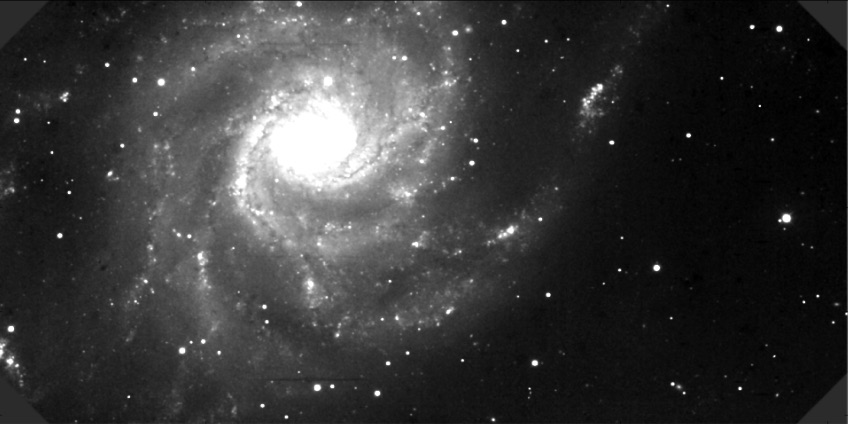}
  \caption{\emph{Top:} A rare science frame with a very extended source (M101) without scatter-light correction. \emph{Bottom:} A sky flat corrected science image preserving the light from extended sources.}
  \label{fig:SL_science_extended}
\end{figure}

\subsection{Cosmic Ray detection, rejection and masking}
\label{s:detrending_cosmics}

\begin{figure}
  \centering
  \includegraphics[width=\linewidth]{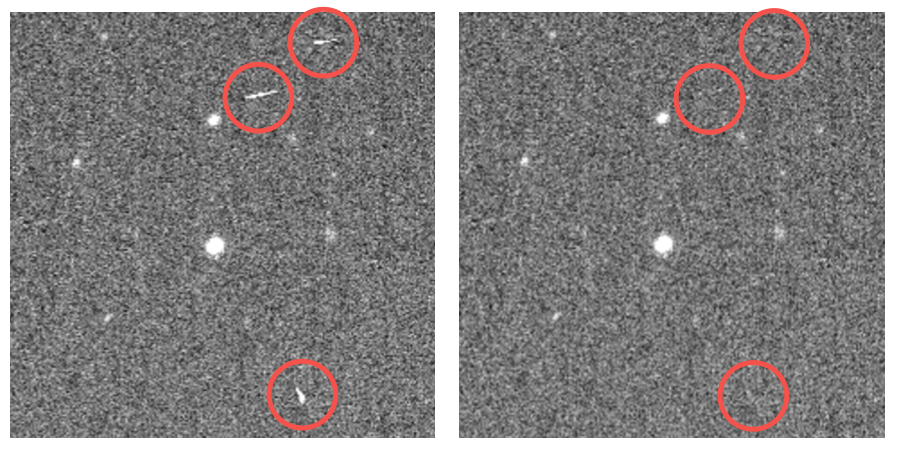}
  \caption{\emph{Left:} A crop of an image without cosmic ray masking. \emph{Right:} The same image crop after the Laplacian filter algorithm, successfully identifying, masking and interpolating the cosmic ray hits in a single-epoch image. Red cicles highlight the location of cosmics before and after detection and masking.}
  \label{fig:cosmics}
\end{figure}

The fully depleted detectors are also very sensitive to cosmic rays impacting the silicon sensitive area and leaving a trace that will impact the photometry if it overlaps with the target source. The typical exposure times for PAUS are between 2 and 3 minutes, causing a reduced amount of cosmic hits in the exposure. However the camera is capable of performing long exposures thanks to its integrated auto-guiding system, and in these cases having a proper cosmic ray identification is critical. In any case, we perform a cosmic ray detection, rejection and masking, following a Laplacian filtering algorithm known as L.A.Cosmic \citep{cosmics-vandokkum2001}. The main idea is to take advantage of the specific sharp profile of the cosmic ray hits, due to the fact that the particle is not blurred by the atmosphere PSF as the rest of the photons in the image and can be highlighted with a Laplacian filtering. For the noise model described in the algorithm, we provide the measurement of the readout noise taken in the overscan regions of the raw image. Figure \ref{fig:cosmics} illustrate how cosmic ray hits over a PAUS image disappear after being identified and masked with neighbouring information.

\subsection{Image Masking}
\label{s:detrending_mask}

In order to keep track of the history of each pixel, we attach a mask to each exposure image, of the same size as the original. The pixels in the mask are mapped into a bitmap where each bit corresponds to a certain issue. Bits 1 to 5 are left for \textsc{SExtractor} \citep{sextractor-bertin1996} flags, and will specify when a detected source is crowded, merged, under a halo, truncated or deblended. Bits 6 to 13 are set at the pixel or image level in the mask image to specify for crosstalk-corrected, detector cosmetics (dead or hot pixels), saturated pixel, rejected cosmic ray, highly vignetted area, close to edge and very high distortion. Other bits are left for later photometry such as pixels or sources affected by scatter-light background, very high extinction, discordant measurements, astrometry issues and noisy background. The full list of flags can be found in Appendix \ref{s:flags}.

To classify the cosmetics in the detector, a flattening process over the master flats is applied and pixels with less than 40\% of the normalized flux are added to the mask and flagged as bad pixels. For each science frame, we also produce the associated weight map, built from the master flat-field, with a very low weight value the pixels with mask values larger than 0. In this way we assign a higher weight for pixels that have a better system response and neglecting the masked pixels.

\section{Astrometric calibration}
\label{s:astrocal}

Once the image has been cleaned and detrended for all known instrumental effects described in \S \ref{s:detrending}, we can proceed to correct for telescope pointing inaccuracies and optical distortions. There are three key elements in this process; the reference catalogue (\S \ref{s:astrocal_gaia}), the calibration of the WCS (\S \ref{s:astrocal_wcs}) and the astrometry matching and correction of the single-epoch images (\S \ref{s:astrocal_singleepoch}). 

\subsection{Gaia reference catalogue}
\label{s:astrocal_gaia}

The reference astrometric catalogue is the publicly available Gaia Data Release 2 (DR2) \citep{gaia-dr2-2018}. This dataset provides the most accurate stellar catalogue, complete between 12 < G < 17 and with a limiting magnitude of G=21. It also includes proper motions, which can be critical to find an accurate astrometric solution in some situations. Nevertheless the observation period of PAUS and Gaia overlap in time, so accurate proper motions corrections are not as critical as with observations that are far apart in time. The whole Gaia DR2 was ingested into the PAUS database, enabling a high-quality astrometric reference for any PAUCam observation in the sky. 

\begin{figure}
  \centering
  \includegraphics[width=\linewidth]{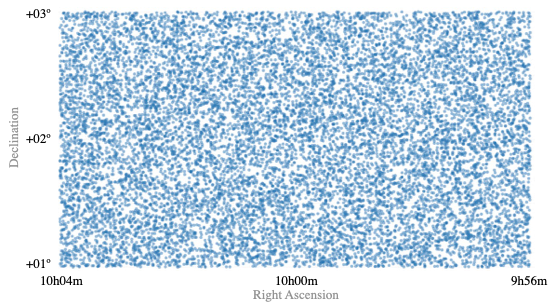}
  \caption{A uniform and dense star coverage in the COSMOS field from Gaia DR2 enabling an accurate reference for astrometric calibration.}
  \label{fig:gaia}
\end{figure}

\subsection{WCS calibration}
\label{s:astrocal_wcs}

\begin{figure}
  \centering
  \includegraphics[width=0.7\linewidth]{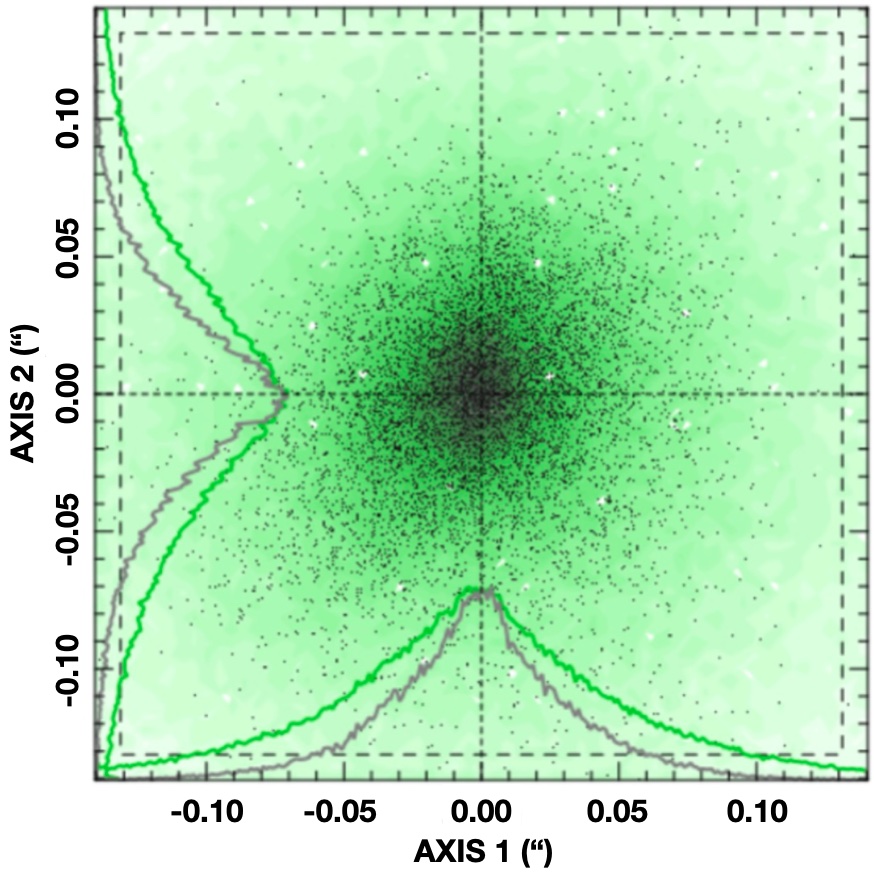}
  \includegraphics[width=0.74\linewidth]{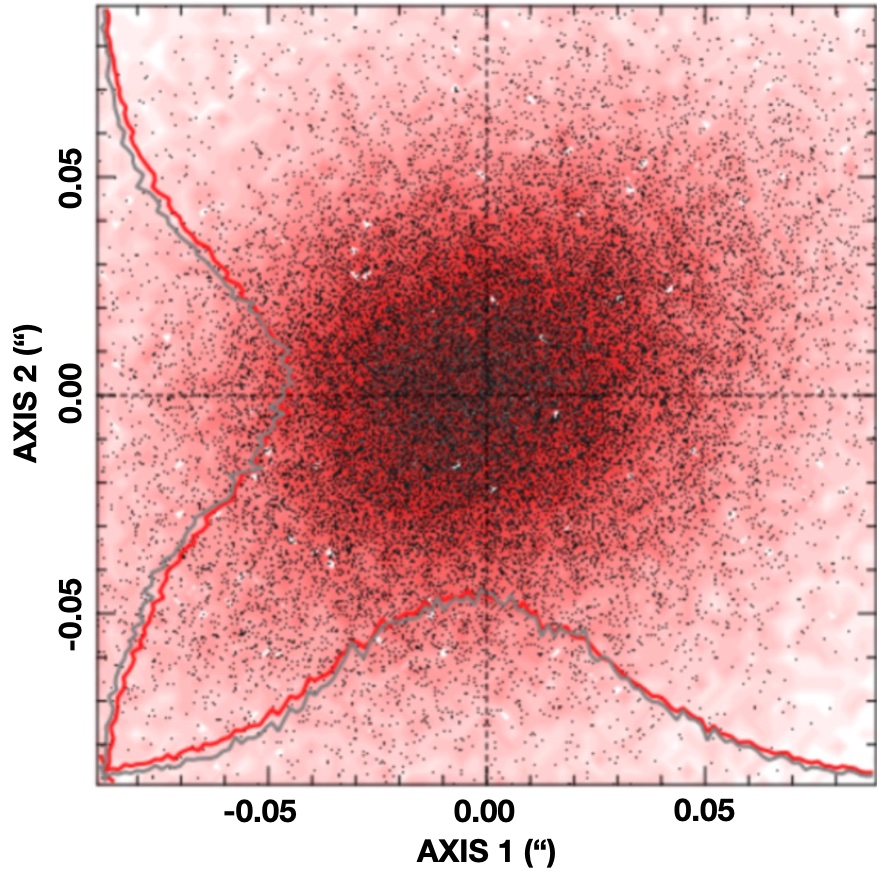}
  \caption{The position residuals corresponding to the difference between the measured position of each source after the astrometric calibration against its reference in the two coordinate axes (1: Right Ascension, 2: Declination). The plot highlights the axis histogram distributions (color: all sources, grey: high SNR sources). \emph{Top (green):} The astrometric residuals between internal overlapping sources. \emph{Bottom (red):} The astrometric residuals against the reference catalogue.}
  \label{fig:astro_residuals}
\end{figure}

The raw mosaic exposures come with a base World Coordinate System (WCS) \citep{wcs-calabretta2002} in its header that approaches the plate scale of PAUCam at the WHT. However this default WCS is not enough to calibrate single-epoch exposures and we need to compute a more accurate WCS solution for the implemented focal plane geometry. Furthermore, the solution for each detector position, rotation and scale needs to be defined independently as the base WCS comes from the mechanical layouts and not from precise measurements of the built focal plane. For this purpose we use \textsc{SCAMP} \citep{scamp-bertin2006} in a particular configuration (\texttt{MOSAIC\_TYPE LOOSE}), giving freedom to each detector to move, scale and rotate around the focal plane for a perfect match between the overlapping dithered exposures. We provide \textsc{SCAMP} all overlapping measurements in the PAUS reference sky location, the 2 deg$^2$ COSMOS field \citep{cosmos-capak2007}, where additional dithers were observed by the survey for validation purposes. This allows us to compute a precise solution with enough stars across the focal plane, accurately determining the position, rotation and scale of the entire detector mosaic. The homogeneity of the reference astrometric catalogue used in the WCS calibration can be seen in Figure \ref{fig:gaia}. Figure \ref{fig:astro_residuals} shows the measured errors in both coordinates when calibrating the WCS with 100 overlapping exposures with respect to the reference Gaia catalogue and internally to PAU. From this validation analysis we estimate an absolute astrometric accuracy of $\sim$40 milli-arcseconds (mas) Root Mean Square (RMS) and an internal consistency of $\sim$30 mas RMS. This process only needs to be computed once (unless the astrometric reference catalogue is updated), producing a calibrated WCS of the PAUCam focal plane that is used during the nominal image processing.

\subsection{Single-Epoch astrometry and PSF modelling}
\label{s:astrocal_singleepoch}

At this stage, we perform the first extraction of sources in the single-epoch image, using \textsc{SExtractor} in a configuration mode specific for astrometry, with extended centroid measurements of windowed positions and a vignette matrix for each source. 

\begin{figure}
  \centering
  \includegraphics[width=0.8\linewidth]{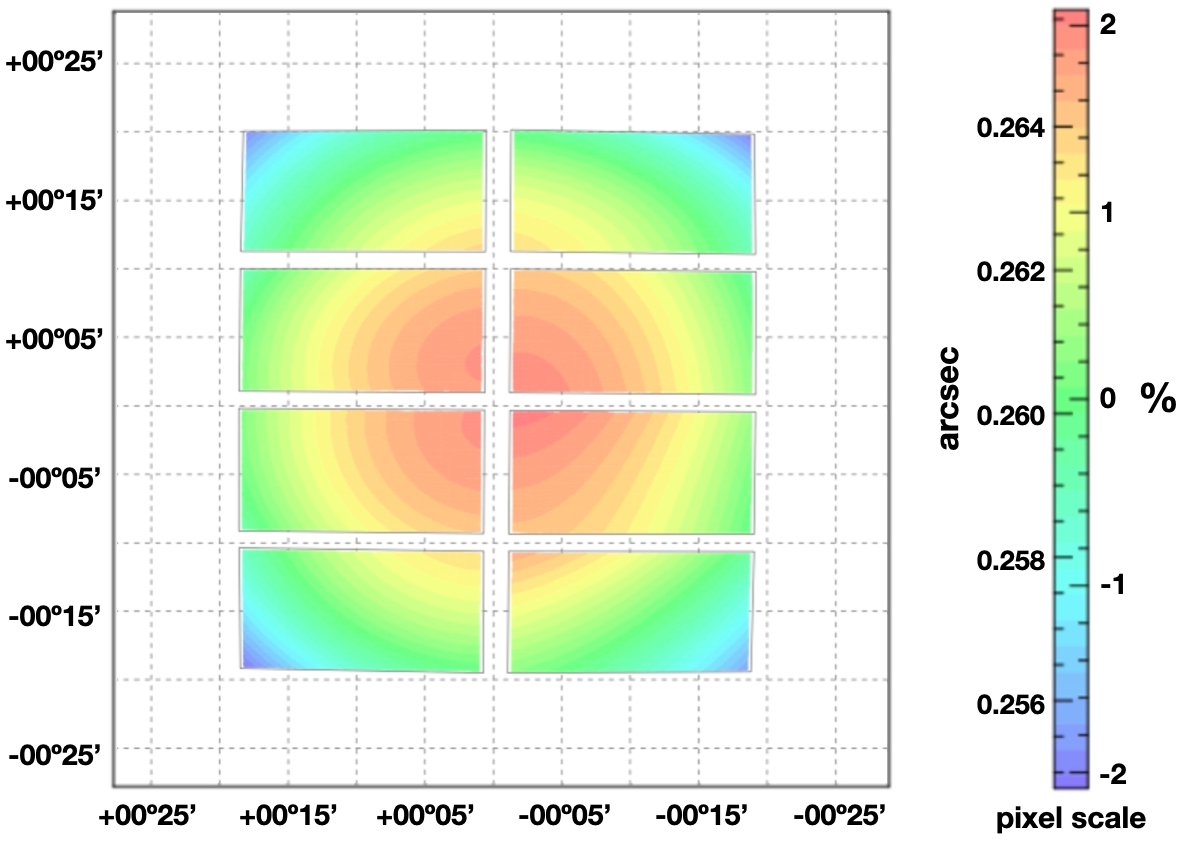}
  \includegraphics[width=0.8\linewidth]{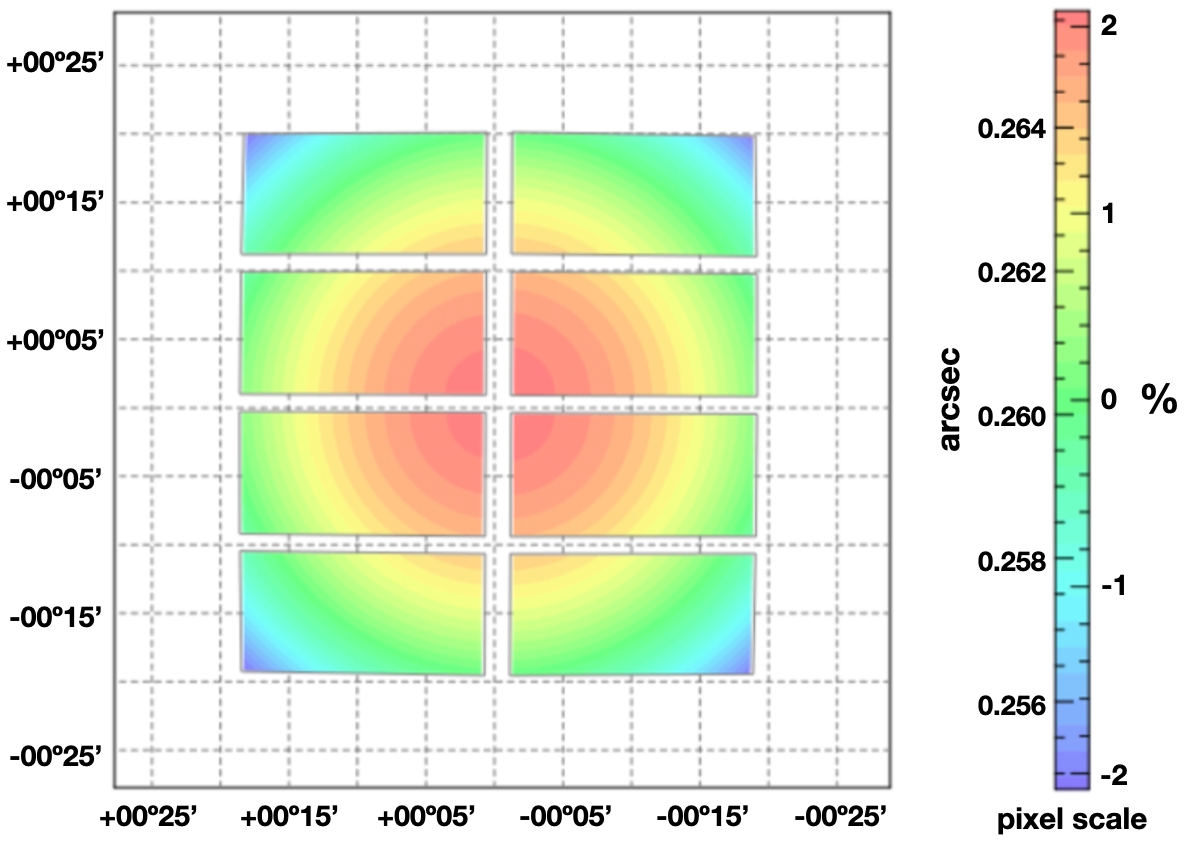}
  \caption{\emph{Top:} The spatial variation of the pixel scale found by \textsc{SCAMP} in a single-epoch exposure. \emph{Bottom:} The spatial variation of the pixel scale found by \textsc{SCAMP} with 100 overlapping exposures.}
  \label{fig:astro_distortion}
\end{figure}

To compute each exposure astrometric solution, we use again \textsc{SCAMP} but in a configuration that takes into account the precalibrated WCS (\texttt{MOSAIC\_TYPE SAME\_CRVAL}), helping the global mosaic solution and increasing the performance even with individual exposures. It provides an updated WCS header for the FITS file that includes the offset and distortion correction under a 3 degree polynomial fit. Even though \textsc{SCAMP} works best with overlapping exposures, we perform the astrometry solution on a single-epoch basis, as it already delivers 50 mas RMS of astrometric accuracy, below from what it is required to perform forced photometry (<100 mas). We have simulated a 50 mas aperture position error over an average galaxy size (1\farcs22) and an average PSF size (1\farcs2 Full Width Half Max or FWHM), resulting into a tolerable 0.02\% error in the flux.  This is only possible when using the base WCS calibration of the focal plane explained previously, serving as a first guess reference for the single-epoch corrections. Even though there are different filters over each CCD in the detector array, the global mosaic solution performs well and no chromatic distortions remain from \textsc{SCAMP}'s solutions. Figure \ref{fig:astro_distortion} illustrates the estimated pixel scale variations across the focal plane, for the WCS calibration run with 100 overlapping exposures and for the individual single-epoch solutions. The difference between the two analyses is minor, indicating a good solution even when using a single exposure. The focal plane exhibits a strong distortion pattern with variations of 4\% in the pixel scale between the center and the edges of the 8 central detectors.

Using the same astrometric catalogue with additional centroiding and profile information, we model the PSF across the focal plane with PSFEx \citep{psfex-bertin2011}, which delivers a variable PSF model that can be reconstructed at any position in the focal plane. When there are not enough stars in a single detector to properly model the variations, the average PSF FWHM value is given. This information is key to obtain accurate aperture photometry as described in \S \ref{s:photometry_aperturescaling}. The current aperture scaling uses the average PSF FWHM at each detector.

\section{Photometric Calibration}
\label{s:photocal}

In this section we briefly describe a key step in the data reduction process: the photometric calibration of the narrow-band images. The detailed description of the process is explained in 
\cite{paucalibration}. {\blue Note that our calibration is not absolute but relative to SDSS photometry, which has been shown to be accurate to 1\% in the bands (g,r,i) overlapping PAUS wavelength coverage (\citealt{2008ApJ...674.1217P}).}

\subsection{Overview}
\label{s:photocal_concept}

In a photometric night, the atmospheric extinction correlates linearly with airmass. Traditionally the calibration of astronomical images is performed taking observations of photometric standards \citep{standards-landolt1992} observed at different values of the airmass to compute the extinction coefficient. After the extinction model of the night is computed, it can be applied to the rest of the science exposures. Due to the non-standard filter set of PAUS and the variety of observing conditions in the survey, we had to design a particular process that allows us to calibrate the fluxes in each image observed. 

The approach presented in \cite{paucalibration} is to infer stellar templates from the SDSS broad band data and compute synthetic narrow-band photometry from the stellar templates. Synthetic photometry is then compared to the PAUS measurements of the same stars to obtain a zero-point (ZP) for each star. Even though the stellar templates inferred from the SDSS broad bands for a particular star can be wrong, combining zero-points from multiple-stars cancel errors of individual star zero-points and delivers accurate narrow-band calibration.

As we compute the photometric zero-point for every PAUS image independently and directly to the already calibrated SDSS synthetic photometry, we could observe under all sky conditions, even in non-photometric nights. This has been essential to maximize the use of available time in the observatory which is always limited and precious.

\subsection{Implementation}
\label{s:photocal_implementation}

The photometric calibration code is implemented inside the \textsc{Nightly} pipeline. It has two main steps: the star photometry and the zero-point calibration. For the first step, we run \textsc{SExtractor} on the instrumentally detrended and astrometrically calibrated images. We choose calibration stars that are moderately bright, comprising magnitudes between 14 and 19, that typically deliver a signal-to-noise ratio higher than 20. For such bright stars we do not need to optimize the aperture with complex and PSF-dependent methods that could be sensitive to the observation conditions or optical distortions in the focal plane. Instead we use a constant large aperture ($\sim$4\farcs radius) that gathers all the light from the star independently from the image PSF, ensuring that the truncated flux left outside the aperture is negligible even in the worst conditions tolerated by the survey. From simulations we verified that for the average survey PSF of FWHM 1\farcs1, the loss in flux is 0.03\% and in the worse case of 1\farcs8 seeing, the loss in flux is below 0.5\%. We tested various configurations of aperture sizes, background modelling and scatter-light correction, and the method using aperture photometry with a large aperture size was the most reliable across the different observing conditions. Once the photometry is processed, we perform a spatial matching with the SDSS DR12 catalogue \citep{sdssdr12-2015}, {\blue as these are the stars of interest for the photometric calibration of the narrow-band images due to their accurate $\sim1\%$ relative calibration }. We make use of the standard flags described in {\blue \textsc{SExtractor} }
documentation\footnote{https://sextractor.readthedocs.io/en/latest/Flagging.html} and only clean measurements (flag=0) are used to determine the star zero-points.

The second step is to compute the zero-point for each star, as well as the combined zero-point for each detector image by combining the individual star zero-points. For this process we provide to the calibration code the narrow-band fluxes measured in the uncalibrated image, attached to the ID of the reference SDSS catalogue with its broad band photometry. The calibration code computes the synthetic narrow-band fluxes for all the stellar templates and compares them to the observed uncalibrated fluxes of the selected stars in each image. It then computes the zero-point for each particular set of star-templates. Finally, {\blue or each narrow-band image},  the code returns a zero-point per star, weighted by the $\chi^2$ value of the stellar template fit and the synthetic broadband fluxes of the stellar templates used. 

The image zero-point is computed by the median of all the stars available. The median is more robust than a SNR-weighted average as it reduces the weight of underestimated errors in the brightest stars that would otherwise dominate and possibly bias the final measurement. Weighting all the stars equally produces a more uniform spatial sampling of the zero-point throughout the detector than inverse-variance weighting that typically determines the global zero-point with just the brightest stars only sampling the detector in a few points. Even though we just use the image zero-point in the calibration of the scientific catalogue, we store in the database all the data for stellar photometry, individual star zero-points and the combined image zero-points for validation purposes. In \S \ref{s:photometry} we detail how to use the zero-points computed to obtain calibrated fluxes and propagate the corresponding error.

\section{Forced Photometry}
\label{s:photometry}

The most important aspect that maximizes photometric redshift accuracy is preserving the colors of its measured bands. The narrow-band images of PAUS deliver a low SNR (<3) at the target magnitude of $i_{\rm AB}=23$ and therefore we need to perform the photometry based on external reference catalogues. With the information from the reference catalogues, we can compute Forced Photometry for the PAUS images, measuring the same fraction of light in each band. This would not be possible with the PAUS images themselves for the faintest objects, as the shape could not be properly estimated on sources with such low SN. From the reference catalogue we establish a consistent location, shape and scale of each object and define an aperture that preserves the flux fraction at all wavelengths. The only constraint for this technique is to have good astrometry accuracy as the positions of the apertures for the sources are defined blindly (no object detection or centroiding is done in the PAUS images). As we defined in \S \ref{s:astrocal}, we have a consistent astrometry at the sub-pixel level, enough for the purpose of Forced Photometry. As PAUS images have different seeing and PSF sizes, it is also important to model the PSF such that apertures are scaled accordingly for a constant flux fraction. This process is described in \S \ref{s:photometry_aperturescaling}.

\subsection{Reference catalogues}
\label{s:photometry_refcats}

Performing a forced photometry technique requires a reference catalogue that overlaps with the images observed by PAUS. There are two main aspects that the reference catalogues must have: the necessary parameters to perform the forced photometry accurately and the complementary galaxy lensing measurements that, in combination with the outstanding redshift accuracy of PAUS, deliver a unique scientific spot. Moreover, the reference catalogue must be complete down to the magnitude limit of PAUS such that the final combined catalogue has no target selection that could bias the cosmological measurements.

The selected PAUS fields are the Canada-France-Hawaii-Telescope Lensing Survey (CFHTLenS) fields W1, W3 and W4, the GAMA G09 field over the Kilo-Degree Survey (KiDS) North field and COSMOS. In the CFHTLenS catalogue \citep{cfhtlens-heymans2012} we have combined state-of-the-art reduction with \textsc{THELI} \citep{cfhtlens-theli-erben2013}, shear measurement with \textsc{lensfit} \citep{cfhtlens-wl-miller2013}, and photometric redshift measurements with PSF-matched photometry \citep{cfhtlens-photoz-hildebrandt2012}. In the case of the KIDS we use its latest release DR4 \citep{kids-dr4-kuijken2019}, also with outstanding cosmological lensing measurements \citep{kids-lensing-kuijken2015}. And finally, in COSMOS we built a merged catalogue from \cite{cosmos-laigle2016} and the Zurich Structure \& Morphology Catalog\footnote{Zurich COSMOS catalogue \url{https://irsa.ipac.caltech.edu/data/COSMOS/gator_docs/cosmos_morph_zurich_colDescriptions.html}} for the accurate shape information. COSMOS has been our main validation sample and, with so many multi-wavelength observations, it provided interesting photo-z tests with more than 70 bands (PAUS + COSMOS).

Furthermore, all these catalogues contain the information to perform the forced photometry measurements such as sky coordinates, a star-galaxy classification, a reference $i_{AB}$ magnitude, the scale of the source deconvolved from its observed PSF, the axis ratio to estimate its intrinsic ellipticity, the position angle and its S\'ersic index (or an equivalent parameter that allows us to infer the S\'ersic profile).

\subsection{Background modelling}
\label{s:photometry_background}

Accurate background subtraction is a key step to achieve precise photometry. The fluxes of the sources we need to measure sit on top of a floor of counts produced by the brightness of the sky, plus residuals of electronic bias and scatter-light. The latter is particularly important due to the configuration of the filter trays in PAUCam, causing significantly more light to be reflected and scattered in the edges of the narrow-band filter glass (see \S \ref{s:detrending_sl}). An additional complication of the scatter-light is that it can produce a non-homogeneous background, that is much harder to estimate and subtract. Underestimating the background will have a stronger impact on faint sources with few counts, as it will add a bias that scales with the size of the aperture.

\begin{figure}
  \includegraphics[width=\linewidth]{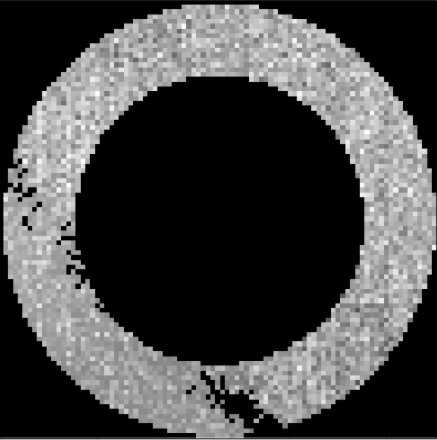}
  \caption{The annulus used to estimate the background of a particular source. It can be seen how pixels present in the image mask are discarded from the annulus and will not enter into the sigma clipping statistics.}
  \label{fig:annulus_mask}
\end{figure}

For all reasons stated above, a careful background estimation needed to be implemented. Amongst the different background subtraction options, we designed an annulus around each source where we wanted to perform forced photometry. This provides an accurate estimate that takes into account large-scale variations of the scatter-light. The annulus had to be placed at a close distance to pick up smaller scale variations, but not too close to introduce flux from the source itself. Taking into account the typical size of sources we aim to measure {\blue (1\farcs22)}, we set a fixed limit for the inner annulus at 30 pixels from the center of the target source. To get enough pixel statistics the outer annulus was set at 45 pixels. All bad pixels that fall into the background annulus are removed from the median statistics. A sample of an annulus with real pixels can be seen in Figure \ref{fig:annulus_mask}. Additionally, to avoid blending sources to affect the estimate, we perform a sigma clipping in the remaining pixels of the annulus, leaving only background free pixels in the average calculation. The median of the available pixels provides our estimate of the background in each source. This average value is multiplied by the area of the aperture and subtracted from the measurement in the main photometry process. The standard deviation and number of pixels used in the estimate is kept as it is used in the flux error estimate.

An alternative Neural Network method to estimate and subtract the background of PAUCam narrow-band images proposed in \cite{bkgnet-cabayol2020} seems to deliver very accurate modelling of the background. It is not yet implemented in production and is planned to be integrated in the main pipeline for future releases. 

\subsection{Aperture scaling}
\label{s:photometry_aperturescaling}

We correct here for the impact of an elliptical PSF on an elliptical aperture for a galaxy. For simplicity, we
present the results in circular coordinates ($a=b=r_0$), but these results can be  extended to elliptical apertures by just re-scaling the coordinate $(x,y)$ units using elliptical coordinates:
\begin{equation}
x'\rightarrow x=  r_0 q^{1/2} \cos{\theta} \,\,\, ;  \,\,\,
y' \rightarrow  y = r_0 q^{-1/2} \sin{\theta}\,,
\label{eq:elliptical}
\end{equation}
where $b \equiv q a$ is the smaller axis of the ellipses {\blue and $\theta$ the polar angle}. This scaling is tested later on in Fig.\ref{fig:aperture_scaling_validation} below.
The goal is to measure the fluxes in an elliptical aperture that corresponds to the same fraction of the total light after taking into account the effects of the PSF at the time and sky location when the image was taken.

\paragraph*{S\'ersic profiles and aperture fluxes}
\label{s:photometry_aperturescaling_sersic}

We assume a S\'ersic circular profile of slope $n$ and scale $r_0$ for the surface brightness distribution:
\begin{equation}
I(r)=I(0) \exp{[-(r/r_0)^{1/n}]}\,.
\end{equation}
The total luminosity in an aperture radius $r=A$ is:
\begin{equation}
L(A) = 2\pi \int_0^{r=A} \mathrm{d}r~r~ I(r)\,.
\label{eq:Aperture}
\end{equation}
For a S\'ersic profile, we can relate $r_0$ to the effective radius $r_{50}$, defined as the aperture which contains half the total light ($L(r_{50})=L(\infty)/2$): $r_0= [(0.86n -0.142) \ln{10}]^{-n} ~r_{50}$.

\paragraph*{Convolved profiles}
\label{s:photometry_aperturescaling_conv}

The observed surface brightness profile $I_{\rm o}(r)$ will result from the convolution of the intrinsic $I(r)$ and the PSF kernel $W_{\rm PSF}$. This is a 2D convolution, so even when the image is originally symmetric, $I(\vec x)=I(x)$,  the convolved image $I_o$ might not be symmetric:
\begin{equation}
I_o(\vec r) = \int_{\rm Image} ~ \mathrm{d}{\vec x } ~ I(x) ~ W_{\rm PSF}({\vec r }-{\vec x })\,.
\end{equation}
For a circular PSF: $W_{\rm PSF}({\vec{r}}-{\vec{x}})= W_{PSF}(|{\vec{r}}-{\vec{x}}|)$
we have:
\begin{equation}
I_o(r) = \int_0^{\infty} \mathrm{d}x  x I(x) \int_0^{2\pi} \mathrm{d}\theta  W_{\rm PSF}\left(\sqrt{r^2+x^2-2xr \cos{\theta}}\right)\,.
\label{eq:convolved}
\end{equation}

\paragraph*{Moffat PSF}
\label{s:photometry_aperturescaling_moffat}

We will use a radial Moffat PSF profile
\begin{equation}
W_{\rm PSF}(r)= \frac{\beta-1}{\pi \alpha} \left[ 1 + (r/\alpha)^2 \right]^{-\beta}
\end{equation}
which has a $ {\rm FWHM}= 2 \alpha \sqrt{2^{1/\beta}-1}$. 

\paragraph*{Effect of seeing on Aperture}

To obtain the same aperture flux {\blue $L(A)$ in Eq.\ref{eq:Aperture}  with a different PSF we need to change the aperture from $r=A$ to $r=A_o$ to account for the change in $I(r)$ to $I_0(r)$. This can be achieved} by solving
\begin{equation}
L(A) = L_o(A_o) \equiv  2\pi \int_0^{r=A_o} \mathrm{d}r~r~ I_o(r)\,,
\end{equation}
where $I_o$ is the convolved profile in   Eq.\ref{eq:convolved}.

\paragraph*{Stellar Apertures}
\label{s:photometry_aperturescaling_stars}

A point source convolved with a Gaussian profile of width $\sigma$ gives a Gaussian $I_o(r)$ and an aperture flux,
\begin{equation}
L(A)= \sqrt{2\pi} \sigma \left[ 1- e^{-0.5 A^2/\sigma^2} ~\right] 
= \sqrt{2\pi} \sigma \left[ 1- e^{-4 A_{F}^2 \ln{2}} ~\right] \,,
\end{equation}
where $A_{F} \equiv \frac{A}{ 2\sigma\sqrt{2\ln{2}}}$ in units of the FWHM. The fraction of light is then
$
\frac{L(A)}{L(\infty)} = 1- e^{-4 A_{F}^2 \ln{2}}
$
which for $A_{F}=1/2$ gives 50\% of light (FWHM). For a Moffat profile the fraction is only a bit smaller.

\paragraph*{Implementation}
\label{s:photometry_aperturescaling_implement}

Computing the convolution integrals for the 2D S\'ersic profile with a Moffat PSF model for each measurement (120 per galaxy on average) is prohibitive as this operation may take several seconds when it is not optimized. We take advantage of the very optimal implementation of the fraction of light radius calculation in \textsc{Galsim} \citep{galsim-rowe2015}. Even when \textsc{Galsim} claims this is only accurate down to a few percent, with these smooth profiles we achieve precision better than 1\% in most cases. This is precise enough as our PSF model and the estimated effective radius will dominate the error budget and will determine the final precision we can reach in estimating the aperture.

\begin{figure}
  \centering
  \includegraphics[width=\linewidth]{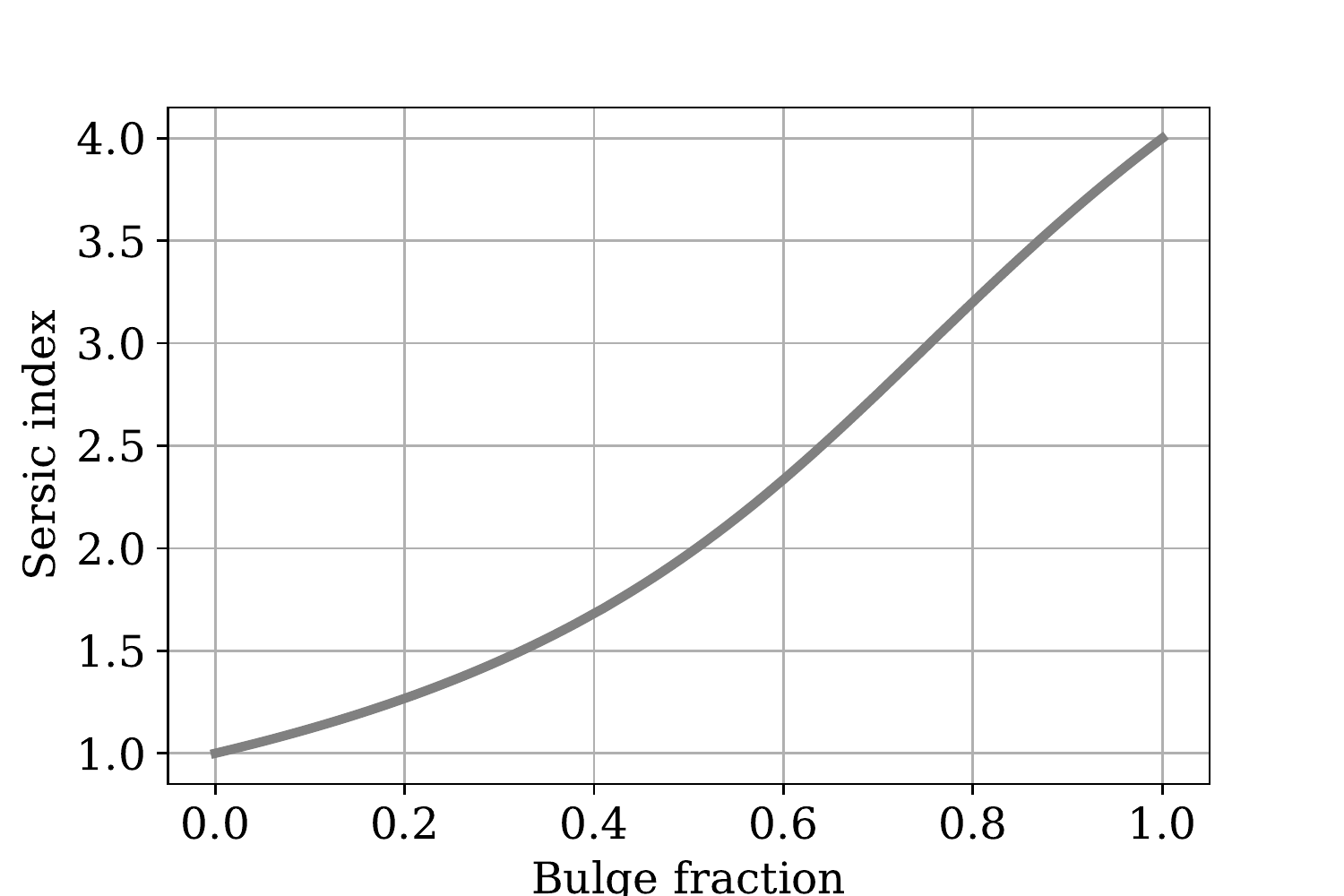}
  \caption{The relation between bulge fraction and S\'ersic index assumed for CFHTLenS and KiDS reference catalogues.}
  \label{fig:btt_to_sersic}
\end{figure}

The S\'ersic profile is generated with the effective radius of the reference catalogue and the S\'ersic index estimate. In case the S\'ersic index is not available in the reference catalogue, we assign an index of 4 to elliptical galaxies and an index of 1 spiral galaxies. In the case of CFHTLenS and KiDS we derive the S\'ersic index from the bulge fraction as shown in Figure \ref{fig:btt_to_sersic} (see \citealt{2019A&A...624A..92K}). 

\begin{figure}
  \centering
  \includegraphics[width=\linewidth]{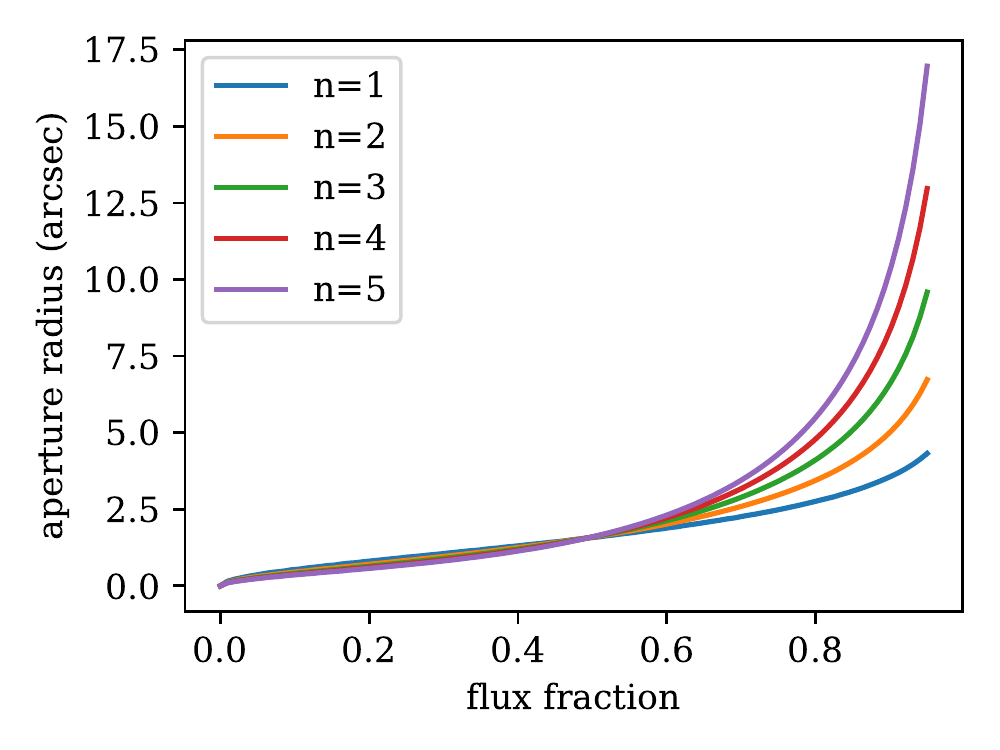}
  \caption{The aperture size radius as a function of flux fraction at various S\'ersic indexes. The sample source is a galaxy with an effective radius $r50$ of 1\farcs5 and a Moffat PSF with a FWHM of 0\farcs7 and a beta parameter of 4.75.}
  \label{fig:aperture_scaling}
\end{figure}

The PSF is modelled as a Moffat profile with a $\beta \sim 4.75$ \citep{moffat-trujillo2001} and the FWHM measured in PSFEx for each detector image. The \textsc{Galsim} function \texttt{calculateHLR} (Calculate Half Light Radius) is applied over the convolved object and it allows to compute either the HLR or a specific flux fraction. Typically we run the photometry over a variety of flux fractions, from 0.5 to 0.9. On average, the signal-to-noise is maximized around 0.625, even if the optimal SNR depends on each source. An example of how aperture size relates to flux fraction can be seen in Figure \ref{fig:aperture_scaling}.

As the r50 is defined as the effective radius on the major axis, we apply the same procedure for the minor axis, multiplying the r50 by the axis ratio defined in the reference catalogue. 

We verified the aperture scaling method, including Eq.\ref{eq:elliptical},
with a simple error-free simulation, rendering the models described previously (S\'ersic profile convolved with a Moffat PSF) at a reference flux and performing aperture photometry. For multiple combinations of PSF FWHM, galaxy scales and S\'ersic indexes, the reconstructed flux was always accurate within less than 1\%. Figure \ref{fig:aperture_scaling_validation} shows an example of a convolved galaxy with a circular and elliptical aperture at 62.5\% of light. Both cases estimate and recover accurately the flux fraction but the elliptical aperture delivers a higher SNR.

\begin{figure}
  \centering
  \includegraphics[width=\linewidth]{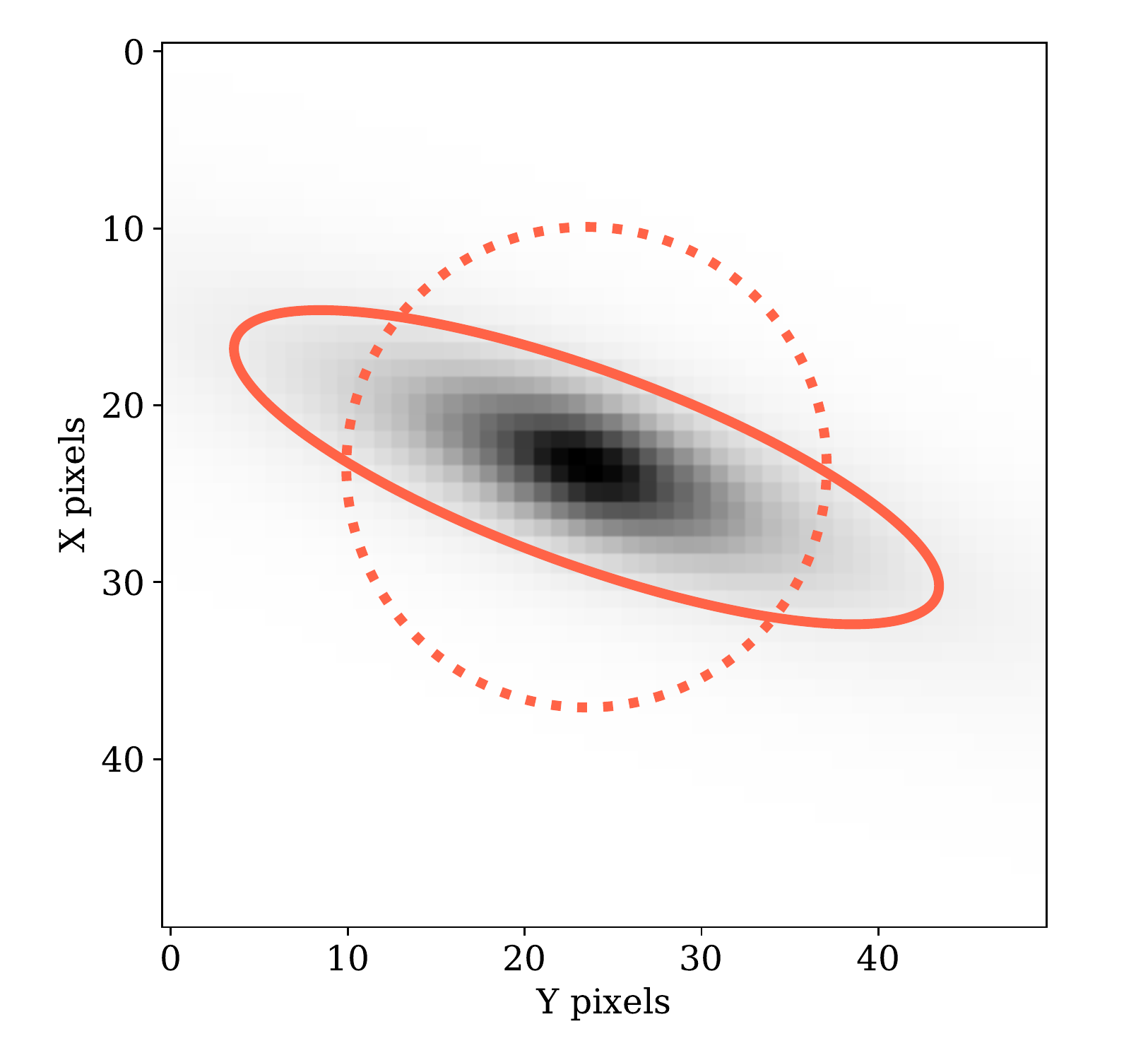}
  \caption{A simulation of an elliptical Sersic profile of 1\farcs9 and b/a = 0.2 convolved with a Moffat profile equivalent to a PSF of FWHM = 1\farcs2. \emph{Dashed line:} A circular aperture containing 62.5\% of light accurate to 0.3\%. \emph{Solid line:} An elliptical aperture at the same flux fraction with equivalent accuracy but with 60\% less area, improving the SNR of the measuremnt.}
  \label{fig:aperture_scaling_validation}
\end{figure}

\subsection{Flux measurement and error estimate}
\label{s:photometry_fluxmeasure}

In the forced photometry process we blindly place the aperture in the image, without any additional centroiding. As the broad-band reference catalogue is significantly deeper than the narrow-band PAUS images, we can rely on good and complete positions and shapes from the reference catalogue. At this stage the narrow-band images successfully passed all astrometry quality cuts and therefore, we can use the calibrated single-epoch WCS solution to determine the pixel positions of the source with sub-pixel precision. 

After computing both major and minor aperture radius from the previous step, we can perform aperture photometry with elliptical apertures. For that we make use of the \textsc{PhotUtils} library \citep{photutils-bradley2020}, an affiliated package of \textsc{Astropy} \citep{astropy:2018}. The aperture flux $f_{tot}$ is measured on the reduced image, still containing the component of the background. The bitwise information of all pixels inside the aperture is propagated to the flux measurement flag.

At the same location we estimate the background with an annulus as described in \S \ref{s:photometry_background}. The flagged pixels in the annulus do not enter into the combined flagging of the source, they are simply ignored in the statistics of the background calculation. The background modelling function estimates the median background flux $\overline{B}$, standard deviation $\sigma_{\rm bg}$ and number of valid samples in the annulus $N_{\rm bg}$. 

The background subtracted flux of the source $f_{src}$ is therefore:
\begin{equation}
{f_{\text{src}}} = f_{\text{tot}} - \overline{B} \cdot N_{\rm ap}\,,
\end{equation} 
where $N_{\rm ap}$ is the area of the aperture in pixels, taking into account the fractional overlap of the aperture and each pixel region.

The model of the error in the measurement $\sigma_{\rm tot}$ is composed by the source flux error ($\sigma_{\rm src}$), the noise in the background ($\sigma_{\rm bg}$) and the uncertainty in the estimation of the background. 

The source flux error is assumed to be a Poisson process, and it is estimated as the variance of the flux inside an aperture. As we perform the measurement over single-epoch images, we assume there are no pixel-to-pixel correlations as it would appear on a remapped image. Therefore the flux error of the source in electrons is:
\begin{equation}
\sigma_{\text{src}} ({\rm e}^-) = \sqrt{g \cdot f_{\rm raw}}\,,
\end{equation}
where $g$ is the amplifier gain in the detector readout system and $f_{\rm raw}$ is the raw detector counts of the source itself. As the measured units are flux rate in e$^-$/s, we need to scale the variance and the error such as:
\begin{equation}
{\sigma_{\text{src}}} ({\rm e}^- /{\rm s}) = \frac{\sqrt{f_{\text{src}} t_{\text{exp}}}} {t_{\text{exp}}} = \sqrt{\frac{f_{\text{src}}}{t_{\text{exp}}}}\,.
\end{equation}

Finally, we compute the total measurement uncertainty by adding in quadrature the flux error, the independent background noise and the background estimate error in the annulus such as:
\begin{equation}
{\sigma_{\text{tot}}} = \sqrt{ \frac{f_{\text{src}}}{t_{\text{exp}}} + \left(N_{\rm ap} + k \frac{N_{\rm ap}^2}{N_{\rm bg}}\right) \sigma_{\rm bg}^2 }\,,
\end{equation} 
where $k=\pi/2$ is the efficiency correction\footnote{https://wise2.ipac.caltech.edu/staff/fmasci/ApPhotUncert.pdf} for the median we used in our background estimation method.

\subsection{Flux co-addition}
\label{s:photometry_coadd}

So far all measurements were made at the single-epoch level. However PAUS observes each area on the sky multiple times for every filter. On average we perform 3 passes in the main fields (W1, W2 and W3) and 5 times in our calibration field (COSMOS - \citealp{cosmos-laigle2016}). Observing the same area multiple times has some advantages, such as covering the gaps between detectors, increasing the signal-to-noise, rejecting outliers, reducing the density of cosmic rays and increasing the dynamic range of the sources, as shorter exposure times will allow brighter stars not to saturate. This is done at the expense of an increased volume of data and a slower observing rate due to a constant readout time of 20s.

Most surveys combine their multiple layers at the image level, which is convenient for cosmic ray rejection (a median average almost completely removes all cosmic hits) but the stacking requires to resample the images, causing correlated noise which is complicated to model. Due to our objective of measuring very low signal-to-noise galaxies, we decided to stack the measurements at the catalogue level, performing all image measurements on individual exposures at their original pixel sampling. The combined measurements are called coadd fluxes.

Before combining the single-epoch aperture measurements, they need to be corrected to a standard system so all fluxes are consistent. Light rate from the same source may vary due to particular observing conditions like variations in the atmospheric extinction on non-photometric nights,  different telescope elevation resulting into different observing airmass or any other effect that varies the transmission with time. For this purpose we have calibrated each image and assigned a multiplicative factor ZP and its corresponding calibration error $\sigma_{\rm zp}$.

The calibrated single-epoch flux is simply defined as:
\begin{equation}
f_{\text{cal}} = f_{\text{src}} \cdot \text{ZP}
\end{equation}
and its calibrated error $\sigma_{cal}$, assuming non-linear error propagation with independent and not negligible variances, we derive:
\begin{equation}
\sigma_{\text{cal}} = \sqrt{\sigma_{\text{src}}^2  \sigma_{\rm zp}^2 + \sigma_{\text{src}}^2  \text{ZP}^2 + f_{\text{src}}^2\sigma_{\rm zp}^2}
\label{eq:caliberror}
\end{equation}

To avoid too small numbers that could require special data types, we added a magnitude offset of 26 in the calculation of the zero-point and thus, one could derive the AB magnitude from the PAUS calibrated flux as:

\begin{equation}
    m_{\rm AB} = -2.5 \log_{10}(f_{\text{cal}}) + 26\,.
\end{equation}

However with narrow-band photometry we are often dealing with sources that have flux close to zero and magnitudes are inconvenient at that level. All the processing and archive of source brightness is done at the flux level.

Now that we have a calibrated flux and its associated error for the individual measurements, we can proceed to combine all the repeated measurements of the same source and band into a coadd flux and error using an inverse-variance weighted average such as: 
\begin{equation}
f_{\text{coadd}} = \frac{\sum f_{\text{cal}_{\rm i}}  \sigma^{-2}_{\text{cal}_{\rm i}}} {\sum \sigma^{-2}_{\text{cal}_{\rm i}}}\,,
\end{equation}
where only non-flagged sources (\S \ref{s:photometry_flagging}) will enter into the combined measurements.

Assuming that the overlapping measurements are independent we estimate the coadd error as:
\begin{equation}
\sigma_{\text{coadd}}^2 = \frac{1}{\sum_{N} \sigma^{-2}_{\text{cal}_{\rm i}}}\,,
\end{equation}
where N is the number of unflagged measurements to be combined.

Additionally we compute the reduced chi-square $\chi^2$ as a measurement of consistency for the multiple measures:
\begin{equation}
\chi_{\text{coadd}}^2 = \sum_{\rm i}^N \frac{(f_{\text{cal}_{\rm i}} - f_{\text{coadd}})^2}{\sigma_{\text{cal}_{\rm i}}^2} / (N-1)\,.
\end{equation}

All three forced photometry coadd parameters are stored in the database for further processing and quality analysis.

\subsection{Flagging}
\label{s:photometry_flagging}

Throughout the whole processing of an image from its original raw state, we identify any possible cause that may affect the confidence of its value. To track each possible cause of problems we use flags both in the \textsc{Nightly} processing and in the \textsc{MEMBA} pipeline. In the image calibration process of the \textsc{Nightly} Pipeline we track the flags at the pixel level. Thus we created a mask image where each pixel contains the flag values of the corresponding pixel in the science image. In order to track all possible flag combinations in a single value, we have mapped each flag condition to a bit in the value of the pixel, allowing for 16 different flags in a 16-bit depth image. In PAUdm we have defined the following image-type flags:

\begin{itemize}
    \item Cosmetics: pixels not responding correctly to light, either hot ones that deliver constant high values or dead pixels that do not react to light inputs. Dust or imperfections in the detector mosaic or filter may appear here too. 
    \item Saturated: pixels with so much flux that reached the ADC limit (18-bit in the case of PAUCam)
    \item Cosmic Rays: pixels identified as cosmics rays in the Laplacian filtering algorithm \citep{cosmics-vandokkum2001}. Even though pixel values are interpolated from the neighbouring ones and cosmic rays may seem to have disappeared, the mask will keep track and the pixel value will not be used for science.
    \item Vignetted: areas in the focal plane with low transmission due to optical vignetting. The default value is set to 40\%.
    \item Crosstalk: pixels contaminated by a strong signal of crosstalk from a related amplifier or detector (\S \ref{s:detrending_ct}).
\end{itemize}

In contrast, in the \textsc{MEMBA} pipeline we perform photometry of sources and flagging will take place at the catalogue level for each source measurement. The flags in the image that overlap the aperture are propagated to the measurement flag. Additionally we have defined the following catalogue-type flags: 

\begin{itemize}
    \item Edge: source too close to the edge (< 80 pixels) or partially out of the image array.
    \item Distortion: source in an area with strong optical distortion (>50 arcmin from the focal plane center) or with an elongated PSF such that flux ratio in the aperture scaling may be inaccurate. 
    \item Scatter-light: source with intense and spatially dependent scatter-light that could compromise the background subtraction. We estimated the presence of scatter-light in the background with two methods: variance ratio and ellipticity ratio. In the first method, we simply compared the variance in the annulus around each source compared to a global variance in the background of the whole image. If the ratio was above a certain threshold (typically 5\%) we flag the source. This method was effective to flag sources in scatter-light areas but was not efficient as it was over-flagging sources that were simply on noisier areas. In the second method, we compute the ellipticity of the background image using the second order brightness moments of the sigma-clipped stamp around each source, defined as:
    \begin{gather}
        q_{xx} = \sum_{xy} I(x,y)(x-\overline{x})^2 \Delta x \Delta y\,,\\
        q_{xy} = \sum_{xy} I(x,y)(x-\overline{x})(y-\overline{y}) \Delta x \Delta y\,,\\
        q_{yy} = \sum_{xy} I(x,y)(y-\overline{y})^2 \Delta x \Delta y
    \end{gather}
    and from these quadrupole moments we build the ellipticity of the stamp as:
    \begin{equation}
        \epsilon = \frac{q_{xx} - q_{yy} -2iq_{xy}}{q_{xx} + q_{yy} + 2 \sqrt{q_{xx}q_{yy}- q_{xy}^2}}\,,
    \end{equation}
    where real component measures deviations from circle along axes and imaginary component along the main diagonals \citep{great08-bridle2009}. {\blue Effectively this measurement estimates gradient variations in the background pixels around the target sources.}
    
     To obtain a reference ellipticity value, we compute the median ellipticity of the whole image scanning the detector in steps of 25 by 25 pixels and we flag those measurements with a background ellipticity larger than 10x the median of the image. This has proved to be an efficient method to track scatter-light residuals, flagging only sources with non-reliable background subtraction. 
\end{itemize}

As described in \S \ref{s:photometry_coadd} we skip all measurements that contain any flag inside its aperture for the combination of the coadd measurement. {\blue Approximately 9\% of measurements are flagged and therefore do not enter into the final coadd average}. The full list of flags and its bit mapping value can be found in Appendix \ref{s:flags}.

\subsection{Survey mask}
\label{s:photometry_mask}

The particular layout of the narrow-band filters in the camera trays and the fact that CCD detectors are separated, leave gaps in the focal plane and result in a non-homogeneous coverage of the sky for each pass band. Additionally there are telescope pointing errors that result in a more in-homogeneous sky coverage. In large cosmology surveys that intend to identify statistical correlations of galaxy positions and densities, it is mandatory to accurately identify how the survey has tiled the sky with its thousands of exposures. 

For this purpose we have built a survey mask with two levels of information. First we generate the \emph{exposure mask}, where we define for each filter how many times we observed each area in the sky for an entire field with a resolution of 5 arc-seconds. The mask is built taking into account variations in the system response from the flat-field and flagged pixels. This will create a complex mask that introduces effects like vignetting, bright stars that saturate or corners in the detector not visible due to mechanical pieces in the optical path.

\begin{figure}
  \centering
  \includegraphics[width=\linewidth]{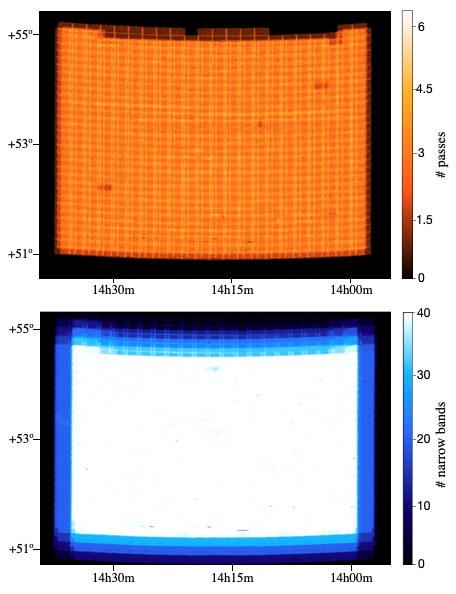}
  \caption{The PAUS Survey mask in the 25 deg$^2$ W3 field. \emph{Top:} The exposure mask of narrow-band NB655 indicating the number of overlapping observations of this filter on the sky. \emph{Bottom:} The bands mask indicating the number of narrow-bands with at least one observation in each part of the sky.}
  \label{fig:survey_mask}
\end{figure}

The second level of mask is the \emph{bands mask} and it is built from the combination of all 40 exposure masks for each band. It represents the same area in the sky as the exposure masks but contains the number of bands available in each location of the sky with one or more effective units of exposure. An example of how both levels of mask can be seen in Figure \ref{fig:survey_mask}.

The survey masks are associated with \textsc{MEMBA} runs, as the resulting coadd catalogue is built with a set of images and this same set is the one used to build the masks. As we are performing forced photometry from an external catalogue, the selection of sources will not depend on PAUS observations and we will need access to the survey mask provided by the external survey. However the final survey mask must be created as an intersection between the external mask and the PAUS survey mask. 
 
We used \textsc{SWARP} \citep{astromatic-bertin2002} from Astromatic\footnote{https://www.astromatic.net/} to remap and build image stacks. We process the science images and their corresponding flat-field maps as the weight map of an entire field with \textsc{SWARP}. We obtain the exposure mask as the resulting combined weight map provided by \textsc{SWARP} in \texttt{"MAP\_WEIGHT"} mode. As a by-product we obtain the science-stacked image even though this is not used as part of the main science processing.

\section{Quality assurance \& validation}
\label{s:validation}

One of the major challenges in PAUdm is the volume and complexity of the data to be processed and analyzed. Contrary to spectroscopic surveys that can obtain a spectrum in a single observation (or stacking a reduced number of individual spectrum), each galaxy in PAUS is composed by more than 120 measurements. Furthermore a single problematic image can impact thousands of galaxies, altering its 40 narrow-band spectrum and causing catastrophic outliers in its photometric redshift determination.

Tuning an algorithm to process a small dataset is simple, as one can manually verify the correct behaviour of the processing and its output result. However with such large and complex dataset we need to build automatic control systems that verify that a particular code or configuration worked for the entire volume of data and raise an alarm or discard the data that did not meet some specific requirement.

\subsection{Quality controls}
\label{s:validation_qc}

At the time of writing this publication, the PAUS data management system has processed more than 7 million images of PAU, and the number is increasing with additional reprocessing of data. The complexity and volume of this set requires an automated data quality control system to ensure that the data products meet the expected requirements. Although the most imaginative and cautious developer will miss the variety of circumstances that data from an observatory can contain. From closed petals of the main mirror, to dust from the Sahara desert or even vapor condensation in the entry window of the camera in extreme weather conditions. These are some of the unpredictable conditions that we must catch to reject bad quality exposures and request observations to be repeated.

With this particular aim, we built a quality control system associated with the \textsc{Nightly} pipeline. We define the following quality control tests with the corresponding tolerance limits in each metric to classify an image as valid:
\begin{itemize}
    \item Readnoise: check that electronic readnoise is under specification. This is measured in the overscan region of each amplifier. The default limit is set to 20$\mathrm{e^-}$.
    \item Flat-field level: check that the flat-field image is illuminated in the correct range of values. Too bright illumination could result into saturation and too faint illumination would increase the noise of the master flat-field. The default range is set between 1000 and 120.000 ADUs.
    \item Saturation: check that the science images do not contain too many saturated pixels. A certain amount of saturated pixels are expected due to bright stars in the field. However too many satureated pixels are indicative of an issue in the exposure time, electronics or target selected. The default limit is 0.1\% of saturated pixels (average is $\sim$0.02\%).
    \item Cosmic rays: check that the cosmic ray detection algorithm does not classify too many cosmic ray pixels. Issues in the electronics or very noisy images may affect the sensitive CR detection algorithm and end up with too many pixels being classified as cosmics. The default limit to reject an image is 1\% of pixels (average is $\sim$0.05\%).
    \item Astrometry: check that the contrast (the ratio of the amplitude of the detected peak to the amplitude of the second highest peak found in the cross-correlation) and $\chi^2$ to the reference catalogue are good enough to ensure \textsc{SCAMP} found a reliable solution. This is critical as we rely on the single-epoch astrometry and images with high extinction may end up with too few stars to deliver a solution. The default limits are contrast greater than 3 and reference $\chi^2$ below 50.
    \item Seeing: check that the average image PSF FWHM measured by PSFEx is below a certain value. Large PSFs reduce the signal-to-noise and limit the target sources of interest. The default limit is 1\farcs8. 
    \item Calibration stars: check that there are enough stars matched with SDSS to be used for the photometric calibration. Images observed under high extinction atmospheric conditions may reduce the number of available stars. The default limit is set to 5 stars and approximately 5\% of images do not meet the required value. 
    \item Zero-point error: check that the estimated error in the photometric zero-point is constrained. An unusually high ZP error may be due to a non uniform response across the detector. The default limit is set to 0.2 (flagging $\sim$3\% of the images).
    
\end{itemize}

The quality controls processed in each job are aggregated and propagated to the parent jobs so quality issues in large processing sets with many dependencies can be tracked easily. 
\begin{figure}
  \centering
  \includegraphics[width=\linewidth]{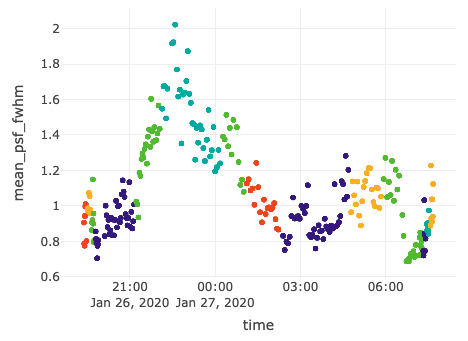}
  \includegraphics[width=\linewidth]{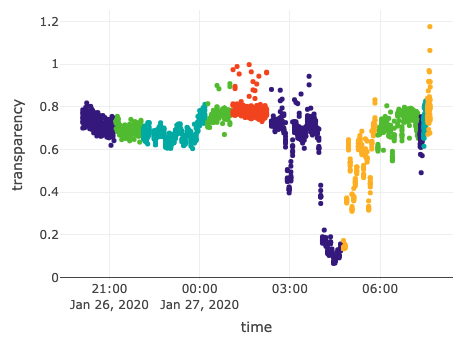}
  \caption{Some of the quality control plots that the nightly report produces as a result of the nightly pipeline processing. This data corresponds to one of the last nights of PAUS before the COVID-19 pandemic break. Each color represents a different narrow-band filter tray. \emph{Top:} Evolution of the atmospheric seeing and how it got quite high before midnight, stabilizing low before the end of the night. \emph{Bottom:} Evolution of the transparency, where some clouds entered at around 4am, reducing the absolute transparency to almost 10\%.}
  \label{fig:nr_plots}
\end{figure}

\subsection{Nightly report}
\label{s:validation_nightlyreport}

Periods of observation typically last for 1 to 2 weeks. During this time it is very important to provide feedback to the astronomers in the observatory in the shortest time. This was one of the key constraints in the design of the PAUS data management system and we managed to process the whole night data set before the next night starts ($\sim$8 hours).

The \textsc{Nightly} report is a web-based application that provides feedback to the astronomers on the quality of the data from the previous night so that observers can reschedule observations that did not meet a certain quality. It has also been critical to identify issues in the camera or telescope that were fixed with minimal delay for the remaining observing run. The \textsc{Nightly} report has default quality limits necessary for the survey and generates a report file that can be ingested directly to the PAUCam control system for re-scheduling targets. The application also displays statistics and evolution plots for each night, with options to adjust the metrics to be analyzed and its time span, such as the ones in Figure \ref{fig:nr_plots}. A total of 35 parameters can be displayed to help PAUS astronomers understand the atmospheric, weather and instrument behaviour from any previous night.

In addition to the quality checks described in \S\ref{s:validation_qc}, the \textsc{Nightly} report displays the processing status of the main blocks in the \textsc{Nightly} pipeline: detrending, astrometry, PSF modelling and photometric calibration. The status in each block can be used as quality cut (i.e. repeating all observations where PSF modelling failed in any detector). Approximately 40\% of the exposures did not meet the image quality requirements imposed by the survey and had to be rejected due to bad weather or any other possible issues.

\subsection{Comparison with SDSS and VIPERS spectra}
\label{s:validation_spectra}

A very reliable reference to validate PAUS measurements are high-resolution spectra from external surveys. The selection of the calibration fields was in fact driven by the overlap with datasets of carefully calibrated spectra. Our main validation field COSMOS contains $\sim$17.000 sources with good reference spectra. Additionally our main fields contain other datasets such as VIPERS \citep{vipers-lefebre2013} in W1 or DEEP2 \citep{deep2-newman2013} in W3, complementing the already extensive set in COSMOS. From the calibrated spectra, we can infer the synthetic narrow-band and perform a direct comparison to the PAUS measurements. Some examples are shown in Appendix \ref{s:app_synth}. Galaxy sources with a spectrum provide accurate spectroscopic redshift estimations which allow us to know where the expected emission and absorption lines are, providing additional confidence on a particular spectral shape.

We use spectroscopic data from SDSS and VIPERS to predict the \textsc{MEMBA} flux measurements. We then compare the predictions with the actual measurements to test \textsc{MEMBA} and the previous data calibration steps.

\begin{figure}
  \centering
  \includegraphics[width=\linewidth]{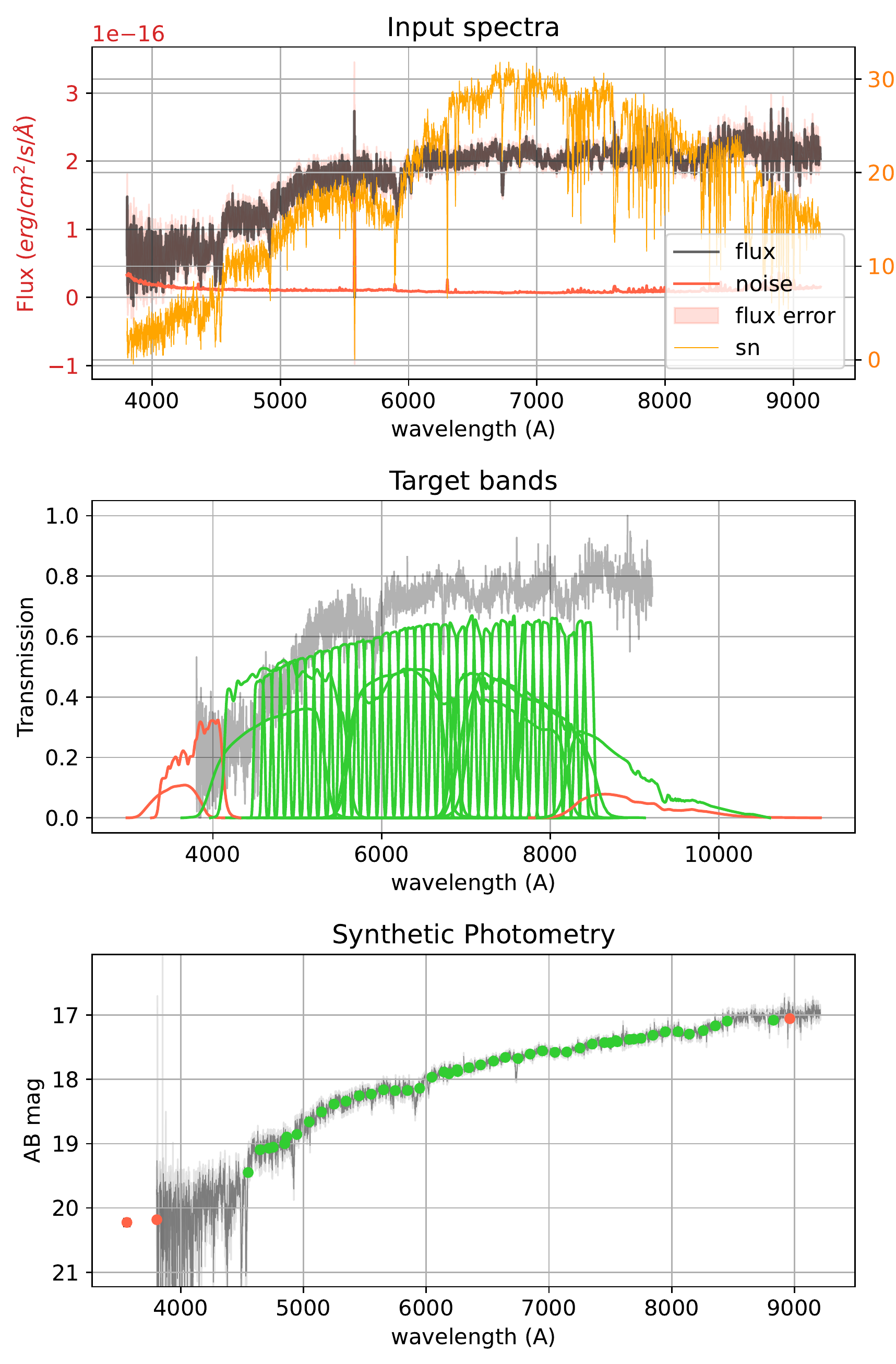}
  \caption{\emph{Top:} A galaxy sample at z=0.14 from SDSS used as input for synthetic photometry. The left axis (in red) represents the flux and its error while the right axis (in orange) represents the SNR. \emph{Middle:} The target bands are the 40 narrow-band set from PAUS plus the two broad band systems from SDSS and CFHT. \emph{Bottom:} The computed synthetic photometry from the hi-resolution spectrum. The bands without enough unmasked samples from the spectrum are marked in red.}
  \label{fig:sdss_input_spectra}
\end{figure}

\subsubsection{Synthetic narrow-band photometry}
\label{s:validation_spectra_synth}

In order to compare spectra with PAUS narrow-band photometry, we need to generate synthetic photometry from the high-resolution spectra. This provides a high quality reference to compare with any passband, especially valuable with the non-standard PAUS filter system. This comparison will only be possible with objects that both are observed by a spectroscopic survey and by PAU.

The first step in the process of generating the synthetic bands is to retrieve and homogenize the spectral data. In our case we have converted all fluxes to a more common $f_\lambda$ with units of erg$/$cm$^2/$s$/$\textup{\r{A}}. Generally, the data set contains the wavelengths where the spectrum is sampled, the fluxes, the noise (or inverse variance) and a mask. Optionally SDSS also includes a measurement of the sky, that allows to identify possible contamination in strong emission or absorption lines. Second, we interpolate the bandpass response $R(\lambda)$ to the sampling of the spectral data, as these two are not necessarily in the same space. Then we mask both the spectral fluxes and the passband with the flagged measurements from the spectrum mask. At this point we can compute the integrated average flux density of the source at the specific passband in erg$/$cm$^2$$/$s$/$Hz such as:
\begin{equation}
\langle F_\nu \rangle = \int \frac{f_\lambda R(\lambda) \lambda^2}{c}  \mathrm{d}\lambda
\end{equation}
and the its associated integrated response:
\begin{equation}
R_{\rm i} = \int R(\lambda) \mathrm{d}\lambda\,.
\end{equation}

Finally we can compute the synthetic magnitudes in the AB system with the following transformation:
\begin{equation}
m_{\rm syn} = -2.5 \left( \log \frac{\langle F_\nu \rangle}{ R_{\rm i} } \right) -48.6\,.
\end{equation}

It is also important for the statistical analysis to estimate the error of each synthetic band. As the flux in the spectrum has been weighted by the response of the transmission, we must weight the noise in the spectrum by the relative transmission throughout the entire passband:
\begin{equation}
\sigma_{f_\nu}^2 = \int  \frac{R(\lambda)^2 \sigma_\lambda^2 \lambda^2 }{c^2 R_{\rm i}^2}  \mathrm{d}\lambda\,,
\end{equation}
where $\sigma_\lambda$ is the noise in the high-resolution spectrum. 
We can approximate the magnitude error such as:
\begin{equation}
\sigma_{m_{\rm syn}} \approx 1.0857  \frac{\sigma_{f_\nu}^2}{\langle F_\nu \rangle/ R_{\rm i} }\,.
\end{equation}

Following the previous procedure we compute the photometry over all VIPERS spectra and all SDSS spectra that overlap with PAUS over the 40 PAUS narrow-band set and the SDSS and CFHT broad band systems. We have flagged all measurements where the overlap between the systems response and the unmasked spectrum is below 70\%. An example of synthetic spectrum with SDSS over PAUS narrow-bands and other broad bands is shown in Figure \ref{fig:sdss_input_spectra}. More examples together with PAUS real observations are shown in Appendix \ref{s:app_synth}.

\begin{figure}
  \centering
  \includegraphics[width=\linewidth]{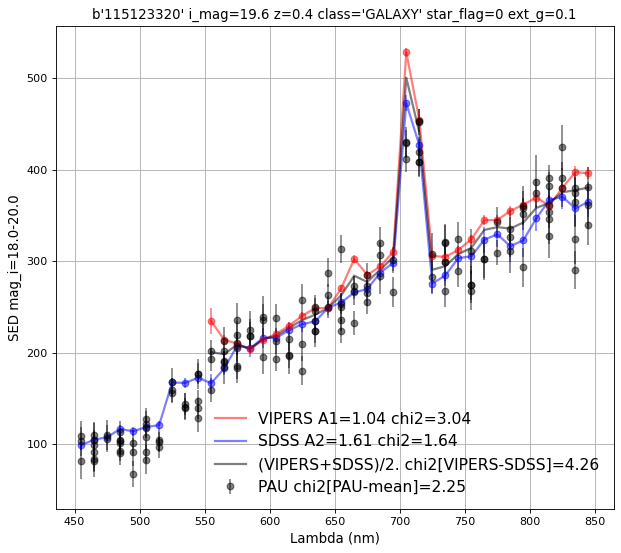}
  \includegraphics[width=\linewidth]{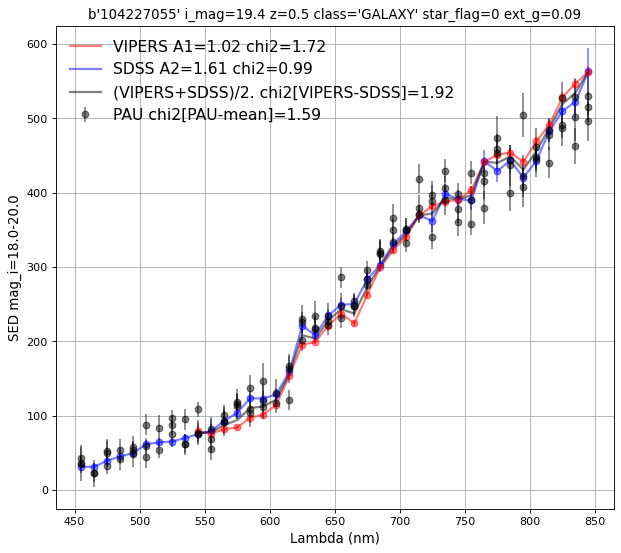}  
  \caption{Example of validation spectrum for the photometric calibration in PAUS (points with errorbars {\blue corresponding to 5 independent exposures before co-ad in \S\ref{s:photometry_coadd}}) with synthetic narrow-band photometry from SDSS (blue) and VIPERS (red) galaxy spectrum. SDSS (VIPERS) spectra have been multiplied by $A_1$ ($A_2$), as shown in the labels, to account for possible differences in the aperture used in each observation. The "chi2" label shows the normalized $\chi^2$ as compared to PAUS data.}
  \label{fig:SEDvipersSDSSpaus} 
\end{figure}

\subsubsection{Re-calibration of spectra}
\label{s:validation_spectra_recalib}

To account for remaining aperture or PSF effects in the measured spectra we use total broad band (BB) photometry in the corresponding reference survey (SDSS or VIPERS) to re-calibrate each individual spectrum.
To do this we first estimate synthetic broad-bands from the spectra, $F_S(BB)$, as shown in previous section. We then use the BB measured flux $F_O(BB)$ to find a multiplicative zero-point, $ZP$,  which is in general different for each BB:
\begin{equation}
ZP(BB) = \frac{F_O(BB)}{F_S(BB)}\,.
\label{eq:Rsdss}
\end{equation}
We use ZP to re-scale each individual spectrum. In the cases where we have 2 (or 3) BB measurements fully within the spectra wavelength coverage we combine them using a fit to a linear (or cubic) function ZP=ZP$(\lambda)$, where $\lambda$ is the mean of the bandpass response $R(\lambda)$. Each synthetic narrow-band $\lambda_{\rm NB}$ from the spectrum is re-scaled by ZP$_{NB}=$ZP$(\lambda_{\rm NB})$. 
{\blue  The mean re-calibration is only a 2\% offset with a 5\% scatter: ZP$=1.02\pm 0.05$. Similar results are found for VIPERS (see  \cite{paucalibration} for more details).}

\subsubsection{Aperture corrections}

Once the SDSS spectra are re-calibrated with Eq.\ref{eq:Rsdss}, we also perform aperture correction of the amplitude of each individual spectrum $(S)$ to the PAUS measurements. This is a fit to a linear constant $A=A(S)$ 

\begin{equation}
    A(S) = \frac{ \sum_{\rm i} f_{\rm PAUS}(S,{\rm i}) \, f_{\rm SDSS}(S,{\rm i}) } { \sum_{\rm i} f_{\rm SDSS}^2(S,{\rm i}) }     
\label{eq:Afit}
\end{equation}
between PAUS raw fluxes $f_{\rm PAUS}$ and SDSS re-scaled synthetic spectral $f_{\rm SDSS}$ (including the spectral recalibration). The sum is over individual PAUS measurements $i$ in a given spectrum $(S)$ and it
uses inverse variance weighting $w_i=1/\sigma_i^2$, where $\sigma_i$ is the joint error (from SDSS and PAUS) added in quadrature.
Typically there are 200 PAUS independent measurements (40 narrow-bands times 5 exposures in COSMOS) for each SDSS spectrum. 

{\blue 
We study the distribution of values of $A$ for different SDSS star calibration spectra and 42420 independent measurements for PAUS run \#955 in COSMOS. 
We find a mean value and scatter of $A=0.98 \pm 0.02$, which indicates that PAUS data is in very good agreement overall with the SDSS calibration within 2\% overall scatter.}

Figure \ref{fig:SEDvipersSDSSpaus} shows a comparison of SDSS, VIPERS and PAUS estimated narrow-band fluxes for two typical SED examples of galaxies (at $z=0.4$ and $z=0.5$) with $i_{AB} \simeq 20$.  One can see the corresponding figure for stars in Fig.12 of 
\cite{paucalibration}.

\begin{figure}
  \centering
  \includegraphics[width=\linewidth]{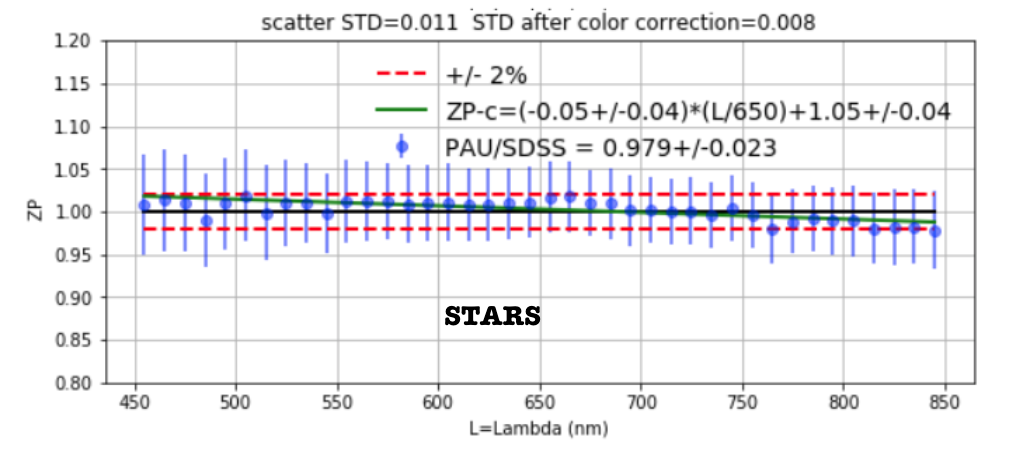}  
  \includegraphics[width=\linewidth]{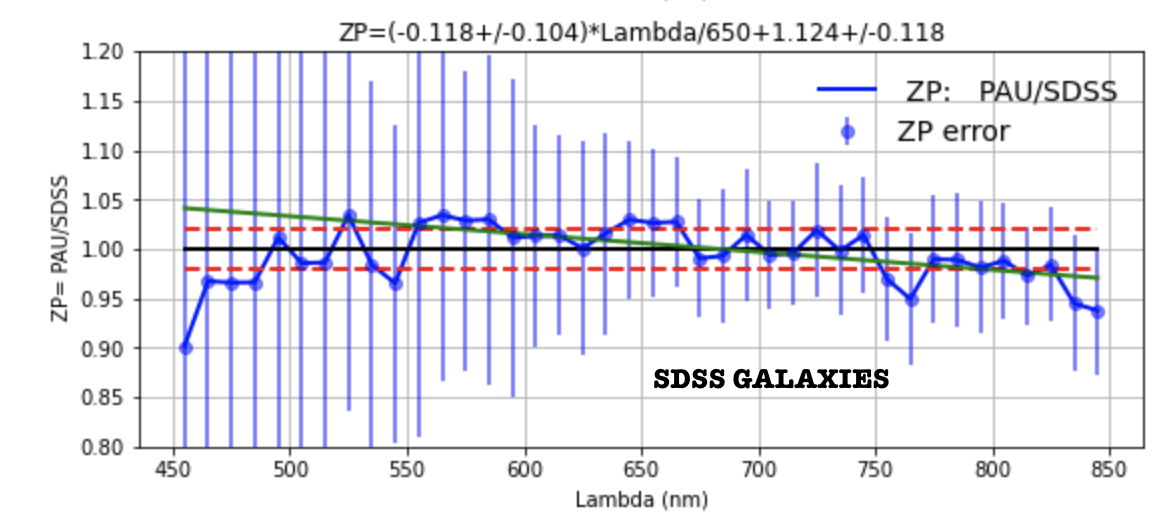}
  \includegraphics[width=\linewidth]{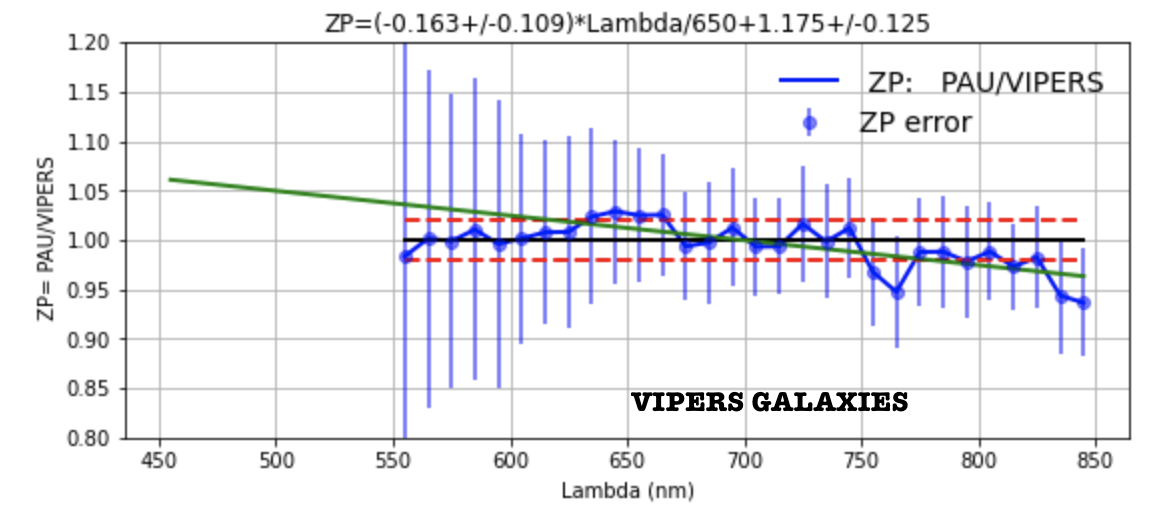}  
  \caption{A validation study of the photometric calibration in PAUS (ZP=PAU/SDSS) using synthetic narrow-band photometry from SDSS stars spectra (\emph{Top}), SDSS galaxy spectra  (\emph{Middle})  and VIPERS galaxy spectra (\emph{Bottom}). Blue errorbars indicate the scatter in the values, which is much larger for galaxies.}
  \label{fig:validation_pau_calib_galaxies} 
\end{figure}

\begin{figure*}
  \centering
  \includegraphics[width=\textwidth]{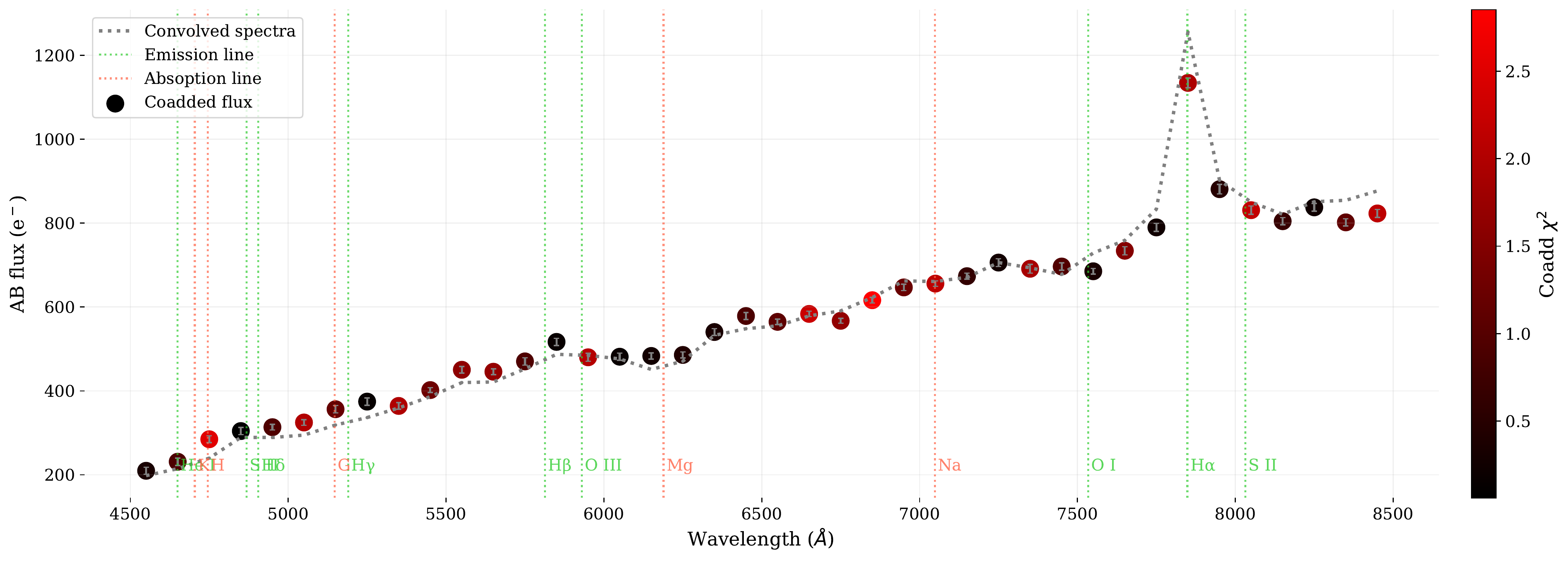}
  \caption{A bright galaxy at z=0.2 from the COSMOS field. The aperture inspector displays the coadd measurements from \textsc{MEMBA} and overlays a synthetic narrow-band photometry from the corresponding SDSS hi-resolution spectrum (when available). With the emission and absorption lines depicted, the redshift solution can be confirmed, specially on emission-line galaxies such as the example above. The redder points correspond to higher $\chi^2$ values, caused by discrepant single-epoch fluxes, suggesting possible issues in the combined measurement.}
  \label{fig:fa_inspector}
\end{figure*}

\subsubsection{Color terms}

We now check for any residual differences as a function of narrow-band wavelength $\lambda$ using galaxies with SDSS and VIPERS spectra. Figure \ref{fig:validation_pau_calib_galaxies} shows the mean and scatter zero-point difference for each narrow-band $\lambda$
\begin{equation}
    {\rm ZP}(\lambda) = \frac{ \sum f_{\rm PAUS}(\lambda) \, f_{\rm SDSS}(\lambda)} { \sum f_{\rm SDSS}^2(\lambda)}     
\end{equation}
between PAUS raw fluxes $f_{\rm PAUS}$ and SDSS or VIPERS re-scaled synthetic spectral $f_{\rm SDSS}$ (including the aperture correction $A$ in Eq.\ref{eq:Afit}). The sum is over all individual PAUS measurements (42420 in total) and uses inverse variance weighting $w=1/\sigma^2$, where $\sigma$ is the join error added in quadrature. 

In the top panel we use SDSS calibration stars which show a much smaller scatter than SDSS galaxies (middle). This is because we use fainter galaxies and also because the aperture correction becomes more important for extended objects.

We find a small residual color tilt between the SDSS and PAUS narrow-band systems when using SDSS star spectra to compare
\begin{equation}
    {\rm ZP}(\lambda) =  1.05 \pm 0.04 - (0.05 \pm 0.04) \left(\frac{\lambda}{650nm}\right) 
\label{eq:tilt}    
\end{equation}
which is consistent with unity within errors.
{\blue This correction is not applied to the data in MEMBA, but we state it here as a potential source of correction that should be applied, if needed.}
The scatter between the 40 bands after correcting for this linear residual slope is just 0.8\%. The scatter increases from 0.8\% to 1.1\% without this linear color correction.

\subsection{Forced aperture inspector}
\label{s:validation_fainspector}

With such a large narrow-band set and with the additional overlapping exposures, each object depends on the correct reduction and calibration of hundreds of images. There are many things that can go wrong, even if all quality tests passed in the processing of an image. It is critical to identify the source of issues that may end up in outliers on PAUS spectra and that the photo-z code could misinterpret as a physical feature of the galaxy, resulting in a catastrophic redshift determination.

We built another quality control web application where PAUS measurements for a particular source are displayed together with the synthetic measurements from a reference spectra. When the redshift is available, we also display the emission and absorption lines in the expected wavelength position. This allows us to visually inspect PAUS data and easily identify outliers or discrepancies with the reference spectrum. The application allows to click on a particular suspicious PAUS band, displaying the measurements that contribute to that band. Clicking again on a single measurement will display the portion of that particular image together with the aperture done in \textsc{MEMBA} and all quality parameters associated to the image or observing conditions. Very rapidly we can identify issues from the final PAUS spectrum to the original image that contributed to every measurement.

As this resulted a very powerful tool and many scientists from the PAUS collaboration contributed to it, we added a reporting system where people inspecting the sources could specify the issue that caused a particular outlier. The list of possible issues grew up as we learned more and more about the PAUS data, and we ended up with a list of 18 possible issues, such as scatter-light, blended source, crosstalk, astrometry issue and more. Even though this is a subjective test that required some training, we could extract valid statistics and correct for multiple systematics that originally caused trouble. An example of a the 40 bands of a galaxy in the Forced Aperture Inspector can be seen in Figure \ref{fig:fa_inspector}.

\subsection{Duplicate observations test}
\label{s:validation_duplicates}

\begin{figure*}
  \centering
  \includegraphics[width=\linewidth]{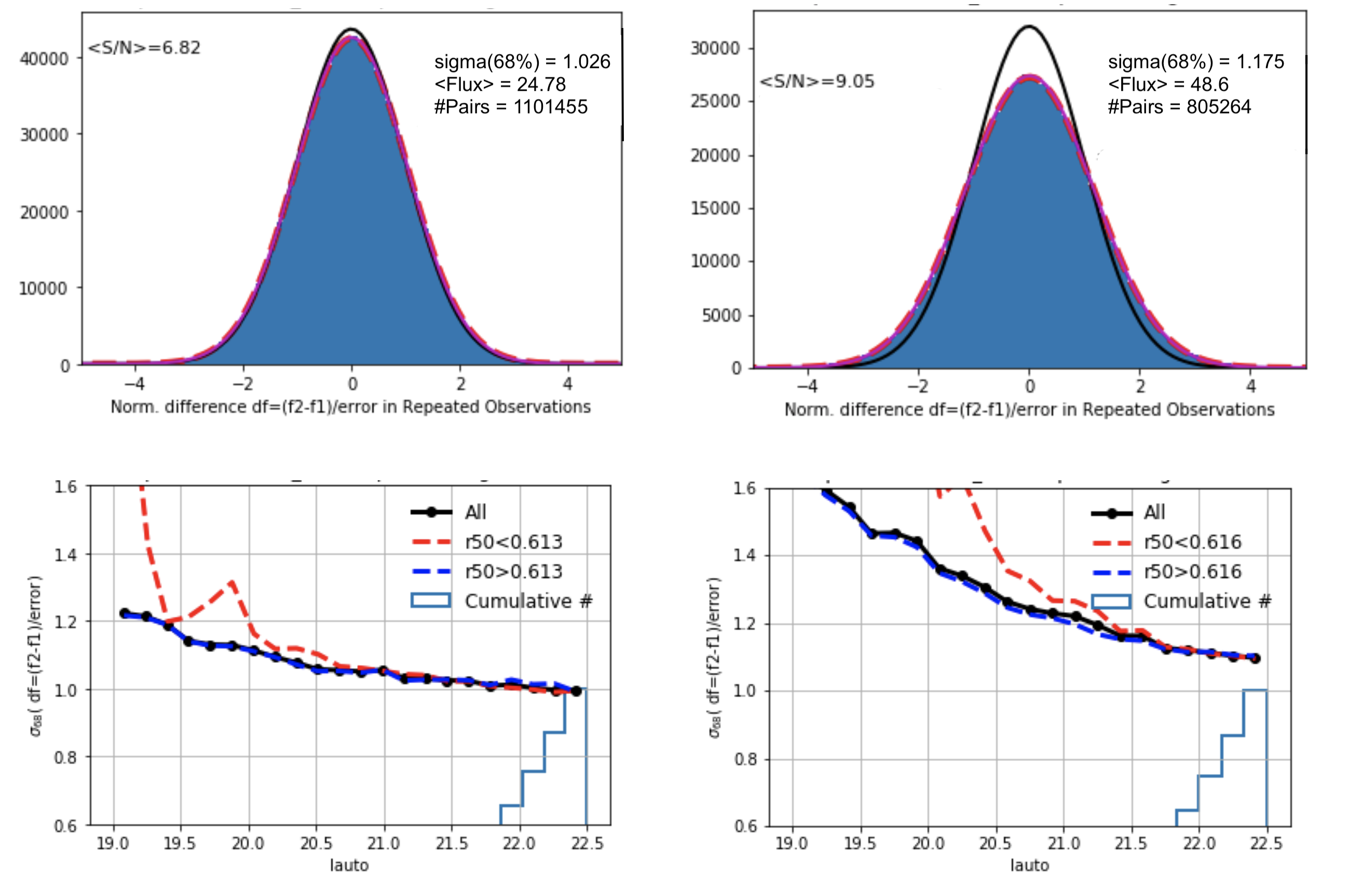}
  \caption{
\emph{Top:} Duplicates for W3 galaxies (run 941) with NB $=575nm$ (\emph{left}) and NB $=755nm$ (\emph{right}). Histograms of values of $df \equiv (f_2- f_1)/\sigma$. There are $\simeq 1.1\cdot10^6$ pairs with mean S/N of 6.82 and mean flux of 24.78  for NB $=575nm$. \emph{Bottom:} $\sigma_{68}$ as a function of the $I_{\rm auto}$ total $i$-band magnitude. Continuous lines correspond to all duplicates. Dashed red line shows values with smaller half light radius $r_{50}<0.616$ arcseconds. The histogram (below the lines) shows the relative number of duplicates which increases sharply at the faint end and dominate the statistics for a full population.}
  \label{fig:dupW3}
\end{figure*}

Repeated exposures over the PAUS fields have been used to validate the final \textsc{MEMBA} photometry. {\blue These repeated measurements correspond to the independent exposures that are co-added in Section \S\ref{s:photometry_coadd}.}
We have between 3-10 independent fluxes for the same object in each narrow-band.
We use these catalogs to build a sample of over a million pairs of duplicate (repeated) measurements of the same object. There are 45 pairs of measurements for the same object and about 40,000 separate objects, before cuts and masking. 
We identified duplicates as objects with the same reference ID (which means apertures with the same position in \textsc{MEMBA}). We select pairs with $SNR>3$. The goal is to test if the uncertainties in fluxes produced by \textsc{MEMBA} are consistent with repeated measurements for the same object as a function of different properties of the object and observation.

The top panels of Figure \ref{fig:dupW3} show two examples of histogram in the values of normalized flux differences
\begin{equation}
df \equiv \frac{f_1- f_2}{\sigma}    =  \frac{f_1- f_2}{\sqrt{\sigma_1^2 + \sigma_2^2}}    
\end{equation}
of duplicate measurements with fluxes  $f_1$ and $f_2$. The error is just added in quadrature from the \textsc{MEMBA} error of  $f_1$ and $f_2$.  
In general, the filled histograms of normalized differences follow a normal distribution (shown by dashed red lines) but the width is typically a bit larger, $\sigma_{68} \simeq 1.028$ and $\sigma_{68} \simeq 1.175$ in these two cases, than unity (black line) which is what we would expect if \textsc{MEMBA} errors were perfectly accurate. 

The bottom panels of Figure \ref{fig:dupW3} show the width of the normalized duplicate distributions, $\sigma_{68}$, as a function of the total broad $i$-band of the reference catalog (W3 in CFHTLS) which we called $I_{\rm auto}$. We can see that there is a strong dependence with $I_{\rm auto}$ which indicates that \textsc{MEMBA} errors are correct at the faint end but are underestimated at the bright end. Similar results are found for other PAUS fields. The machine learning background modelling prototype presented in \cite{bkgnet-cabayol2020} seems to improve this trend for brighter sources, suggesting there is a background subtraction problem in higher SNR sources.

We have seen similar tendencies in the photometry of duplicates for other broad band surveys. This happens for both galaxies or stars and using aperture photometry or also \textsc{SExtractor} total magnitudes (e.g. in the PAUS zero-point calibration with SDSS stars).
We speculate that these results may come from a small biases in zero-point calibration values, which have spatial variations (see e.g. Fig.8 in \cite{paucalibration}). These variations
affect more strongly the brighter fluxes and are not accounted for by error propagation of zero-points to flux in Eq.\ref{eq:caliberror}. A zero-point bias of only 1\% can produce up to 20\% mis-estimation in the flux errors of bright galaxies, which have larger fluxes and smaller relative errors. 

{\blue
In Figure \ref{fig:dupW3nbALL} we show $\sigma_{68}[df]$ for all other narrow-band wavelengths in PAUS. Duplicates are split in 6 magnitude bins: $17.5<I_{auto}<18.5$, $18.5<I_{auto}<19.5$, $19.5<I_{auto}<20.5$, $20.5<I_{auto}<21.5$, $21.5<I_{auto}<22.5$ and $22.5<I_{auto}<23.0$ according to the galaxy total i-band magnitude. The labels show the median value of 
$I_{auto}$ in each bin.
The standard deviation is typically larger than $\sigma_{68}=1$ for brighter galaxies which indicates that flux errors are underestimated in those cases. As explained above this is most likely because of zero point and PSF variations within inidividual CCDs, which are not accounted for in the error estimate. For scientific applications and simulations we interpolate the values in Figure \ref{fig:dupW3nbALL} to correct for this flux error underestimation. But note that the majority of the galaxies are faint and the most interestingly new science cases involved the faintest galaxies.
\begin{figure}
  \centering
  \includegraphics[width=\linewidth]{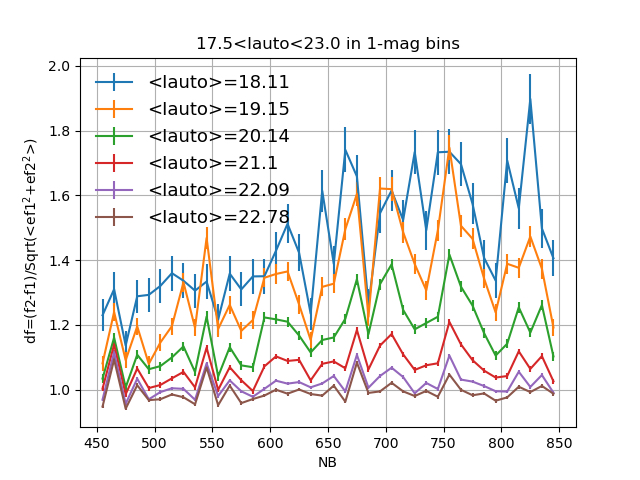}
  \caption{\blue Normalized error (corresponding to the width $\sigma_{68}$ of the histograms of Figure \ref{fig:dupW3} for the latest run 1012 of MEMBA in W3) in $df \equiv (f_2- f_1)/\sigma$  as a function of narrow-band wavelength. Each line corresponds to the $\sigma_{68}[df]$ values for duplicates of galaxies 
  in 6 separate magnitude bins within $17.5<I_{auto}(AB)< 23.0$ in steps of $\Delta I_{auto}(AB) =1$ (the label shows the meadian in the bin). Errorbars correspond to $2-\sigma$ estimates base on the total number of pairs in each bin.}
 \label{fig:dupW3nbALL}
\end{figure}

We have also looked for similar trends of $\sigma_{68}[df]$ as a function of different data reduction inputs (such as zero points and PSF of individual CCDs, airmass, galaxy type, Sernic index, sky background, sky noise, CCD position, scatter light, exposure time, observation time, etc) and find no significant trends. This indicates that the data reduction, the calibration and the aperture corrections that we have applied work very well in general. We do find an error under estimation for small galaxies, with small $r_{50}$ (half light radius), as shown in the bottom panels of Fig.\ref{fig:dupW3}. This is most likely caused by PSF
variations within a given CCD, which are not taken into account in our data reduction.
 }

{\blue 
\section{Photo-z performance}
\label{s:photo-z}

Photometric redshift (or Photo-z)
performance of PAUS data is presented in separate papers (e.g. see \citealt{bcnz-eriksen2019,Eriksen:2020,photoz-alarcon2021,photoz-soo2021}, Navarro-Giron\'es etal in preparation), but it has been a key requirement for the PAUS data reduction  presented here. 

\begin{figure}
  \centering
  \includegraphics[width=\linewidth]{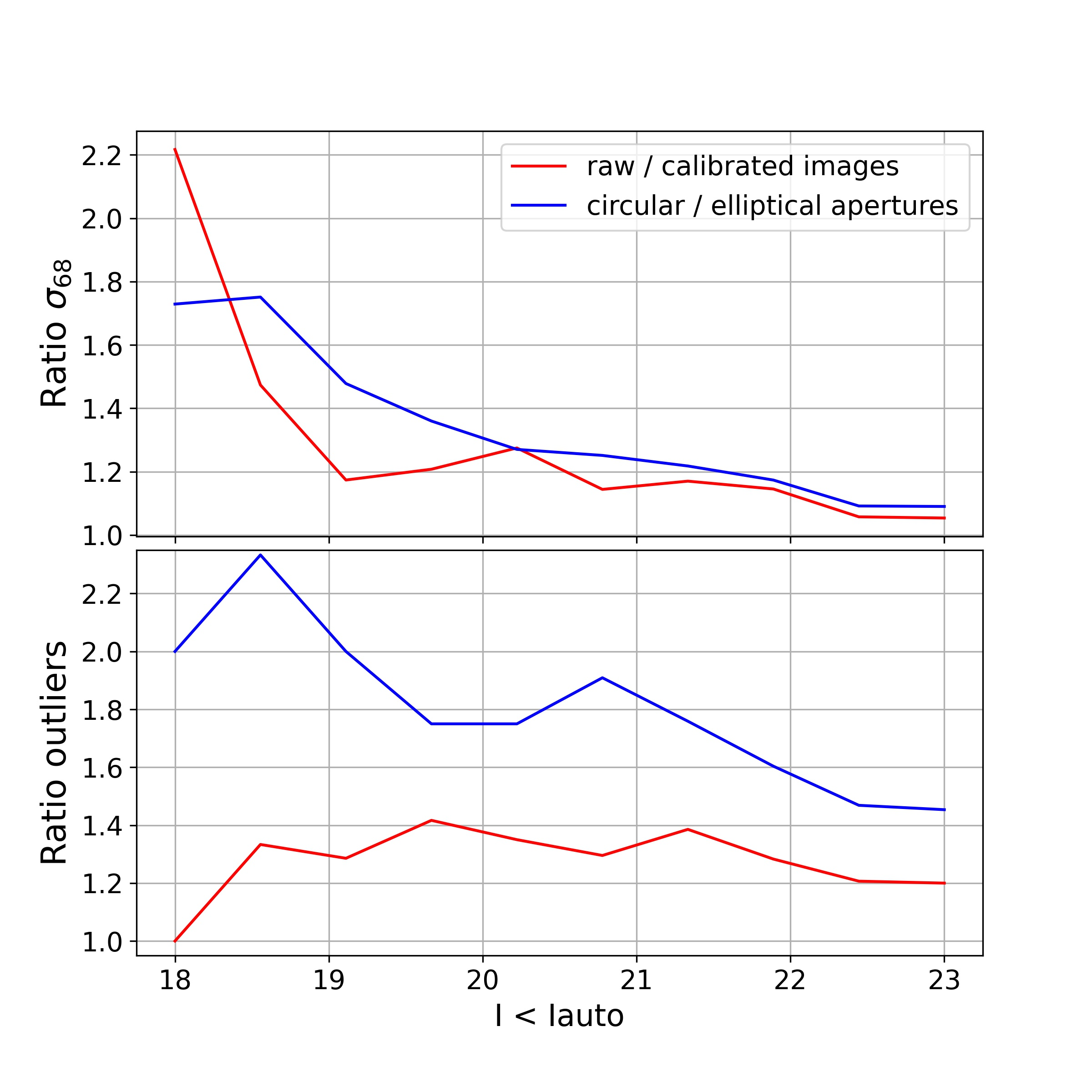}  
  \caption{ \blue TOP: Ratio of the Photo-z performance, given by $\sigma_{68}$ in the scatter of spectroscopic to photometric redshifts in a COSMOS validation sample, as a function of total cumulative i-band magnitude $I<I_{auto}$. The red line compares the ratio from raw images compared to that of calibrated images. The blue line compares the ratio using circular apertures instead of elliptical ones. BOTTOM: the corresponding ratios for the outlier fraction.}
  \label{fig:photo-z} 
\end{figure}

As an illustration of the data reduction improvements, we include some plots of Photo-z performance.
The accuracy of the Photo-z ($z_p$) estimation is defined as the  scatter $\sigma_{68}[dz]$ of the redshift difference $dz=(z_s-z_p)/(1+z_s)$, as compared to an spectroscopic $z_s$ validation sample. We also define the outlier fraction as the fraction of galaxies which has $dz>0.02$. For more details \citealt{bcnz-eriksen2019,Eriksen:2020,photoz-alarcon2021,photoz-soo2021}.

Fig.\ref{fig:photo-z} shows the relative improvement in the Photo-z performance as a function of accumulate $I_{auto}$ magnitude bins.  We compare the ratio of Photo-z performance (top accuracy and bottom outlier fraction) using raw versus calibrated images and also the ratio of using circular versus elliptical apertures. In both cases we see a significant improvement of the Photo-z performance, which validates these steps of the data reduction.

Fig.\ref{fig:ValidateExp} shows 
that the accuracy of $\sigma_{68}[dz]$ 
increases with the number of exposures that we use in the MEMBA coadds. The difference between the top and bottom lines is about a factor 4 in the number of exposures. This translates into an improvement of a factor of 2 in S/N of the NB flux measurements, which is equivalent  to a shift of one magnitude (galaxies that are one magnitude brighter have about 2 times larger S/N). The change in $\sigma_{68}[dz]$ shown in Fig.\ref{fig:ValidateExp} is caused by the degradation in the S/N of the NB measurements as we consider fainter galaxies. The difference in $\sigma_{68}[dz]$ between the top and bottom line roughly correspond to a shift of about one magnitude in the curves, as expected from the corresponding increase in the S/N. This agreement validates our method for adding different exposures, as explained in Section \S\ref{s:photometry_coadd}. It also allows us to predict what type of Photo-z performance we can expect for surveys with different exposures (or collecting power) using our pipeline.

\begin{figure}
  \centering
  \includegraphics[width=0.9\linewidth]{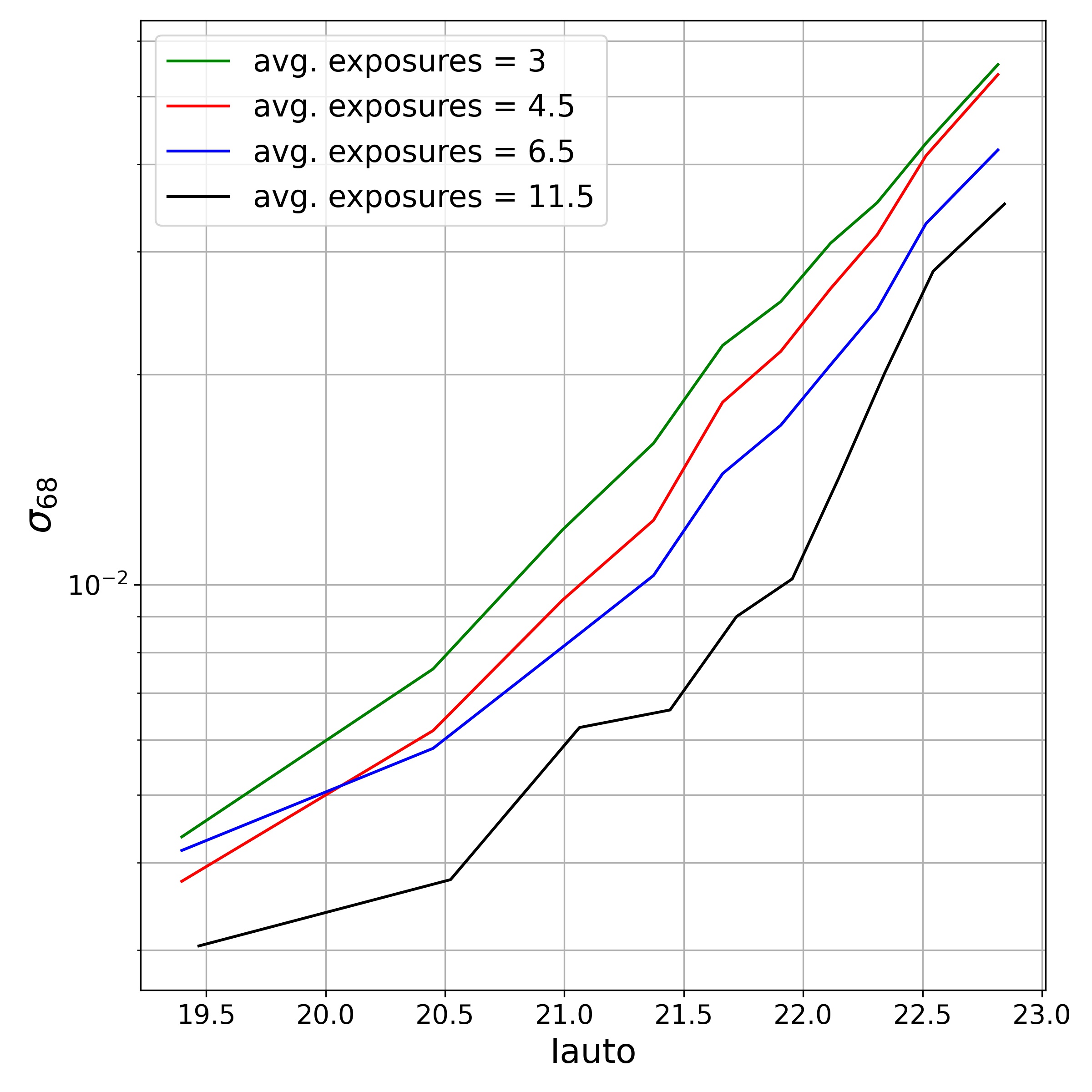}  
  \caption{ \blue Photo-z accuracy ($\sigma_{68}[dz]$) in  $I_{auto}$ magnitude bins for the COSMOS field using different number of repeated exposures.
  The bottom line has twice the S/N in the measured fluxes than the top line. This corresponds to a shift of one magnitude in sensitivity, in good agreement with the results shown.}
  \label{fig:ValidateExp} 
\end{figure}

}

\section{Known limitations \& further work}
\label{s:limitations}

Since the first light of PAUCam we have been constantly improving both \textsc{Nightly} and \textsc{MEMBA} algorithms, continuously improving our understanding of the systematics and the instrument behaviour. Even though we reached outstanding photometry {\blue (see \S\ref{s:validation_spectra}) }
and photo-z accuracy {\blue(see \S\ref{s:photo-z}),}
there are still steps in the overall process that can be improved. However in this paper we refer only to those algorithms that were implemented in the main pipeline and that delivered scientific results in already published papers. We plan to improve and present further algorithms with the corresponding upgrade in scientific results once these are stabilized and validated.

Regarding the flux calibration step in the \textsc{Nightly} pipeline, we know the current dome flat-fields do not reproduce the sky illumination accurately. Therefore the detector response after  the flat-field calibration is not homogeneous. A possible solution is processing sky flats and leave the dome flats for small-scale pixel variations and not to correct the large-scale vignetting and illumination patterns. This has the difficulty of the scatter-light as an additive component which complicates the sky flat processing. The solution proposed in \S \ref{s:detrending_sl_science} has only been applied to particular studies involving extended objects such as the M101 of Figure \ref{fig:SL_science_extended}. Once we verify the fluxes of the main target galaxies (smaller and fainter down to $i_{\rm AB}<23$) are preserved, we will implement the sky-flat scatter-light subtraction to the main processing.

The background modelling implemented in \textsc{MEMBA} is a simple but reliable method. Due to the complex varying background in PAUS images resulting from the flat-field and scatter-light residuals, a more complex background estimator that understands trends of the background can provide significant benefits. This is especially important for low SN sources where small residuals of the background can bias the measurement. A machine learning technique has been studied and published in \cite{bkgnet-cabayol2020} and will soon enter into the main processing of PAUdm. There is also an innovative method developed for PAUS \citep{lumos-cabayol2021} that provides flux estimates from neural network algorithms, improving the accuracy and increasing the signal-to-noise of the measurements. 

The areas in the focal plane with most distortion are currently flagged. There is a possibility to increase the survey efficiency by accurately modelling the PSF at the edges of the focal plane and adapt the apertures to include a larger area of the unvignetted mosaic.

Modelling the PSF and computing the astrometric solution is done at the single-epoch level, independently for each exposure, for simplicity and because it delivers good enough precision. However more stable and accurate solutions can be obtained by computing astrometry and PSF models with multiple overlapping exposures, at the expense of complicating the \textsc{Nightly} processing adding dependency between single-epoch reductions. The astrometric precision would only improve marginally at the sub-pixel level, which could have some benefits for particular applications. But a more accurate and stable varying PSF modelling could improve the photometry and increase the area efficiency, contrary to the simple model we currently use. 

Similar to the previous point, the photometric calibration delivers zero-points independently for each detector image. There are algorithms such as the {\"U}bercalibration that compute a global solution from the overlapping measurements between images, homogenizing the calibration and ensuring a flat response across wide area in the sky. However, as we calibrate against SDSS stars and those have been globally calibrated in a similar process, we would expect this improvement to be minor for PAU, assuming a good match with SDSS photometry.

\section{Summary \& Conclusions}
\label{s:summary}

The PAU data management system described here has been able to provide the most accurate photo-z catalogues available down to $i_{\rm AB}<23$, dealing with very particular aspects of the narrow-band photometry and specifics of the instrument. Custom algorithms were designed to calibrate narrow-bands down to 1\% accuracy. Subtle systematic effects had to be modelled and corrected such as the specialized processing to deal with scatter-light residuals caused by the unusual filter tray disposition of PAUCam. The technical implementation presented has also been challenging due to the large volume and complex data set for this survey. It has been key to orchestrate the processing and metadata around a powerful database with flexibility to modify and extend the processing as needed and allowing very complex analysis that enforced the scientific exploitation of the data. Although some of the adopted solutions are bound to the infrastructure of the data center, it can be adapted to different surveys or hardware configurations with similar volumes of data (below 100 TB and $10^{10}$ database entries). 

The current PAUdm implementation has some limitations that we are currently working to improve. Even under these limitations, the photometric catalogues published by PAUS deliver the most precise photo-z down to  $i_{\rm AB}<23$. PAUS data is available in the Early Data Release \citep{bcnz-eriksen2019}, in the PAUS+COSMOS photo-z catalog \citep{photoz-alarcon2021} and in \cite{photoz-soo2021}. 
{\blue Fig.\ref{fig:validation_pau_calib_galaxies} shows that our pipeline reduction delivers an inter-band photometric calibration of 0.8\% across the 40 narrow-band set.
Fig.\ref{fig:ValidateExp}  shows how the stacking (or co-add) of independent PAU measurements results in a photometric redshift accuracy that scales as expected with the number of exposures. This provides a direct validation of the whole pipeline presented in this paper. It also indicates that the pipeline is optimal as it saturates the performance that can be achieved for a given signal-to-noise in the input data.
  }
These measurements open new windows in various astronomy scientific areas as published in \cite{galaxyfeatures-stothert2018}, \cite{galaxyprops-tortorelli2021}, \cite{pau-ia-johnston2021}, \cite{lymanalpha-renard2021}, \cite{kids-photoz}, \cite{Renard4000} or \cite{PAUS-Euclid2022}
with its large-scale imaging survey of narrow-band photometry where each pixel is a low-resolution spectrum.

\section*{Acknowledgements}
\label{s:ack}

The PAU Survey is partially supported by MINECO under grants CSD2007-00060, AYA2015-71825, ESP2017-89838, PGC2018-094773, PGC2018-102021, SEV-2016-0588, SEV-2016-0597, MDM-2015-0509, PID2019-111317GB-C31 and Juan de la Cierva fellowship and LACEGAL and EWC Marie Sklodowska-Curie grant No 734374 and no.776247 with ERDF funds from the EU Horizon 2020 Programme, some of which include ERDF funds from the European Union. IEEC and IFAE are partially funded by the CERCA and Beatriu de Pinos program of the Generalitat de Catalunya. Funding for PAUS has also been provided by Durham University (via the ERC StG DEGAS-259586), ETH Zurich, Leiden University (via ERC StG ADULT-279396 and Netherlands Organisation for Scientific Research (NWO) Vici grant 639.043.512), Bochum University (via a Heisenberg grant of the Deutsche Forschungsgemeinschaft (Hi 1495/5-1) as well as an ERC Consolidator Grant (No. 770935)), University College London, Portsmouth support through the Royal Society Wolfson fellowship and from the European Union's Horizon 2020 research and innovation programme under the grant agreement No 776247 EWC.  The results published have been also funded by the European Union's  Horizon 2020 research and innovation programme under the Maria Skłodowska-Curie (grant agreement No 754510), the National Science Centre of Poland (grant UMO-2016/23/N/ST9/02963) and by the Spanish Ministry of Science and Innovation through Juan de la Cierva-formacion program (reference FJC2018-038792-I). The PAUS data center is hosted by the Port d'Informaci\'o Cient\'ifica (PIC), maintained through a collaboration of CIEMAT and IFAE, with additional support from Universitat Aut\`onoma de Barcelona and ERDF. P.R. is supported by National Science Foundation of China (grant No. 12073014).

\section*{Data Availability}
 
The data underlying this article are available in the PAUS website, under the Data Releases section, at \url{https://pausurvey.org}. The available catalogues include the Early Data Release (EDR) catalogue and the PAUS+COSMOS photo-z catalogue. The EDR corresponds to PAUS data obtained in the COSMOS field, described and used in \cite{bcnz-eriksen2019}. The PAUS+COSMOS photo-z catalogue contains accurate and precise photometric redshifts in the ACS footprint from the COSMOS field for objects with iAB<23 combining all 40 PAUS bands with 26 broad-bands from the COSMOS2015 catalogue \citep{photoz-alarcon2021}. Further data releases are expected with the photometry presented in this article.




\appendix

\section{Operation and technical performance}
\label{s:app_operation}

The PAUS data management system has been designed to operate in the infrastructure at the Port d'Informaci\`o Cient\'ifica (PIC). In this section we describe the technical aspects of the main scientific pipelines and the tools required to operate the pipeline under the available infrastructure. The solution presented below was not the original design as both the infrastructure and the project required changes since the beginning of the operation. A more technical and infrastructure-oriented description of PAUdm is described in \cite{paudm-ops-tonello2019}. This section also includes updated data and pipeline flows from the ones presented in the technical paper.

\subsection{Archive}
\label{s:app_operation_archive}

The PAUS camera produces $\sim$300GB of raw data per observing night. These are mostly FITS files that contain the exposure images and additional metadata in its header. The data are processed in the main pipelines where more sub-products are generated, multiplying the volumes of raw data. The actual size of the raw archive (until 20A observations) is 42 TB and it is safely archived as a two-copy tape and a third copy on disk. The processed data are significantly larger due to the increased bit depth and the various sub-products generated per exposure (science, mask, weights, PSF models, etc). For the reduced data we have a single copy on disk, except for published releases where we include an additional copy on tape. The raw archive tree organizes the data in observation sets, following the schema in the PAUCam temporary archive at the observatory. The reduced tree adds an additional layer to account for multiple reprocessing of the same data, that we call productions.

Even though most of the access to the archive system is provided by the nodes in the computing farm, we wanted to make available both raw and reduced data to the PAUS Collaboration. For that purpose we set up a WebDAV server that allows web access to the entire archive in a user-friendly format.

\subsection{Database}
\label{s:app_operation_db}

With such a large and complex dataset where millions of galaxies, measurements and images are related, it has been key to set up a relational database that tracks all the information and metadata of the survey.  We have chosen a PostgreSQL database running on a powerful 12-core 96GB server in a twin configuration for reliability and performance. The database can be accessed by the pipelines via an object relational mapper (ORM) for better integration and reliability under the pipeline environment. It is also accessed by the different web applications such as the \textsc{Nightly} report and the forced aperture inspector described in previous sections. As PAUS is a large collaboration, we make available via the PAUdm website a dynamic view of the database to browse the schema and perform simple queries. Additionally, for development and validation purposes, it can be accessed via Python notebooks under a read-only role, dumping queries directly into dataframes with all the flexibility and potential that these objects provide. 

The database model was designed to allow for reprocessing of data at any level, tracked by the \texttt{production} table. Careful constraints were set on each table to ensure unique entries under each production set. We defined 4 main pipelines: The Pixel Simulation, where we produce survey and pixel image simulation for development and assessment of performance. The \textsc{Nightly} pipeline (\S \ref{s:app_operation_flow_nightly}) is where the main image reduction and calibration happens. The \textsc{Nightly} can process input productions from the pixel simulation or real observations from PAUCam. The \textsc{MEMBA} pipeline (\S \ref{s:app_operation_flow_memba}) where we perform forced photometry over the \textsc{Nightly} images. And finally the Photo-z pipeline, a wrapper to \textsc{BCNz} \citep{bcnz-eriksen2019} where we estimate photometric redshifts from \textsc{MEMBA} photometry. Each pipeline can be processed independently, allowing to process multiple times a given input production. For instance, one can process different aperture photometry in \textsc{MEMBA} under various configurations with the same set of image reduction made by the \textsc{Nightly} pipeline. 

In addition to the data-related tables, we have a survey strategy database, synchronized with the one at the observatory, containing all information about fields, exposure status and observation progress. These tables are used throughout the night to schedule targets that need to be observed by PAUCam. It is also necessary in the regular PAUdm processing to select the exposures associated to a survey field and that have been classified as valid exposures.

Finally, the database have all data operation tables with information about job configuration, status and their dependencies. The use of these is described in the next subsection (\S \ref{s:app_operation_bt}). Additionally the whole database schema with the most relevant columns in each table can be found in Appendix \ref{s:dbschema}.

\subsection{Processing}
\label{s:app_operation_bt}

The high volume of data and its complex analysis requires processing the pipelines in a data center with enough computing power such as at the Port d'Informaci\'o Cient\'ifica (PIC). This is a High-Throughput Computing data center and therefore we had to split each pipeline into smaller tasks than can be processed independently with limited consumption of memory and CPU time. Consequently, a pipeline may result in hundreds or even thousands of jobs, with its configuration and inter-dependencies, that need to be launched, monitored and operated. 

For that purpose we designed a job orchestration tool named Brownthrower (BT) that gives us the flexibility to operate the pipelines at PIC. With BT we can create jobs, add dependencies between them so certain jobs do not begin processing until others are complete, share configuration between jobs and monitor the status of a pipeline and its sub jobs. To process these jobs we submit to the computing farm via HTCondor a set of pilot jobs that are continuously grabbing free jobs to be processed (in status Queued and without pending dependencies) until all jobs are being processed. 

A \textsc{Jupyter Lab} web service was recently set at PIC, running over actual nodes from the computing farm (with memory up to 32GB) or even GPU nodes. This service has been of great help to develop and deploy new algorithms, as well as for validation and quick test purposes.

\subsection{Data flow and orchestration}
\label{s:app_operation_flow}

\begin{figure*}
  \centering
  \makebox[\textwidth][c]{\includegraphics[width=\textwidth]{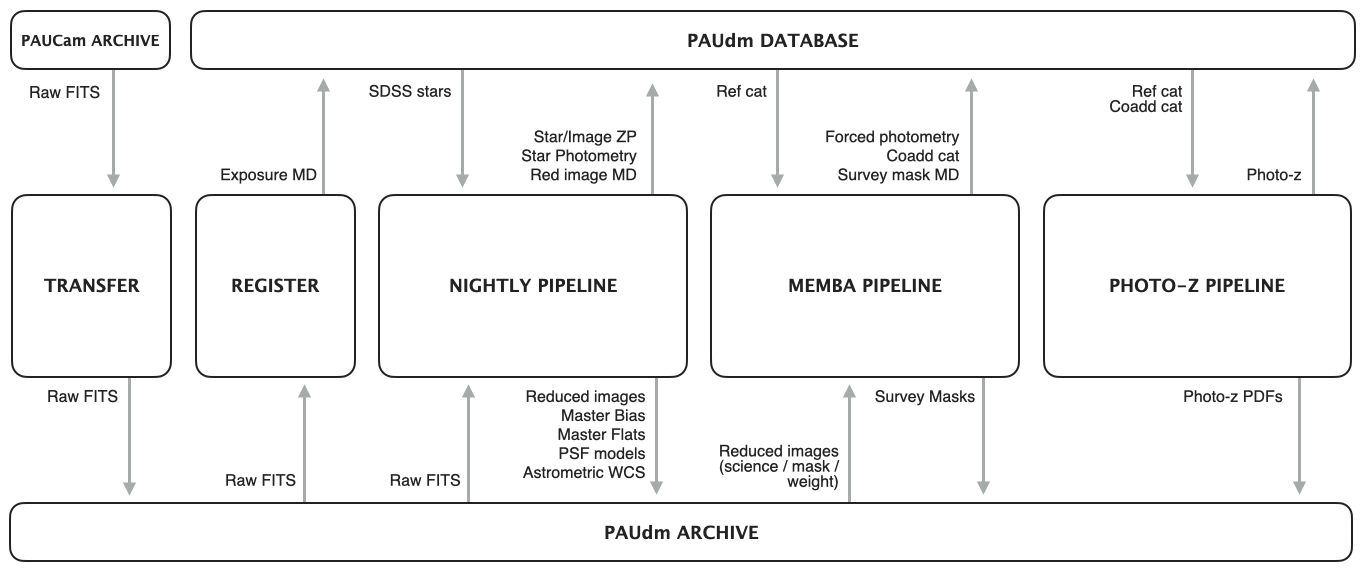}}%
  \caption{A simplified schema of the PAUS data flow, where interactions between pipeline, storage and database are shown.}
  \label{fig:paudm_data_flow}
\end{figure*}

The data flow in PAUdm is divided between the main PAUdm archive, where large files are stored and the database, where metadata and information that requires complex selections is uploaded to the database, orchestrated by the different pipelines. A summary of the data flow is depicted in Figure \ref{fig:paudm_data_flow}. In the first place, the raw data are transferred from the observatory on La Palma to the archive set in Barcelona at PIC (detailed in \cite{paudm-ops-tonello2019}). Immediately, the exposure metadata is registered in the database. Next, the \textsc{Nightly} Pipeline begins its image calibration and archives the clean images, its PSF models and the astrometric solutions in WCS. Photometric measurements and their calibration ZPs are uploaded to the database. Once enough sky area has been processed by the \textsc{Nightly}, \textsc{MEMBA} can begin the galaxy photometry and reingest its coadd catalogue once all measurements have been done. \textsc{MEMBA} is also in charge of producing survey masks and thus, it stores them in the archive. Finally, the photo-z pipeline obtains \textsc{MEMBA}'s measurements and computes the photo-z for each galaxy. The photo-z values and estimated errors enter the database while the large redshift probability distribution files are archived in the storage. 

\subsubsection{Nightly Pipeline}
\label{s:app_operation_flow_nightly}

This is the pipeline in charge of the image calibration, as described in \S \ref{s:detrending}. It begins with a set of raw exposures, including flat and bias calibration images. It ends up with the science exposures astrometrically and photometrically calibrated, ready to perform the flux measurements.

\begin{figure*}
  \centering
  \includegraphics[width=\textwidth]{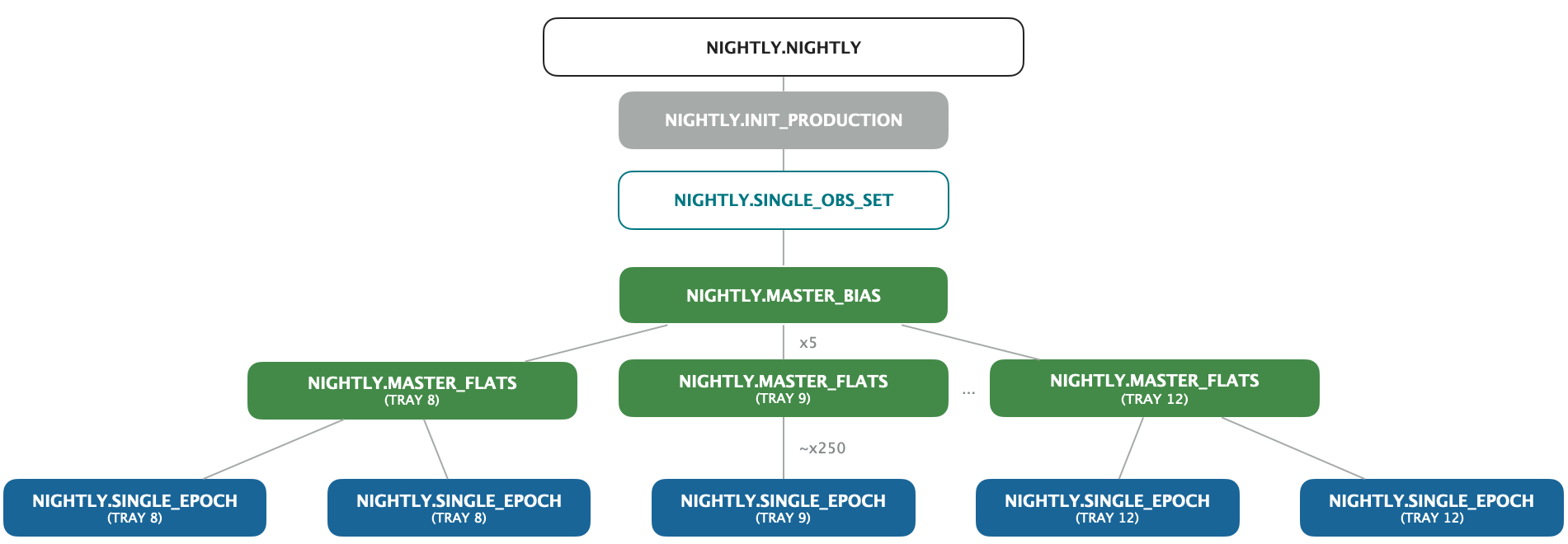}
  \caption{The dependency chart of the \textsc{Nightly} pipeline. Empty boxes define jobs creating sub jobs while filled boxes refer to jobs processing data at the computing farm. The dependency at the filter level starts with the master flat-fields and the single-epochs are related with its own masters.}
  \label{fig:nightly_pipeline}
\end{figure*}

The \textsc{Nightly} pipeline has two main steps. First we have the production of master bias and master flats and secondly the single-epoch reduction of science images. During observation periods, we process the pipeline in batches of observation sets. Typically an observation set contains the exposures from a single night. However, as PAUCam is also available in the WHT as a community instrument, it allows to have more than one observation set per night in cases where observations belong to different surveys. When we operate in observing mode, the \textsc{Nightly} pipeline tree starts with the master bias, then with the master flats of each filter tray and associated to each tray the corresponding sky images. This means that the image calibration won't start until the jobs of master bias and master flat are successfully completed. The pipeline tree can be seen in Figure \ref{fig:nightly_pipeline}.

There is a second mode to operate the \textsc{Nightly} pipeline meant to process entire fields from multiple nights. This is used when we improve the \textsc{Nightly} code and want to reprocess a given subset. First we process all calibration frames (master bias and master flats) from the nights with science exposures that we plan to process. At a second stage, once all the calibration frames are available, we analyze all the science images in parallel.

The pipeline has evolved significantly since the beginning as we had to deal with heterogeneous data sets such as very cloudy skies, observation sets without calibration images, saturated flats, etc. A major effort had to be done to automatically detect any possible situation (detailed in \S \ref{s:validation_qc}) and either correct for it or classify the faulty data as invalid.

To allow for the 8-hour rapid feedback during an observation period, we launch 50 BT pilots to the computing farm that process all images in time. Reprocessing entire fields involves a much greater set of images and therefore we increase the number of pilots to 100. Master bias and master flat jobs are processed in 10 minutes while single-epoch exposure reduction can take up to 20 minutes per job.

The \textsc{Nightly} pilots require intense I/O access to the archive system to retrieve raw data and ingest reduced images and for this reason we do not allow more than 100 \textsc{Nightly} pilots to run in parallel. 

\subsubsection{MEMBA Pipeline}
\label{s:app_operation_flow_memba}

\begin{figure*}
  \centering
  \includegraphics[width=0.8\textwidth]{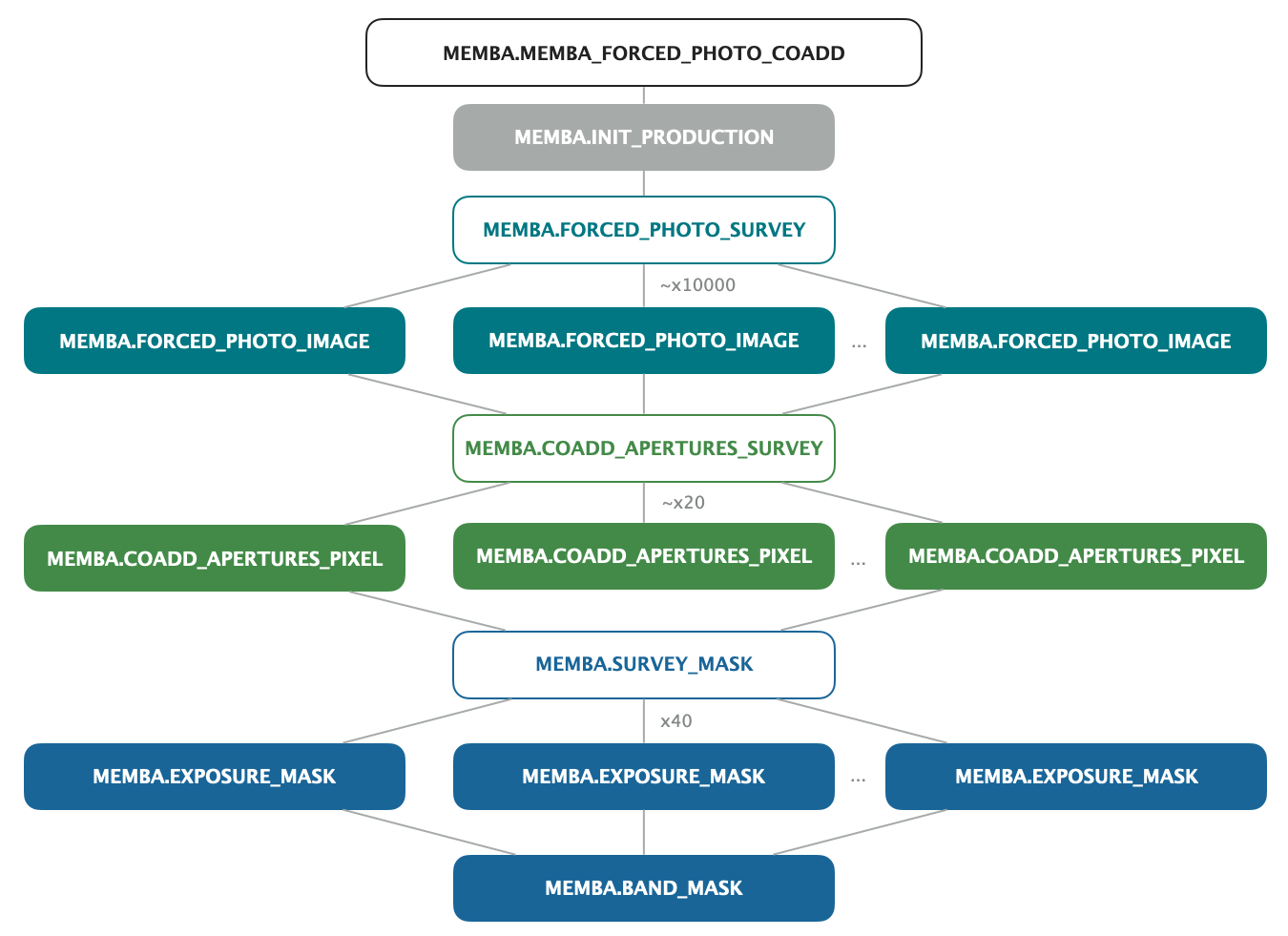}
  \caption{The dependency chart of the \textsc{MEMBA} pipeline. Empty boxes define jobs creating sub jobs while filled boxes refer to jobs processing data at the computing farm. The parallelization takes place at three levels: forced photometry, coadds and survey masks.}
  \label{fig:memba_pipeline}
\end{figure*}

The Multi-Epoch and Multi-Band Analysis pipeline is intended to perform the photometry over the reduced images across the different bands and overlapping exposures from different epochs. It can be divided into three main steps: the forced photometry, the coadd catalogue and the production of survey masks. The pipeline tree can be seen in Figure \ref{fig:memba_pipeline}.

Each forced photometry job takes care of running the photometry of a single detector image. It will load the reference catalogue, the corresponding reduced image and mask and will upload the measurements to the database once completed. 

The coadd jobs are divided in different areas in the sky. We select the areas by \textsc{HEALPix} pixels \citep{healpix-gorski2005} of N side 128. This approach allows us to limit the load of each job, retrieving only the overlapping measurements of a reduced area, defined by the pixel size. Coadding a larger field would only increase the number of jobs but those will always be constrained in memory and processing time.

The dependency of jobs is set between the forced photometry tasks and the coadds, as the latter requires the photometry to be complete at all bands and layers prior to the combination. On the contrary, mask jobs can be processed independently as these do not require inputs from the forced photometry or coadd tasks.  

As \textsc{MEMBA} jobs do not have such intense access to the archive and interact mostly with the database, we can increase the number of parallel pilots up to 200. The CPU time of a \textsc{MEMBA} run is dominated by the forced photometry jobs. The largest fields are made of > 30.000 images. Each job lasts approximately 15 minutes, resulting in a total process time of one day for every 10 deg$^2$.

\section{Database Schema}
\label{s:dbschema}

\begin{figure*}
  \centering
  \includegraphics[width=0.7\textwidth]{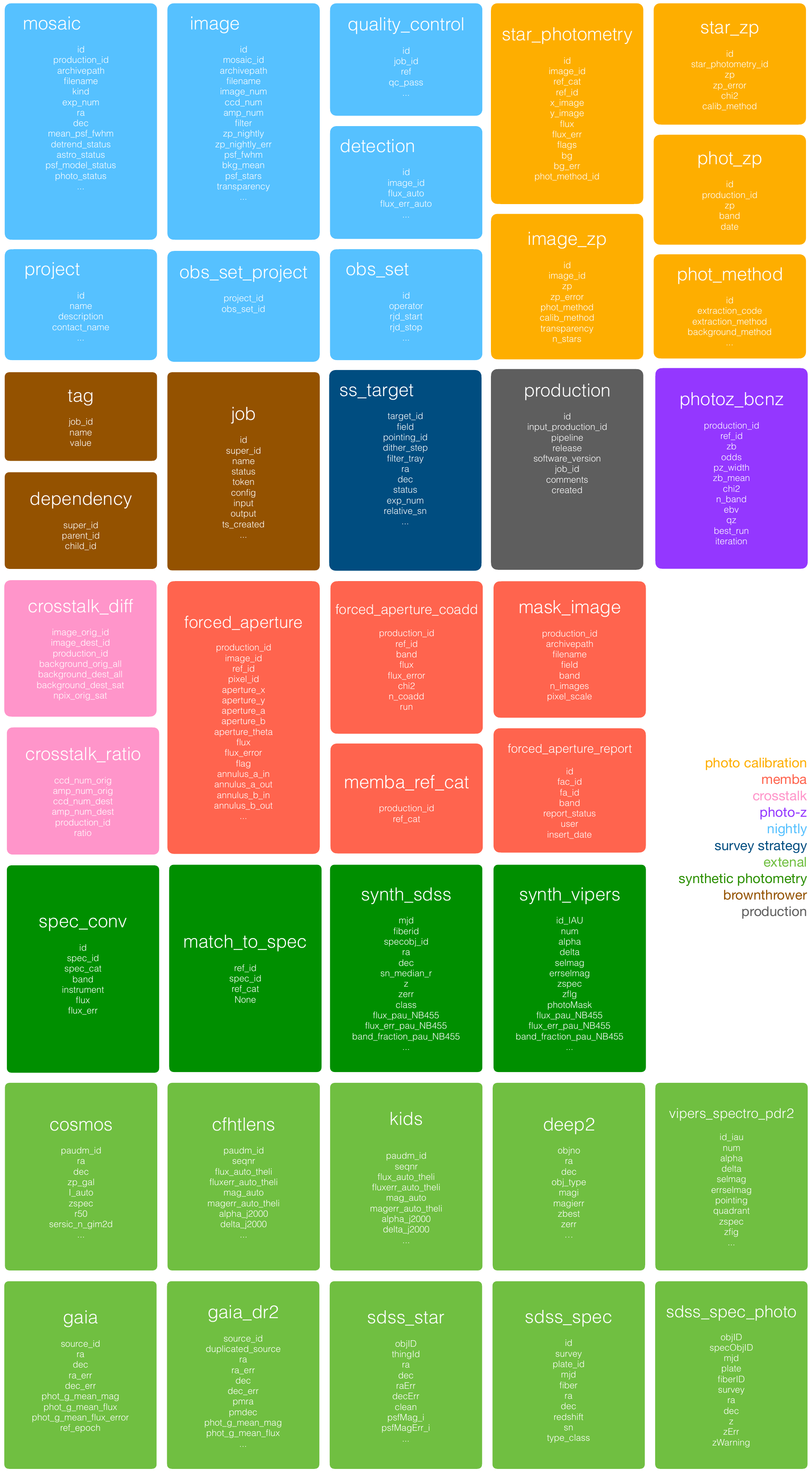}
  \caption{PAUS data base schema. Inside each table with show some example columns. Tables
  are grouped under the same color.}
\label{fig:schema}
\end{figure*}

\begin{figure*}
  \centering
  \includegraphics[width=0.9\textwidth]{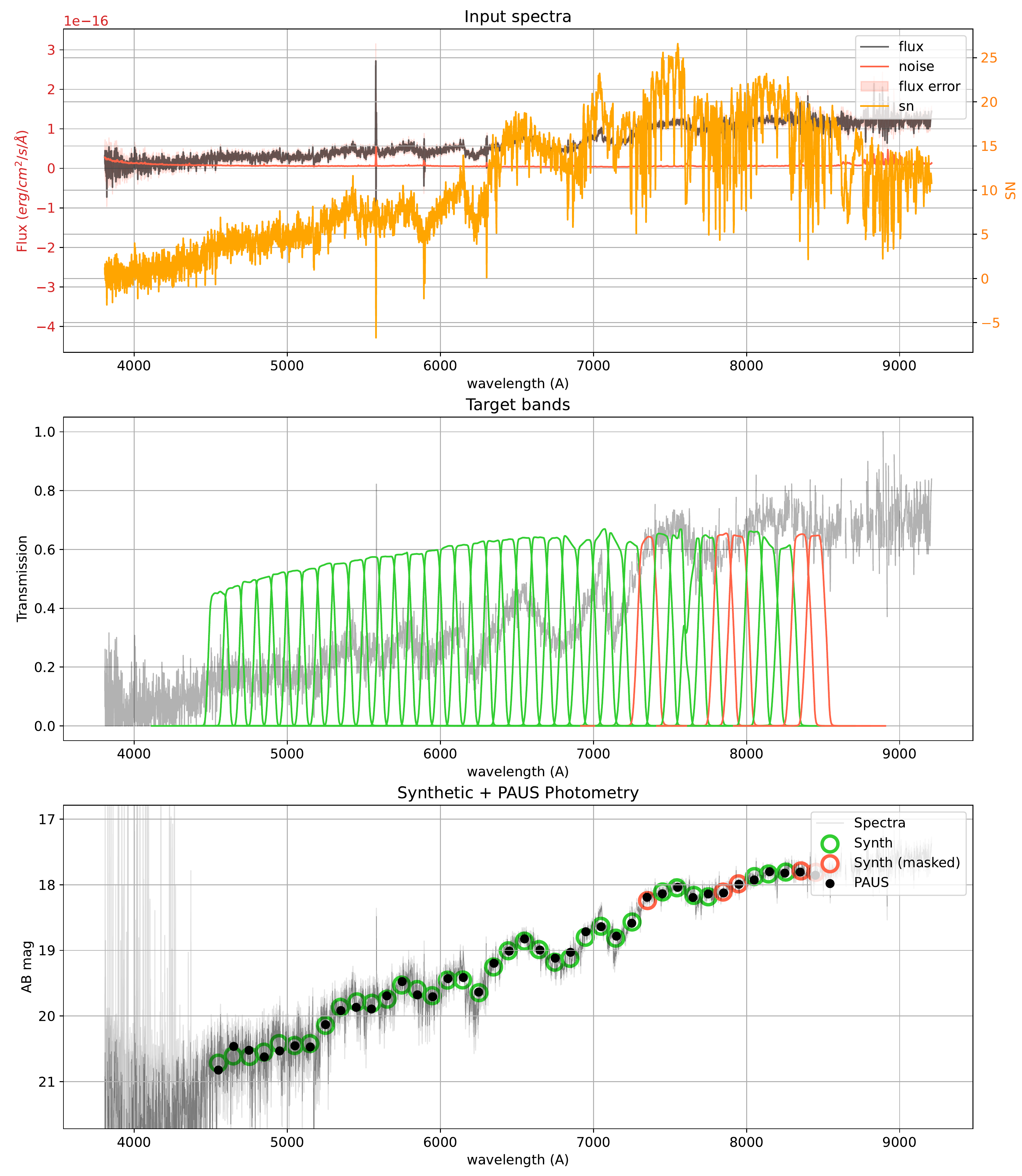}
  \caption{A M3 star observed by PAUS (COSMOS-79081) with reference SDSS Spectrum (Plate 501 - MJD 52235 - FiberID 461 - SNR 8.95).}
  \label{fig:C1}
\end{figure*}

{\blue
In this appendix we describe the PAUS database schema. In 
Table \ref{tab:schema} we group different tables in each pipeline, which are given by different colors in Fig.\ref{fig:schema}. 

\begin{table*}
{\blue
\begin{tabular}{lll}
\hline \\
\textbf{Pipeline} & \textbf{Database} & \textbf{Description} \\
\hline \\
Survey Strategy\\
        & \textbf{ss\_target} &  Survey Strategy targets from observations \\
\textsc{Nightly} \\            
        &\textbf{detection} & detections on the image after the \textsc{Nightly} data reduction \\
        & \textbf{image} & list of images associated to the mosaics (CCD and single amplifier images) \\
        &\textbf{mosaic} & list of mosaic exposure images (raw and reduced) \\
        & \textbf{obs\_set} & list of observation sets registered in the database \\
        & \textbf{obs\_set\_project} & Projects associated to observation sets \\
        & \textbf{project} & List of projects associated to PAUCam observations \\
        & \textbf{quality\_control} & quality control entries \\
Crosstalk \\
        & \textbf{crosstalk\_diff} & Crosstalk differences between raw images \\
        & \textbf{crosstalk\_ratio} & Crosstalk ratios between amplifiers \\        
Photometric \\ Calibration \\ 
        & \textbf{image\_zp} & image zero-point measurements for each photometry-calibration method \\ 
         & \textbf{phot\_method} & Method used during photometry for image calibration \\
        & \textbf{phot\_zp} &  Photometric zero-points per CCD\\
        &\textbf{star\_photometry} & photometry for each star in the \textsc{Nightly} photometry \\
        & \textbf{star\_zp} &  zero-point measurements for each star in the \textsc{Nightly} photometry \\   
\textsc{MEMBA} \\
        & \textbf{forced\_aperture}& Single-epoch forced aperture photometry \\
        & \textbf{forced\_aperture\_coadd} & coadd forced aperture fluxes per band \\
        & \textbf{forced\_aperture\_report}&  Reports for forced aperture inspector \\
        &\textbf{mask\_image} &  list of mask images (band and field) \\
        &\textbf{memba\_ref\_cat} & Reference catalogue used in each \textsc{MEMBA} production \\
 Photo-z \\
        & \textbf{photoz\_bcnz}&  Photometric redshifts from BCNz code \\
External \\
        &\textbf{cfhtlens} &  External CFHTLenS catalogue for forced photometry \\
        &\textbf{cosmos} &  zCOSMOS (DR3). Redshifts for forced photometry and validation \\
        &\textbf{deep2} &  The DEEP2 DR4 redshift catalog \\
        &\textbf{gaia\_dr2} &  Gaia DR2 stellar catalogue \\
        & \textbf{kids} &  KiDS KV450-G9 reference catalogue \\
        &\textbf{sdss\_spec}&  SDSS Spectra catalogue \\
        & \textbf{sdss\_spec\_photo} &  SDSS DR12 (Spec\_Photo view). Spectrum for forced photometry and validation \\
        & \textbf{sdss\_star} &  SDSS DR12 (Star view). Stars for simulation and calibration \\
 Synthetic \\ Photometry \\  
        &\textbf{match\_to\_spec}& Match table between forced aperture catalogues and spectra catalogues \\
        &\textbf{spec\_conv}& Convolved fluxes derived from spectra in external surveys (SDSS, COSMOS and DEEP2) \\
        & \textbf{synth\_sdss}& Synthetic photometry over SDSS spectra (over COSMOS and W1) \\
        &\textbf{synth\_vipers}& Synthetic photometry over VIPERS spectra (over W1) \\
Brownthrower \\ (Operation tables) \\ 
        &\textbf{dependency} & Tracks the dependency between Brownthrower jobs \\
        &\textbf{job} &  Tracks the list of Brownthrower computing jobs (Operation table) \\
         & \textbf{tag} & Contains tags for Brownthrower jobs (Operation table) \\
Production \\         
             & \textbf{production} & Tracks the different processing production runs for all pipelines \\
\hline
\end{tabular}
\caption{DATABASE SCHEMA.}
\label{tab:schema}
}
\end{table*}

\section{Synthetic spectra against PAUS photometry examples}
\label{s:app_synth}

In this section we include some interesting reference synthetic spectra samples (Fig.\ref{fig:C1}-\ref{fig:C4}) against the PAUS narrow-band measurements after all the processing described in this study. This highlights only a star, a galaxy and a QSO but it illustrate the possibilities of PAUS and the validation with this synthetic reference method.
}

\begin{figure*}
  \centering
  \includegraphics[width=0.9\textwidth]{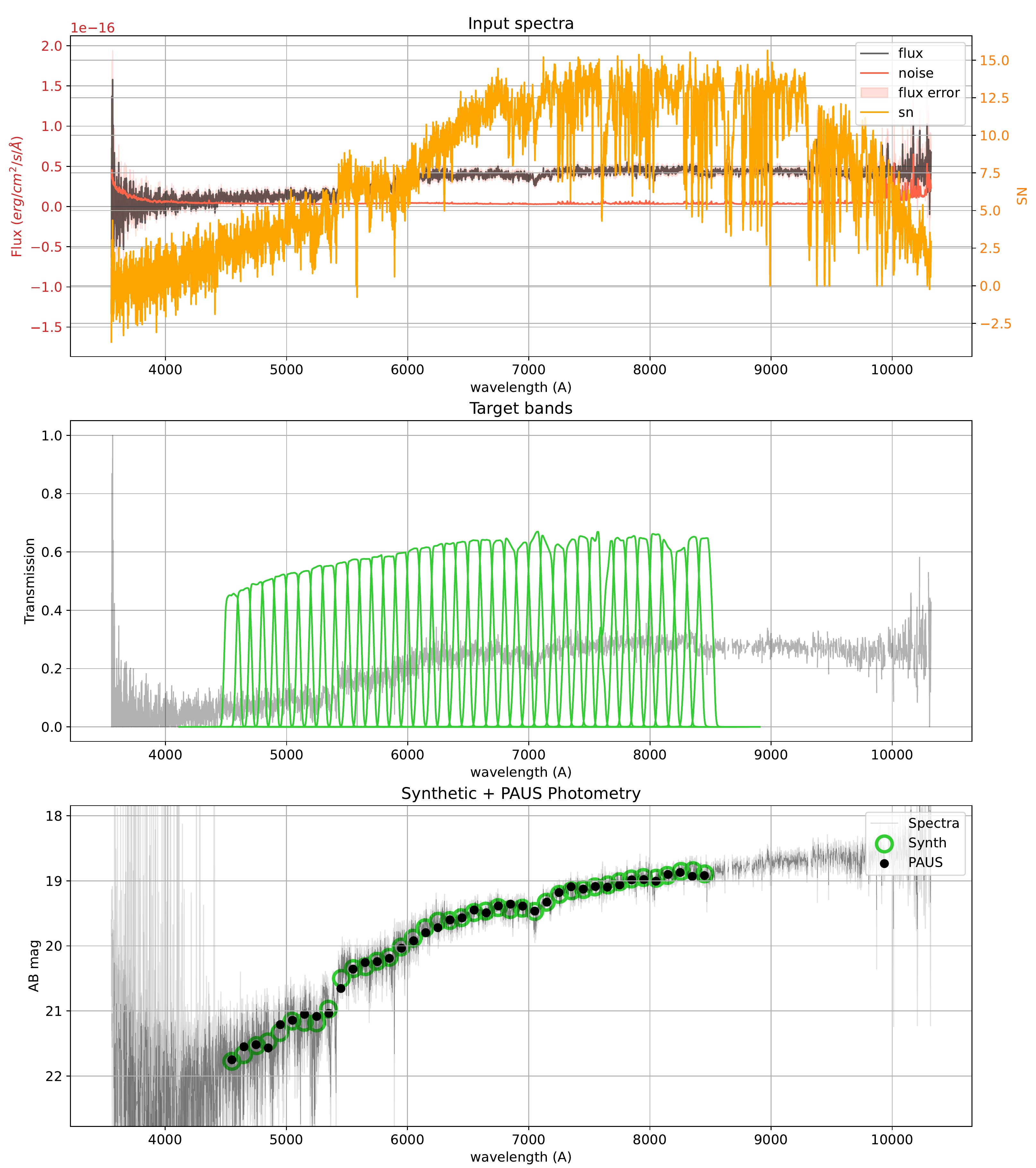}
  \caption{A red galaxy observed by PAUS (COSMOS-3956) with reference SDSS Spectrum (Plate 4737 - MJD 55630 - FiberID 58 - SNR 8.37) at redshift 0.362.}
\end{figure*}

\begin{figure*}
  \centering
  \includegraphics[width=0.9\textwidth]{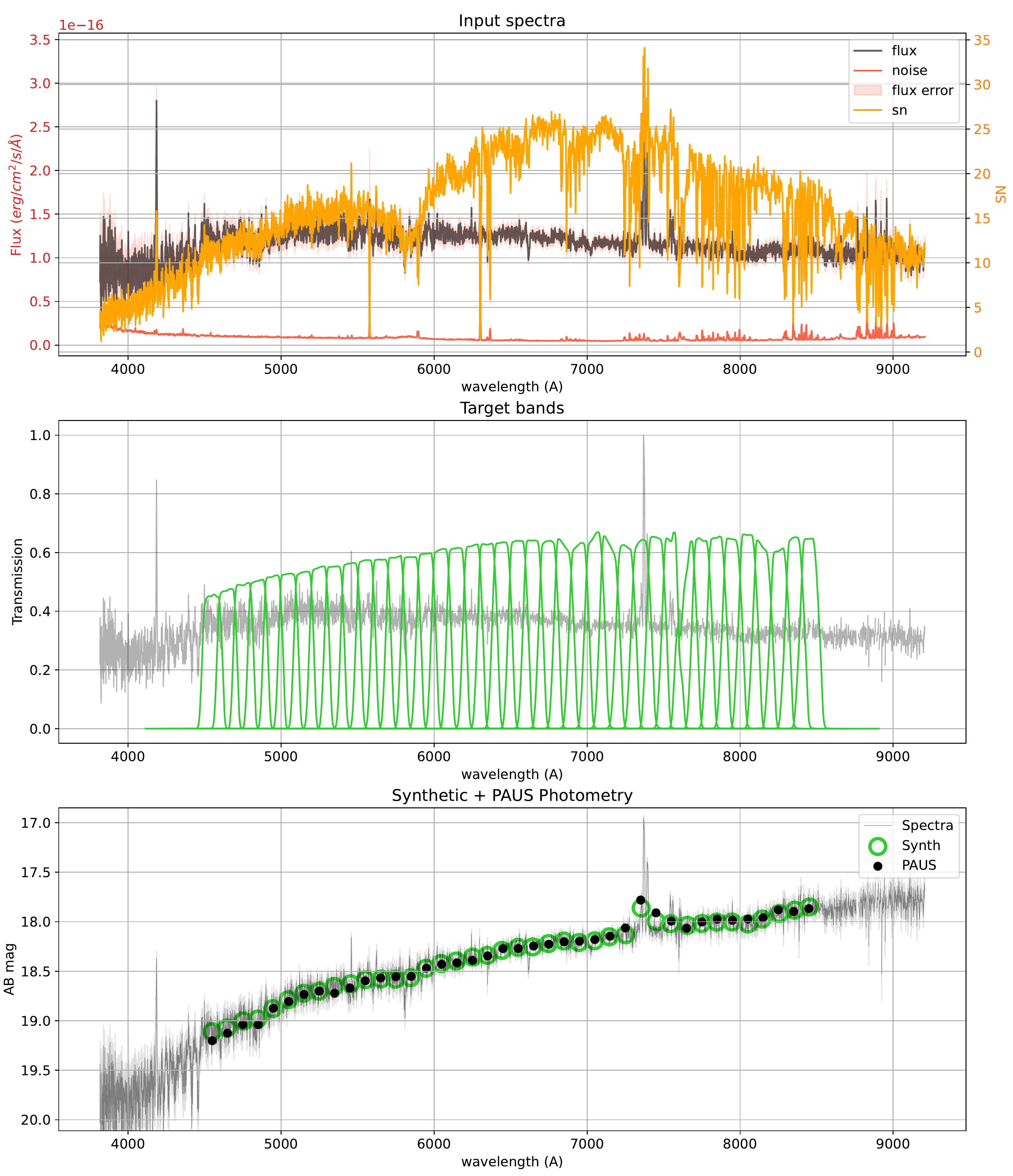}
  \caption{An H$\alpha$ star-forming galaxy observed by PAUS (COSMOS-67024) with reference SDSS Spectrum (Plate 500 - MJD 51994 - FiberID 587 - SNR 19.41) at redshift 0.122.}
\end{figure*}

\begin{figure*}
  \centering
  \includegraphics[width=0.9\textwidth]{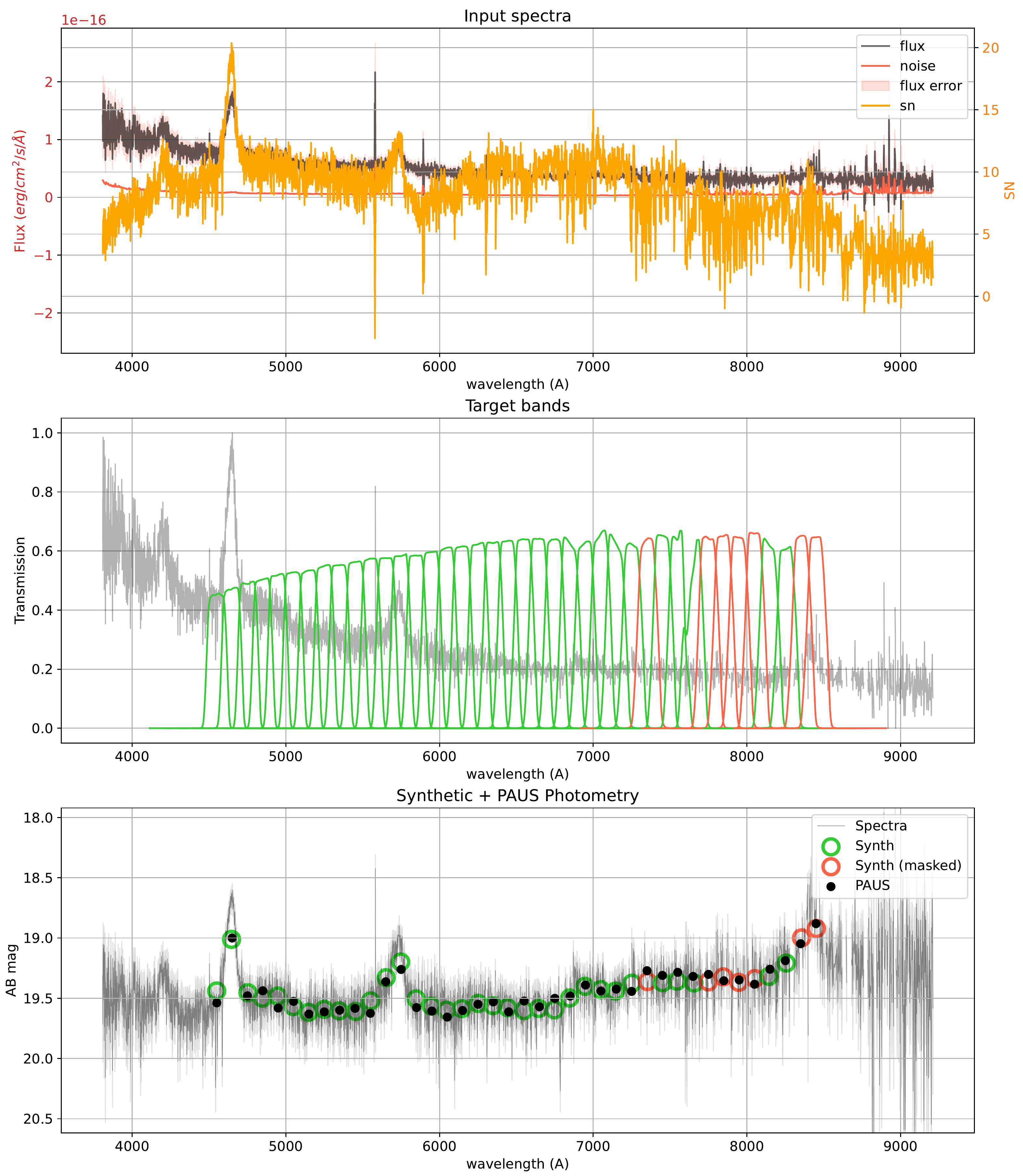}
  \caption{A QSO observed by PAUS (COSMOS-80935) with reference SDSS Spectrum (Plate 501 - MJD 52235 - FiberID 462 - SNR 9.31) at redshift 2.00.}
  \label{fig:C4}  
\end{figure*}

\clearpage

\section{Flagging}
\label{s:flags}

In this section we describe in Table \ref{tab:D1} the list of flags that any source can have throughout the whole data processing of the PAUS data management system. Each flag is assigned to a bit such that with a single integer we can obtain the unique list of flags affecting each source.

\begin{table}
\begin{tabular}{llll}
\hline
\textbf{Flag}   & \textbf{Value (bit)} & \textbf{Origin}    & \textbf{Level} \\
\hline
Crowded         & 1 (1)                & SExtractor         & Source         \\
Merged          & 2 (2)                & SExtractor         & Source         \\
Halo            & 4 (3)                & SExtractor         & Source         \\
Truncated       & 8 (4)                & SExtractor         & Source         \\
Deblended       & 16 (5)               & SExtractor         & Source         \\
Crosstalk       & 32 (6)               & Nightly Mask       & Pixel          \\
scatter-light & 64 (7)               & Nightly Mask       & Pixel          \\
Extinction      & 128 (8)              & Nightly Photometry & Image          \\
zero-point       & 256 (9)              & Nightly Photometry & Image          \\
Cosmetics       & 512 (10)             & Nightly Mask       & Pixel          \\
Saturated       & 1024 (11)            & Nightly Mask       & Pixel          \\
Cosmics         & 2048 (12)            & Nightly Mask       & Pixel          \\
Vignetted       & 4096 (13)            & Nightly Mask       & Pixel          \\
Discordant      & 8192 (14)            & MEMBA Photometry   & Source         \\
Edge            & 16384 (15)           & MEMBA Photometry   & Source         \\
Distortion      & 32768 (16)           & MEMBA Photometry   & Source         \\
Noisy           & 65536 (17)           & MEMBA Photometry   & Source         \\
Astrometry      & 131072 (18)          & MEMBA Photometry   & Source         \\
\hline
\end{tabular}
\caption{The list of flags used across the PAUS data management system. The table specifies the flag reason, the assigned value and bit, the software/pipeline origin and the level at which the flag is defined.}
\label{tab:D1}
\end{table}


\bsp	
\label{lastpage}
\end{document}